\documentclass[a4paper,12pt,titlepage,oneside]{kuphdthesis}

\usepackage{url}
\usepackage[bottom]{footmisc}
\usepackage[caption=false,font=footnotesize]{subfig}
\usepackage{amsmath}
\usepackage{amssymb}
\usepackage{amsfonts}
\usepackage{breakcites}
\graphicspath{{./Figures/}}

\usepackage{xcolor}
\usepackage{comment}
\usepackage{pdfpages}
\DeclareMathOperator*{\dprime}{\prime \prime}
\DeclareMathOperator\erf{erf}
\newcommand\ddfrac[2]{\frac{\displaystyle #1}{\displaystyle #2}}
\usepackage{longtable}

\usepackage{xspace}
\usepackage{epstopdf}

\newcommand \COMMENT  [1] {}       % comment out the contents

\begin{document}

\Title{Electro-Mechanical Contact Interactions Between Human Finger and Touchscreen Under
Electroadhesion} \Author{Easa AliAbbasi} \Year{November 22, 2023}
\Program{Computational Sciences and Engineering}

\TTitle{Doktora Tez Ba\c{s}l{\i}\u{g}{\i}} \TYear{22 Kasım 2023}
\TProgram{Hesaplamalı Bilimler ve M\"{u}hendislik}

\Signature{Prof. Dr. Cagatay Basdogan (Advisor)}
\Signature{Prof. Dr. Philippe Lefevre}
\Signature{Assist. Prof. Dr. Michaël Wiertlewski}
\Signature{Assoc. Prof. Dr. Sedat Nizamoglu}
\Signature{Assoc. Prof. Dr. Mehmet Sayar}

\prelimpages
\titlepage

\thesissignaturepage
\dedication{To my dad, mom, sister, and brother \\
To my beloved fiancée}
\abstract{
Electroadhesion is a promising technology with potential applications in robotics, automation, space missions, textiles, tactile displays, and some other fields where efficient and versatile adhesion is required. However, a comprehensive understanding of the physics behind it is lacking due to the limited development of theoretical models and insufficient experimental data to validate them. In this thesis, we have developed an electro-mechanical model to estimate the magnitude of electrostatic forces between human finger and touchscreen under electroadhesion. We also measured the friction forces between the finger and touchscreen to infer the magnitude of electrostatic forces experimentally. The model is in good agreement with the experimental data and showed that the change in magnitude of the electrostatic force is mainly due to the leakage of charge from the Stratum Corneum layer of the skin to the touchscreen at frequencies lower than 250 Hz and electrical properties of the Stratum Corneum at frequencies higher than 250 Hz. In addition, we proposed a new and systematic approach based on electrical impedance measurements, where skin and touchscreen impedances are measured and subtracted from the total impedance to obtain the remaining impedance in order to estimate the electrostatic forces between the finger and the touchscreen. This approach also marks the first instance of experimental estimation of the average air gap thickness between human finger and voltage-induced capacitive touchscreen. Moreover, the effect of electrode polarization impedance on electroadhesion was investigated. Precise measurements of electrical impedances confirmed that electrode polarization impedance exists in parallel with the impedance of the air gap, particularly at low frequencies, giving rise to the commonly observed charge leakage phenomenon in electroadhesion. We also investigated tactile perception by electroadhesion for DC and AC voltage signals applied to the touchscreen using ten participants with varying finger moisture levels. Our study showed that the voltage detection threshold for an AC signal was significantly lower than that of the corresponding DC signal and we explained this discrepancy by charge leakage at lower frequencies again. We have also observed that the participants with a moist finger had significantly higher threshold levels than the rest of the participants, which is supported by our electrical impedance measurements. Finally, we aimed to investigate the effect of touchscreen's top coating layer on our tactile sensing with and without electroadhesion, and within the time frame of this thesis, we have focused on the latter only. Hence, we first performed psychophysical experiments to quantify human tactile discrimination ability of touchscreen surfaces coated with different materials, followed by multiple physical measurements. The results showed that coating material has a strong influence on our tactile perception and human finger is capable of detecting differences in surface chemistry due to, possibly, molecular interactions. In conclusion, the findings of this thesis provide new insights into the physics of finger-touchscreen interactions under electroadhesion and have implications for the design of robotic systems and haptic interfaces utilizing this technology.
}
\oz{
Elektroadezyon, robotik, otomasyon, uzay görevleri, tekstil, dokunsal ekranlar gibi çok yönlü ve etkili yapışmanın önemli olduğu alanlarda potansiyel uygulamaları olan umut verici bir teknolojidir. Bununla birlikte, teorik modellerin sınırlı gelişimi ve bunları doğrulamak için yetersiz deneysel veri üretilmesi nedeniyle, arkasındaki fizik hala tam olarak anlaşılamamıştır. Bu tez çalışmasında, elektroadezyon altında parmak ile dokunmatik bir ekran arasındaki elektrostatik kuvvetlerin büyüklüğünü tahmin etmek için elektro-mekanik bir model geliştirdik. Elektrostatik kuvvetlerin büyüklüğünü deneysel olarak anlamak için parmak ile dokunmatik ekran arasındaki sürtünme kuvvetlerini de ölçüyoruz. Geliştirdiğimiz model, deneysel verilerle iyi bir uyum içindedir ve elektrostatik kuvvetin büyüklüğündeki değişimin temel olarak parmağın Stratum Corneum katmanından 250 Hz'den düşük frekanslarda dokunmatik ekrana yük sızıntısından ve dokunmatik ekranın elektriksel özelliklerinden kaynaklandığını göstermektedir. Ayrıca, parmak ile dokunmatik ekran arasındaki elektrostatik kuvvetleri elektriksel empedans ölçümlerine dayanan bir yöntem ile tahmin etmek amacıyla, toplam empedansdan parmağın ve ekranın empedanslarının çıkarıldığı ve dolayısıyla kalan empedansın hesaplandığı yeni bir yaklaşım öneriyoruz. Bu yaklaşım, parmak ile iletken katmanına voltaj uygulanan kapasitif dokunmatik ekran arasındaki ortalama hava boşluğu kalınlığının deneysel tahminini yapan ilk çalışmadır ki bu kalınlık elektrostatik kuvvetlerin kestirilmesi için önemlidir. Ayrıca elektrot polarizasyon empedansının elektroadezyon üzerindeki etkisi araştırılmıştır. Empedans ölçümlerimiz, elektrot polarizasyon empedansının, özellikle düşük frekanslarda, hava aralığının empedansına paralel olarak var olduğunu ve yukarıda bahsi geçen yük sızıntısına (parmaktan ekran yüzeyine) yol açtığını göstermektedir. Ayrıca, bu tezde, değişen parmak nem seviyelerine sahip on katılımcıyı kullanarak dokunmatik ekrana uygulanan DC ve AC voltaj sinyalleri için dokunsal algıyı da araştırıyoruz. Çalışmamız, bir AC sinyali için voltaj algılama eşiğinin, karşılık gelen DC sinyalininkinden önemli ölçüde düşük olduğunu göstermektedir. Bu tez kapsamında geliştirdiğimiz elektrik modelimiz bu farklılığın, düşük frekanslardaki yük sızıntısından dolayı meydana geldiğini işaret etmektedir. Ayrıca parmakları nemli olan katılımcıların geri kalan katılımcılara göre önemli ölçüde daha yüksek eşik seviyelerine sahip olduklarını gözlemledik ve bu gözlem de elektriksel empedans ölçümlerimizle desteklenmektedir. Son olarak, dokunmatik ekranın üst kaplama katmanının elektroadezyonlu ve elektroadezyonsuz dokunma duyumuz üzerindeki etkisini araştırmayı amaçladık ve bu tezin zaman çerçevesi içerisinde ikincisine odaklandık. Bu nedenle, ilk önce farklı malzemelerle kaplanmış dokunmatik ekran yüzeylerinde insanın dokunsal ayrım yeteneğini ölçmek için önce psikofiziksel deneyler, ardından çoklu fiziksel ölçümler yaptık. Deneysel sonuçlar, kaplama malzemesinin dokunsal algımız üzerinde güçlü bir etkiye sahip olduğunu ve insan parmağının, muhtemelen moleküler etkileşimler nedeniyle yüzey kimyasındaki farklılıkları tespit edebildiğini göstermektedir. Sonuç olarak, bu tezin bulguları, elektroadezyon altında parmak-dokunmatik ekran etkileşimlerinin fiziğine yeni bakış açıları sağlamakta ve elektroadezyon kullanan robotik sistemlerin ve dokunsal arayüzlerin tasarımına yönelik çıkarımlarda bulunmaktadır.
}
\acknowledgments{
I am deeply honored to extend my heartfelt gratitude to the individuals who have played an integral role in shaping the course of my Ph.D. thesis. However, no words, no matter how eloquent, can adequately convey the depth of my appreciation for their impact on my academic journey.

First and foremost, I am profoundly indebted to my esteemed advisor, Prof. Dr. Cagatay Basdogan, whose unwavering support and expert guidance have been instrumental throughout my doctoral pursuit. Reflecting on the nascent stages of my research, when the intricacies of haptics seemed daunting and my responsibilities as a teaching assistant weighed heavily, his persistent belief in my abilities served as a beacon of encouragement. With his tireless mentorship, he instilled in me the essential qualities of a Ph.D. scholar, continually reinforcing the notion that ``Easa, you are the captain of your ship, and I am merely a lighthouse." Over the course of less than four challenging years, his constant alignment towards excellence and his support, even in the face of setbacks, proved pivotal. I recall with gratitude our lengthy deliberations and collaborative problem-solving sessions, where his door was always open, even amidst his demanding schedule. I remain profoundly grateful for his unparalleled guidance and mentorship, which have been paramount in shaping me into a proficient scientist in my chosen field.

Furthermore, my sincere appreciation extends to the esteemed Prof. Dr. Philippe Lefevre, Prof. Dr. Michael Wiertlewski, Prof. Dr. Sedat Nizamoglu, and Prof. Dr. Mehmet Sayar for graciously accepting the role of my thesis jury committee. Their dedication and substantial time investment in evaluating my research and providing invaluable insights during our discussions have significantly enriched the scholarly depth of my work. Their constructive feedback has been invaluable in honing the quality and rigor of my research. I am eternally grateful for the contributions of these remarkable individuals, without whom my academic journey would not have been as enriching or as transformative.

I am profoundly grateful for the invaluable contributions of Prof. Dr. Ørjan Grøttem Martinsen, whose pivotal role in my Ph.D. journey remains indelible. During the second phase of my thesis, my academic pilgrimage led me to the esteemed University of Oslo, Oslo, Norway, where I had the privilege of serving as a visiting scholar for a transformative period of two months. Despite my relative unfamiliarity, Prof. Martinsen extended his support, graciously investing his precious time to accompany me on each step of my academic odyssey. With an ever-present smile, he consistently bolstered my perseverance, instilling confidence in the significance of my research endeavors. To me, he epitomized the epitome of a remarkable mentor, welcoming me into his office with warmth and generosity. His profound wisdom and affable demeanor shone brightly in our interactions, fostering an environment of intellectual growth and camaraderie. The spirit of benevolence radiating from him extended gracefully to every member of the Oslo Bioimpedance and Medical Technology Group, creating an environment where every individual found joy and fulfillment in their collaborative efforts. My heartfelt appreciation also extends to all the members of the lab for their exceptional hospitality, which made my time there an enriching experience.

Moreover, I extend my sincerest gratitude to Prof. Dr. Fred-Johan Pettersen for his invaluable guidance during my tenure at the University of Oslo and Oslo University Hospital. His profound expertise and hands-on experience in the realm of experimental work left an enduring impression on me. Our engaging discussions were always fruitful, nurturing a space for intellectual exploration and scholarly growth. I remain profoundly grateful for his mentorship, which significantly enriched my academic journey.

During this transformative period, I had the privilege of crossing paths with one of the most distinguished and esteemed scientists globally, Dr. Bo Persson. Reflecting on our interactions, I am filled with profound admiration for his warmth and generosity in sparing valuable moments to delve into our findings. His meticulous attention to detail was evident as he devoted an extensive amount of time to meticulously reviewing my first paper, providing invaluable feedback that significantly elevated the caliber of my research.

My heartfelt appreciation extends to the esteemed Prof. Dr. Edward Colgate for his invaluable collaboration on my paper and his commitment to engaging in insightful discussions, both virtually and in person, during the IEEE World Haptics Conference 2023 held in Delft, Netherlands. His scholarly contributions and willingness to share expertise have significantly enriched the depth and scope of my work, leaving an enduring mark on my academic journey.

Furthermore, I am profoundly grateful to Prof. Dr. Kerem Pekkan and Prof. Dr. Levent Beker for entrusting me with the opportunity to collaborate on their esteemed projects: blood oxygenator and e-skin, respectively. Their support and open-door policy at their laboratories were instrumental, providing me with the necessary resources and guidance for conducting my experiments with utmost precision and rigor.

On a deeply personal note, I would like to express my profound gratitude to my family, whose unwavering support has been the cornerstone of my journey. My parents, through their sacrifices and nurturing guidance, have shaped the person I am today, and their immeasurable contributions defy expression. My siblings, not only my kin but also my closest confidants, have provided persistent support and companionship throughout this demanding pursuit.

Moreover, I am immensely grateful for the tireless support of my love and my soulmate, Arefeh Moradi. Her enduring kindness has been a source of immeasurable strength, propelling me toward a brighter future despite the miles that separate us. I consider myself truly fortunate to have her unwavering love and support as a guiding light in my life.

I am indebted to every member of the Robotics and Mechatronics Laboratory (RML) for fostering an environment of warmth and collaboration throughout the past years. It has been an honor and a privilege to work alongside each exceptional individual, and I am grateful for the friendship and support that each member, including Alireza Madani, Berke Ataseven, Berk Guler, Ehsan Khorram, Enes Ulas Dincer, Feras Kiki, Jahangier Ahmad, Mohammad Reza Alipour Sormoli, Muhammad Muzammil, Pouya Pourakbarian Niaz, Sara Hamdan, Volkan Aydingul, Yahya Mohey Hamad Al Qaysi, Dr. Yusuf Aydin, and Zaid Rassim Mohammed Al Saadi, had graciously extended to me. Their voluntary assistance during my experiments have been pivotal in the successful completion of my research endeavors.

I would also like to extend my heartfelt gratitude to my dear friends at Koc University, whose genuine support has been a source of immense comfort and joy during my time in Istanbul. To Fariborz Mirlou, Taher Abbasiasl, Anil Akseki, and all others who have touched my life during this chapter, I extend my sincere appreciation for their endless friendship. My sincere appreciation goes to Seyed Morteza Hoseyni for his invaluable efforts in performing mechanical vibration measurements.

I acknowledge the financial support provided by the Scientific and Technological Research Council of Turkey (TUBITAK) for supporting the first year of my study and the Graduate School of Sciences and Engineering of Koc University for the last three years. I also acknowledge the academic visit to the University of Oslo, which was supported by Prof. Martinsen, the Erasmus student exchange program, and Koc University.
}
\tableofcontents
\listoftables
\listoffigures
\abbreviations{
\begin{longtable}{lp{1cm}l}
	1D     & & One Dimensional\\
	2AFC   & & 2-Alternative Forced-Choice\\
    2D     & & Two Dimensional\\
	AC     & & Alternating Current \\	
	AFM    & & Atomic Force Microscopy\\
	ANOVA  & & Analysis of Variance\\
	AR     & & Augmented Reality\\
	CoF    & & Coefficient of Friction \\	
	CPE    & & Constant Phase Element\\
	DAQ    & & Data Acquisition\\
	DC     & & Direct Current\\
	DI     & & Deionized \\	
	EA     & & Electroadhesion \\
    EDA    & & Electrodermal Activity \\
	EDL    & & Electric Double Layer\\
	Eq     & & Equation\\
    FFT    & & Fast Fourier Transform\\
	Fig    & & Figure \\	
	HSD    & & Honestly Significant Difference\\
	IR     & & Infrared\\
	ITO    & & Indium Tin Oxide \\	
	JKR    & & Johnson-Kendall-Roberts\\
	MDS    & & Multi-Dimensional Scaling \\	
	MR     & & Magnitude ratio \\	
	NaCl   & & Sodium Chloride\\
	PC     & & Personal Computer\\
    PID    & & Proportional-Integral-Derivative\\
	PMMA   & & Polymethyl Methacrylat\\
	PPS    & & Percent Phase Synchronization\\
	RA     & & Rapidly Adapting\\
    RMS    & & Root Mean Square\\
	Ref    & & Reference\\
	SA     & & Slowly Adapting\\
    SC     & & Stratum Corneum\\
	SD     & & Standard Deviation\\
	SiN    & & Silicon Nitride\\
	SiO2   & & Silicon Dioxide\\
	SiOxNy & & Silicon Oxynitride\\
	TiOx   & & Titanium Oxide \\	
	TS     & & Touchscreen\\
	VR     & & Virtual Reality\\
	ZrOx   & & Zirconium Oxide\\
\end{longtable}
}
\textpages
%////////////////////////////////////////////////////////////////////////////////////////////////////////////////////////////////////////////////////////////////////

\chapter{Introduction} \label{chapter:Introduction}
Surface haptics is an innovative technology that has improved the way users interact with digital devices and surfaces, enhancing the user experience by providing realistic tactile feedback \cite{basdogan2020review}. It enables the simulation of tactile sensations on various smooth surfaces, such as touchscreens, touchpads, and interactive displays, mimicking the sensation of texture, resistance, or physical interaction with digital objects. By integrating surface haptics into digital devices, users can experience a heightened sense of immersion and engagement in digital worlds, contributing to more intuitive and lifelike interactions.

Surface haptics technology has many applications in gaming, virtual reality (VR), augmented reality (AR), and various other interactive digital environments. In the gaming industry, for instance, surface haptics can provide gamers with a more immersive and realistic gaming experience by allowing them to feel the impact of in-game events and interactions through their fingertips. In AR and VR applications, surface haptics can significantly enhance the sense of presence and realism, making virtual environments more convincing and engaging. It has continued to advance, with ongoing research and development aimed at refining the technology and exploring new applications across different sectors. With its potential to redefine user interactions with digital interfaces, surface haptics remains a promising area of innovation, fostering the evolution of more intuitive and lifelike user experiences in the digital realm.

This technology relies on a combination of sophisticated techniques, including the use of vibrations, electrostatic forces, and mechanical mechanisms, to create the illusion of physical texture and resistance on flat and featureless surfaces. These techniques are carefully engineered to deliver precise and responsive haptic feedback, enabling users to feel the sensation of interacting with virtual objects, buttons, and interfaces, as if they were tangible physical entities.

Among all actuation techniques, electrostatic actuation appears to be the most promising one. An electrostatic attractive force can be generated between two solids having different electrical potentials and this phenomenon is known as electroadhesion. In practical terms, touchscreens with electroadhesion technology find applications in various industries, including consumer electronics, automotive, gaming, and VR. In consumer electronics, this technology can be integrated into smartphones, tablets, and other touchscreen devices, providing users with a more intuitive and realistic touch experience. In the automotive sector, it can contribute to the development of advanced touch-based infotainment systems, improving the usability and safety of in-vehicle touchscreens. Moreover, in gaming and VR, electroadhesion technology can elevate the level of immersion, allowing users to feel more connected to the virtual environments and enhancing the overall sensory experience.

\section{Problem Definition and Approach}\label{sec:ch1_ProblemDefinitionApproach}
\begin{figure}[!t]
\centering
\includegraphics[width=0.5\linewidth]{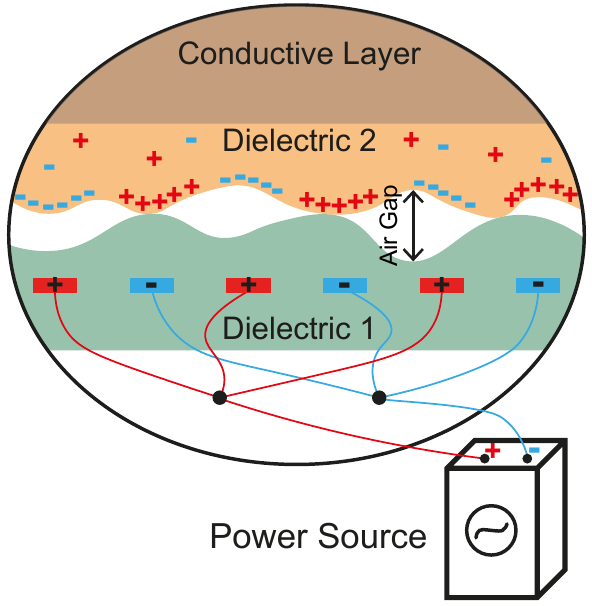}
\caption{Schematic representation of electroadhesive devices. Dielectric 1 is in contact with dielectric 2 and there is an air gap between them due to their surface roughness. The electrodes embedded in dielectric 1 are connected to the power source, and hence induced electrical charges are accumulated at the interface of dielectric 2. The opposite charges at the interfaces of the dielectrics generate electrostatic forces and attract the dielectrics to each other.}
\label{fig:ch1_Electroadhesion}
\end{figure}

Electroadhesion refers to the attraction between two dielectric materials due to an applied electric field \cite{shintake2018soft, levine2021materials, rajagopalan2022advancement}. As shown in Fig. \ref{fig:ch1_Electroadhesion}, the high voltage applied to the electrodes embedded in dielectric 1 results in the accumulation of electrical charges. Once dielectric 1 contacts dielectric 2, induced charges with opposite polarities accumulate at the interface of dielectric 2. The opposite charges at the interfaces of the dielectrics attract each other and result in electrostatic forces due to the air gap between them.

The electrostatic forces between the contacting surfaces can be calculated using the normal component of Maxwell’s stress tensor \cite{cheng1989field}:
\begin{equation}\label{eq:ch1_MaxwellTensor}
    F_e=\frac{1}{2} \varepsilon_0 A_{app} \frac{1}{u^2} \Delta V_{gap}^2
\end{equation}
\noindent where, $\varepsilon_0$, $A_{app}$, $\Delta V_{gap}$, and $u$ are the permittivity of free space ($8.854\times10^{-12}$ F.m$^{-1}$), apparent contact area, the voltage at the air gap, and average air gap thickness, respectively. Eq. \ref{eq:ch1_MaxwellTensor} indicates that the magnitude of electrostatic force is proportional to the square of the voltage at the air gap and inversely proportional to the square of its thickness. It is worth mentioning here that the voltage term in Maxwell’s stress tensor is the voltage at the air gap and not equal to the voltage applied to the electrodes, as the voltage drops at the dielectric layers must be accounted for. Furthermore, even if the contacting surfaces appear smooth, they have finite roughness at different length scales, resulting in a nonuniform air gap between the dielectrics \cite{persson2001theory}. Since this gap typically varies from a few nanometers to a few micrometers across the dielectrics, there is currently no direct experimental approach to measure the voltage at the air gap and the thickness of this gap to infer the electrostatic forces using Eq. \ref{eq:ch1_MaxwellTensor}. 

This thesis aims to provide a comprehensive insight into the fundamental science of contact interactions between a human finger and a voltage-induced touchscreen under electroadhesion through sophisticated theoretical modeling and multiple experimental techniques. At first, the thesis focuses on studying the frequency-dependent behavior of electrostatic forces between the human finger and a touchscreen under electroadhesion. The motivation for this phase of the thesis stemmed from the discussion articulated in \cite{ayyildiz2018contact}, which highlighted the phenomenon of charge leakage. As human finger touches a voltage-induced touchscreen, some amount of electrical charges transfer from the Stratum Corneum (SC) layer of the skin to the surface of the touchscreen. These charges reduce the electric field in the air gap between the finger and the touchscreen, resulting in low electrostatic forces. The charge leakage phenomenon is underestimated in most modeling studies. Sirin et al. \cite{sirin2019electroadhesion} argued that charge leakage mostly happens at frequencies lower than 30 Hz since electrical charges have less time to accumulate at the interfaces in higher frequencies. However, there exists almost no model to consider or calculate the amount of charge leakage under electroadhesion. In addition, the frequency-dependency of the electrical properties of the SC is not properly taken into account in earlier modeling studies \cite{meyer2013fingertip,vezzoli2014electrovibration,vardar2017effect,forsbach2021rigorous}. 

Although modeling the role of charge leakage under electroadhesion is essential for estimating the electrostatic forces, understanding its underlying physics is necessary for an enhanced design of touchscreens. On the other hand, direct measurement of the electrostatic forces is not trivial due to the nature of the contact problem. Hence, in the second phase of this thesis, we focus on the electrical interactions between the finger and the touchscreen by utilizing precise electrical impedance measurements. This study presents a milestone in inferring electrostatic forces between the finger and touchscreen under electroadhesion by proposing a novel approach based on the measurement of electrical impedances. The proposed approach explains various physical mechanisms behind electroadhesion and enables the calculation of average air gap thickness and electrical capacitance between the finger and touchscreen. To the best of our knowledge, this is the first study reporting an experimental and systematic approach to infer the magnitude of electrostatic forces.

Our proposed approach enables the development of a straightforward yet precise electrical circuit model for the interface between the fingertip and the touchscreen under electroadhesion. This model facilitates a deeper understanding of the physics behind the interactions at the interface, such as the formation of the electric double layer (EDL) leading to the phenomenon of charge leakage and the electrode polarization impedance. Moreover, the model helps us to understand the difference between DC and AC stimulations. Hence, in the first part of the third phase of this thesis, we investigate human tactile perception of electroadhesion for DC and AC stimulations and explain our results by the proposed electrical circuit model.

Electroadhesion is a sophisticated phenomenon with over thirty-three different factors affecting the magnitude of electrostatic forces \cite{guo2015investigation}. In terms of a human finger in contact with a voltage-induced touchscreen, the hydration level of human finger has noticeable effects on the frictional forces between the finger and the touchscreen and also the magnitude of electrostatic forces. Adams et al. \cite{adams2007friction} showed that in the presence of water, the sliding of skin becomes irregular, and the frictional forces increase. This increase in friction was described by the increase in real contact area, which is a result of water plasticization on the SC. However, the increase in frictional forces due to the presence of water is limited to a threshold where the asperities of the fingertip fill with water and the air gap diminishes. Under electroadhesion, this is even more important since there are approximately 500 eccrine sweat glands in a 1 $cm^2$ human's fingertip area \cite{tripathi2015morphology}. All these sweat glands are actively working and they are very sensitive to the charges on the skin. According to Eq. \ref{eq:ch1_MaxwellTensor}, the electrostatic forces acting on the finger are inversely dependent on the air gap thickness. Still, at some point, the gap can be replaced with water and ease the transfer of charges, resulting in a diminished electric field. Note that, as much as the real contact area increases (air gap decreases), transferring electrical charges increases as well. Hence, the second part of the third phase of this thesis sheds light on the effect of finger moisture on electroadhesion. Using tactile perception experiments, we investigate the voltage thresholds of ten participants with varying moisture levels. Furthermore, we perform electrical impedance measurements to facilitate a better understanding of the electrical interactions between a moist human finger and the touchscreen.

All the aforementioned efforts are aimed at enhancing the user's ability to perceive tactile stimuli. However, there has been relatively less emphasis on comprehending the specific tactile sensations experienced when interacting with the surface of a touchscreen. Touchscreens typically exhibit very low surface roughness, on the order of nanometers \cite{shultz2018electrical}, making them very smooth to the touch. When a finger makes contact with a touchscreen, the tactile perception involves both the sensation of the smooth surface's friction and the modified frictional properties of the surface due to electroadhesion. In 2017, Vardar et al. \cite{vardar2017effect} investigated the latter by studying the effect of different input voltage waveforms on the human perception of electroadhesion on touchscreens. Through psychophysical experiments, they initially established the thresholds for detecting electroadhesion stimuli generated by both sinusoidal and square voltages across a range of frequencies. Their results indicated that participants were more sensitive to square wave stimuli than sinusoidal waveforms at fundamental frequencies below 60 Hz. They suggested that this discrepancy in perceived sensations at lower frequencies can be attributed to the frequency-dependent electrical properties of human skin and the tactile sensitivity of the participants. In addition, they argued that the Pacinian channel was the primary sensory channel involved in detecting the square wave input signals used in their study. Despite these findings, there has been a notable lack of research investigating the scientific aspects of human tactile perception of a touchscreen's smooth surface without electroadhesion. Hence, the final phase of this thesis centers on this aspect, employing various surface characterization techniques to gain a deeper understanding of the science of touch on extremely smooth surfaces and the remarkable precision of human tactile perception in encoding the world around us.

\section{Contribution}\label{sec:ch1_Contribution}
In this thesis, we investigate the electro-mechanical contact interactions between a human finger and a touchscreen in four phases. The purpose of this thesis is to provide a fundamental understanding of the physics behind these interactions through theoretical and experimental approaches. The findings of this thesis could also serve as a reference to enhance the design of touchscreens.

In the first phase of the thesis, we develop an electro-mechanistic contact model to calculate the magnitude of electrostatic forces. Since these forces are the result of the electrical field in the air gap, we consider the fundamental laws of electromagnetism (Gauss law, Maxwell-Wagner effect, Kirchhoff's law, Ohm's law, and charge continuity equation) in the electrical model. In this modeling, including the charge leakage and frequency-dependent electrical properties of the SC are two important challenges. In addition, the air gap between the finger and the touchscreen is nonuniform due to the multi-scale roughness of the skin, which makes it difficult to calculate the contact area and hence the contact pressure. We use Persson's contact theory to consider these nonuniformities. Since this theory considers roughness in all magnifications, it requires sophisticated numerical approaches to calculate the real area of contact using a probabilistic approach. The outcome of the model is the electrostatic pressure acting on the finger with a variable air gap estimated by the probability distribution of interfacial separations. On the other hand, we measure the frictional forces between the finger and the touchscreen with and without electroadhesion using a custom-made mechatronics set-up and estimate the magnitude of electrostatic forces. Our theoretical model is in good agreement with the experimental data and shows that the change in magnitude of the electrostatic force is mainly due to the leakage of charges from the SC to the touchscreen at frequencies lower than 250 Hz and the electrical properties of the SC at frequencies higher than 250 Hz.

The work presented in the second phase of this thesis facilitates the estimation of electrostatic forces between the human finger and the voltage-induced touchscreen under electroadhesion based on electrical impedance measurements. We propose a novel approach that takes the individual electrical impedances of finger skin and touchscreen, and when they are coupled (total) as the input and returns the electrostatic forces as a function of frequency as the output. The magnitude of this force depends on the thickness of the air gap between the surfaces. Since all surfaces have finite roughness, there is always an air gap between them, which varies from a few nanometers to a few micrometers across the contact area and makes it difficult to measure the electrostatic forces directly by some other methods. To our knowledge, this is the first experimental approach in literature for measuring electrostatic forces between the finger and the touchscreen under electroadhesion based on electrical impedances. In addition, for the first time, we report the average thickness and electrical capacitance of the air gap between the finger and the touchscreen under electroadhesion. We also present the first demonstration of the effect of electrode polarization impedance under electroadhesion. We show that electrode polarization impedance occurs at the contact region due to the formation of EDL and adversely affects the impedance measurements. We provide a method to eliminate this noisy impedance in order to estimate the electrostatic forces correctly. This phase of the thesis provides a deep and scientific understanding of the electroadhesion phenomenon by explaining the physical mechanisms involved in electrical interactions between the human finger and the voltage-induced touchscreen, such as dielectric polarization, dispersion, EDL, charge transfer, charge leakage, and electrode polarization.

During the third phase of this thesis, we first investigate the effect of input voltage signal type (DC vs. AC) on the tactile perception of electroadhesion using psychophysical experiments. We show that human tactile detection of electroadhesion is notably less pronounced under DC stimulations compared to AC stimulations, and our electrical circuit model proposed in the second phase of the thesis explains this discrepancy. Second, we study the effect of moisture on the electrostatic forces between the finger and the touchscreen. It is known that the human body maintains a constant perspiration to regulate its temperature. Furthermore, the application of a voltage signal to the skin's surface (as in the case of electroadhesion) can alter the rate of perspiration \cite{grimnes1983skin}. Consequently, as the finger slides across a touchscreen stimulated by a voltage signal, sweat can accumulate within the interface between the finger and the touchscreen, subsequently affecting the magnitude of the electrostatic forces. Our findings reveal an increase in the voltage detection threshold for participants with moist fingers. In addition, our electrical impedance measurements demonstrate that the introduction of a 0.9\% Sodium Chloride (NaCl) layer into the interface between the finger and the touchscreen reduces the electrical impedance of the sliding finger on the touchscreen in comparison to the nominal condition of no artificial liquid at the interface. This suggests that as moisture fills the air gap between the finger and the touchscreen, it creates a path for the electrical charges to transfer from one interface to the other. Hence, the transfer of charge diminishes the magnitude of the electric field in the air gap, consequently leading to a reduction in the magnitude of the electrostatic forces.

In the final phase of this thesis, we aim to understand the science of human tactile perception during interactions with the smooth surface of touchscreens without electroadhesion. Typically, touchscreens are made of a conductive layer (ITO: Indium Tin Oxide) sandwiched between two insulating layers (SiO$_2$: Silicon Dioxide). Both the conductive layer and the top insulating layer are produced using coating techniques, approximately 1 $\mu$m and 250 nm in thickness, respectively \cite{ayyildiz2018contact}. Additional finishing processes are applied to render their surfaces smoother. Hence, a touchscreen's surface is considered a coated smooth surface. To explore human tactile perception of these smooth surfaces, we prepare samples with different coating materials as the top layer and perform psychophysical experiments with them. Furthermore, we employ various surface characterization methods to establish connections between the physical measurements conducted on these surfaces and our tactile perception, thereby advancing our understanding of the adhesive contact mechanism that influences our tactile perception. The results from our psychophysical experiments and physical measurements indicate that the coating material significantly affects our tactile perception of extremely smooth surfaces, revealing the human finger's capability to detect variations in surface chemistry, potentially attributed to molecular interactions.

\section{Outline}\label{sec:ch1_Outline}
Including the introduction chapter, this thesis is presented in seven chapters. The organization of the manuscript is as follows:

Chapter 2: We provide a thorough literature review covering the history of electroadhesion and its advantages and challenges. 

Chapter 3: First, we introduce the electrical model that enables us to calculate the electric field at the air gap and the electrostatic attraction force between the finger and touchscreen, where we also utilize Persson's multi-scale contact theory to consider the variable air gap. Following, the experimental set-up and procedure for frictional force measurements to infer electrostatic forces are explained. The experimental results are reported together with the numerical solutions of the model and are discussed at the end of the chapter.

Chapter 4: We present our experimental procedures for the electrical impedance measurements. Then, we report the experimental results and discuss them. We develop a simple electric circuit model for the interface between the finger and the touchscreen under electroadhesion and propose a systematic approach to calculate the magnitude of electrostatic forces using only the electrical impedance measurements. Finally, we validate the electrostatic forces inferred from our electrical impedance approach using electrostatic forces inferred from frictional force measurements.

Chapter 5: First, we introduce the experimental set-up and the procedure for our psychophysical experiments that investigate the effect of voltage type and finger moisture on tactile perception under electroadhesion. Then, we present the results of our threshold detection experiments for DC and AC stimulations using participants with varying finger moisture levels. We also present our experimental measurements of the electrical impedances of the sliding finger on the touchscreen when 20 $\mu$L of NaCl is placed into the interface between the finger and the touchscreen. We utilize our electrical circuit model proposed in Chapter \ref{chapter:ExperimentalElectrostaticForcesFromImpedance} to justify the results of our psychophysical experiments for the differences between DC and AC stimulations. We then discuss the differences in the threshold levels of participants using the results of our electrical impedance measurements.

Chapter 6: We start by explaining our sample preparation process and the procedure utilized for the psychophysical experiments that investigate the effect of the top coating material of the touchscreen on our tactile perception. Using the 2-alternative forced-choice (2AFC) method, we display the coated surfaces in pairs side by side to eight participants and ask them to explore the surfaces with their index fingers and select the surface that feels more ``resistive" to sliding. Then, we present our experimental set-up and the procedure followed for the measurement of the dynamic coefficient of friction (CoF) between the index finger of seven participants and the sample surfaces. In the following, we introduce contact angle measurements performed by the Sessile drop test to calculate the surface free energies and contact angle hysteresis of the sample surfaces. Finally, we talk about our Atomic Force Microscopy (AFM) measurements for surface roughness. The results of our tactile perception experiments and the physical measurements are reported and discussed at the end of the chapter.

Chapter 7: This chapter serves as the culmination of the thesis, providing a comprehensive synthesis of the findings and contributions while offering recommendations for potential avenues of future research.

%/////////////////////////////////////////////////////////////////////////////////////////////////////////////////////////////////////////////////////

\chapter{Background} \label{chapter:Background}

\section{History of Electroadhesion}\label{sec:ch2_HistoryElectroadhesion}
\begin{figure}[!t]
\centering
\includegraphics[width=1\linewidth]{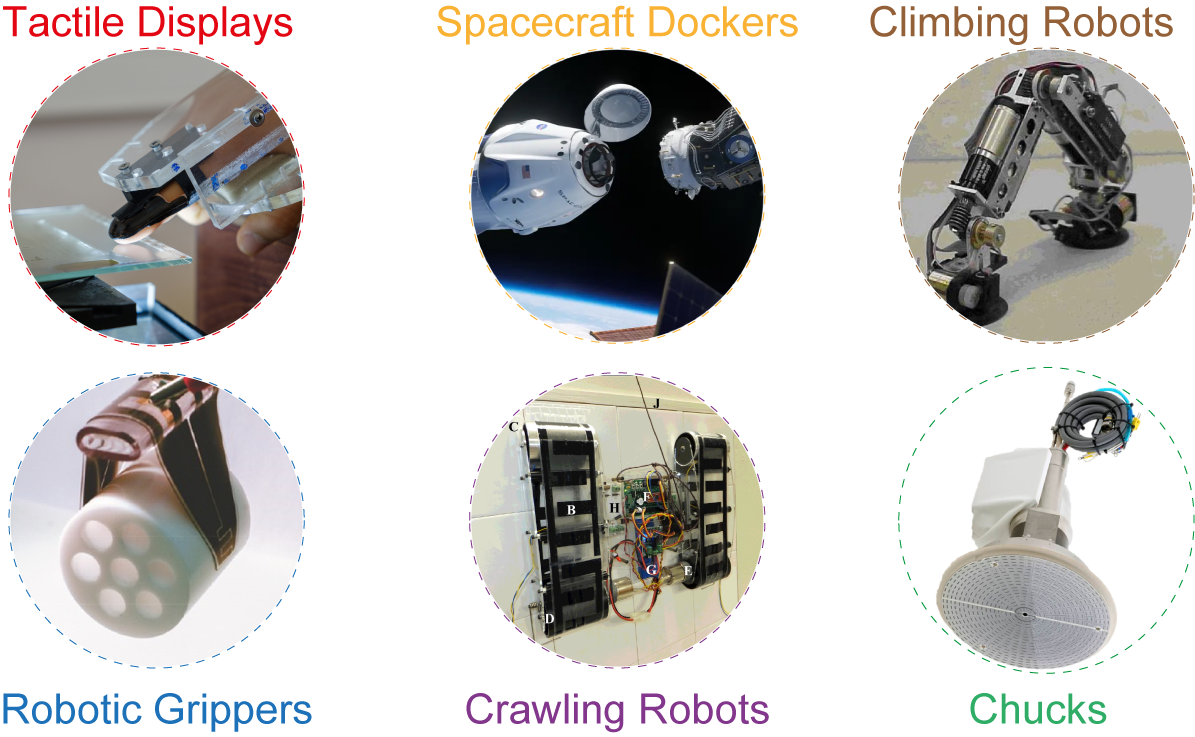}
\caption{The electroadhesion technology has applications in tactile displays, spacecraft docking, climbing and crawling robots, robotic grippers, electrostatic chucks, etc. The images are reproduced with permission: Copyright 2023, Wiley \cite{shintake2016versatile}, Copyright 2023, Wiley \cite{xiao2015advances}, and Copyright 2023, Elsevier \cite{koh2016hybrid}.}
\label{fig:ch2_Applications}
\end{figure}

The discovery of electroadhesion can be traced back to 1923 when Frederik Alfred Johnsen and Knud Rahbek \cite{johnsen1923physical}, two Danish engineers, observed an adhesive force between a brass plate and a slab of limestone resting on a voltage-induced conductor surface. Later, Mallinckrodt et al. \cite{mallinckrodt1953perception} accidentally discovered in the 1950s that there was a sense of adhesion when a human finger explored a smooth metal surface coated with an insulating layer and connected to an AC source. This phenomenon was attributed to the principle of parallel-plate capacitors, wherein the metal surface and the human finger were treated as two parallel plates of a capacitor attracting each other. Since then, broad ranges of applications of electroadhesion have been introduced in various fields \cite{shintake2016versatile, nakamura2017modeling, ayyildiz2018contact, kim2020low, basdogan2020review, yang2021recent,  rajagopalan2022advancement, liu2022electrically, wang2022control, liu2022switchable} (see Fig. \ref{fig:ch2_Applications}). Electroadhesion became commercially viable for industrial applications in the 1960s with the introduction of electrostatic plotters \cite{monkman2003electroadhesive}. Around the same time, NASA started to employ electroadhesion technology in space missions for material handling \cite{beasley1971development} and later controllable earth orbit grappling and soft docking \cite{bryan2015innovative, leung2015validation}. After a decade, electroadhesion entered the semiconductor industry with applications in material handling by using the devices commercially known as electrostatic chucks \cite{wardly1973electrostatic, yatsuzuka2001electrostatic, asano2002fundamental}. Starting from the 1990s, electroadhesion was first embedded into the end effectors of robots for gripping composites and fibrous materials like fabric \cite{monkman1989principles} and carbon fibers \cite{monkman1995robot, zhang1999modeling} and then later crawling and wall climbing robots \cite{yamamoto2007wall, prahlad2008electroadhesive, wang2014crawler, wu2018paper}. More recently, it was utilized in touch surfaces of mobile devices for displaying haptic feedback to the user with potential applications in many new domains (see the review of surface haptics in \cite{basdogan2020review}).

\section{Advantages of Electroadhesion}\label{sec:ch2_AdvantageousElectroadhesion}
The research on electroadhesion has experienced a growth rate of approximately 10$\%$ in annual publications over the past 60 years \cite{rajagopalan2022advancement}. It has successfully entered various fields due to its unique benefits, which are difficult or even impossible to achieve through some other techniques, such as magnetic adhesion. For instance, electroadhesive devices can adhere to both conductive and insulating surfaces. They are compatible with vacuum environments since the gap between the contacting surfaces under electroadhesion is typically filled with air and its permittivity is almost equal to that of the vacuum. Additionally, electroadhesion consumes ultra-low energy in some applications compared to its alternatives due to driving a small amount of current (see, for example, robotic insect application in \cite{graule2016perching}). Some of the electroadhesive devices are also lightweight and less complex compared to some mechanical devices utilizing vacuum pumps and motors for the same purpose (e.g., material handling applications).

\section{Electroadhesion in Tactile Displays}\label{sec:ch2_ElectroadhesionTactileDisplays}
As mentioned in Section \ref{sec:ch2_HistoryElectroadhesion}, Mallinckrodt et al. \cite{mallinckrodt1953perception} accidentally discovered that if human finger moves gently on a smooth metal surface, which is coated with an insulating layer and connected to a 110 volts alternating power supply, there is a sense of adhesion. They explained the cause of adhesion by the parallel-plate capacitor principle and reported that this sensation vanished when the power supply was turned off. Years later, Grimnes \cite{grimnes1983electrovibration} named this phenomenon as “electrovibration”. Researchers started to show further interest in this topic when Strong and Troxel \cite{strong1970electrotactile} developed the first surface display for blind that utilized the electroadhesion technology to provide them with tactile feedback. Their display was composed of an array of small electrodes coated with an insulating surface. They observed that the applied voltage had significant effects on the intensity of touch sensations, whereas the applied current did not have such effects. To study the effect of spatial resolution and information transmission capacity, Tang and Beebe \cite{tang1998microfabricated,beebe1995polyimide} developed a more sophisticated electroadhesion display based on the design similar to that of Strong and Troxel \cite{strong1970electrotactile}. They performed experiments with visually impaired people and reported their tactile stimulus detection thresholds and also their detection rates in line separation and pattern recognition. The participants were able to recognize basic tactile patterns (circle, triangle, square) by tactile exploration with an accuracy of approximately 70\%. More recently, Bau et al. \cite{bau2010teslatouch} introduced TeslaTouch, a commercial surface capacitive touchscreen with a conductive layer (ITO) underneath an insulator layer (SiO$_2$), that can display tactile feedback to a sliding finger when a voltage signal is applied to the conductive layer. Since the human finger interacting with the touchscreen also has conductor and insulator layers, an electrostatic attraction force is generated between the finger and the touchscreen. The insulator layer in finger is SC, which is mainly composed of dead cells \cite{adams2007friction}. In fact, this layer is not a perfect insulator and can partially prevent the electrical charges from passing. Yamamoto and Yamamoto \cite{yamamoto1976dielectric} showed that both electrical resistivity and permittivity of the SC are highly dependent on the frequency of applied voltage.

\subsection{Modeling}\label{sec:ch2_Modeling}
In order to estimate the electrostatic forces between human finger and a touchscreen under electroadhesion, parallel-plate capacitor models have been utilized (see the review in \cite{vodlak2016multi}). Meyer et al. \cite{meyer2013fingertip} developed a model based on this principle and showed the linear relationship between the magnitude of electrostatic force and the square of the applied voltage amplitude. They also reported that the inferred electrostatic force increases with the frequency of the voltage signal. Shultz et al. \cite{shultz2015surface,shultz2018electrical} made a number of important contributions, including the observation that the electrical impedance of the air gap (which relates to leakage current) depends critically on the motion of the finger across the surface. They performed experimental measurements of the impedance at the air gap and supported their results with a circuit model. Basdogan et al. \cite{basdogan2020modeling} also utilized a parallel-plate capacitor model to estimate the average normal pressure due to electroadhesion and the work done by electroadhesion. The estimated work was then used in the well-known Johnson-Kendall-Roberts (JKR) contact model to investigate the increase in tangential friction force and, hence the inferred electrostatic force due to electroadhesion. They showed that the results of the model matches well with the experimental data collected by a custom-made tribometer.

A recent work by Heß and Forsbach \cite{hess2020macroscopic} utilized Shull's compliance method \cite{shull2002contact}, which is a generalized version of JKR theory, to model contact interactions between human finger and touch surface under electroadhesion for large deformations. Assuming pressure-controlled friction, a model for the sliding electro-adhesive contact was developed, which adequately imitates the experimental data reported in Basdogan et al. \cite{basdogan2020modeling}. Nakamura and Yamamoto \cite{nakamura2017modeling} investigated the decrease in electrostatic force under DC input voltage using an electro-mechanical model. The mechanical part was based on a simple mass-spring system and simulated the fluctuations in air gap thickness during sliding. The electrical model was based on the Johnsen-Rahbek effect, which states that an insulator has a finite resistivity and electrical charges can travel inside it \cite{watanabe1993relationship}. The authors concluded that electrostatic force decreases if the ratio of electrical permittivity to conductivity of an insulator is higher than that of air. Persson \cite{persson2018dependency} developed a model for the electrostatic attraction forces between two surfaces having a potential difference based on his multi-scale contact mechanics theory \cite{persson2001theory,persson2006contact}. Recently, he generalized this model to include charge leakage and sweat accumulation at the air gap \cite{persson2021general}. Ayyildiz et al. \cite{ayyildiz2018contact} and Sirin et al. \cite{sirin2019electroadhesion} investigated the electroadhesion between human finger and a touchscreen using Persson's theory and showed that the main cause of the increase in friction force (and hence the electrostatic force) is due to the increase in the real contact area of finger.

\subsection{Perception}\label{sec:ch2_Perception}
In terms of tactile perception of electroadhesion, the number of studies are limited. Kaczmarek et al. \cite{kaczmarek2006polarity} explored the sensitivity of the human finger to the polarity of the voltage signal. Their results showed that the participants perceived negative or biphasic pulses better than positive ones. Bau et al. \cite{bau2010teslatouch} measured the tactile thresholds for different stimulation frequencies. They showed that the threshold voltage followed a U-shaped curve as a function of frequency, as observed in earlier vibrotactile studies. Vardar et al. \cite{vardar2017effect} investigated the effect of input voltage waveform on tactile perception of electroadhesion. Their psychophysical experiments showed that the participants were more sensitive to square voltage signals than sinusoidal ones for frequencies lower than 60 Hz. They analyzed the collected force and acceleration data in the frequency domain by considering the human tactile sensitivity curve and the results suggested that the Pacinian channel was the primary psychophysical channel in detecting the electroadhesion stimuli caused by all the square-wave signals displayed at different frequencies. Later, Vardar et al. \cite{vardar2018tactile} investigated the effect of masking on the tactile perception of electroadhesion. They measured the masked thresholds of sinusoidal electroadhesion bursts (125 Hz) of nine participants under two masking conditions: simultaneous and pedestal. They observed that the tactile thresholds were elevated as linear functions of masking levels for both masking types, but the masking effectiveness was higher with pedestal masking than simultaneous. In order to demonstrate the practical application of masking on touchscreens, they compared the perceived sharpness of virtual edges separating two textured regions displayed with and without various types of masking stimuli. The results showed that sharpness perception depends on the local contrast between background and foreground stimuli, which varies as a function of masking amplitude and activation levels of frequency-dependent psychophysical channels.

In the area of tactile perception of surfaces, most of the earlier studies have focused on textured surfaces. These studies have identified the perceptual dimensions of texture perception as rough/smooth, hard/soft, sticky/slippery, and warm/cool \cite{hollins2000individual, tiest2006analysis} though the link between these dimensions and the physical properties has not been fully understood yet. Among these perceptual dimensions, the rough/smooth dimension is considered the most significant one. 

The initial studies by Hollins et al. \cite{hollins2001vibrotactile} using multi-dimensional scaling (MDS) techniques and then neuronal recordings acquired from SA1, RA, and Pacinian afferents of Rhesus macaques by Weber et al. \cite{weber2013spatial} revealed that the mechanism underlying roughness perception is different for micro and macro textures where the threshold for inter-element spacing was determined as approximately 200 microns (see a review in Klatzky and Lederman \cite{klatzky2010multisensory}). It appears that spatial cues play a dominant role in tactile perception of macro textures while the temporal ones in micro textures.

Although the roughness perception of macro textures can be investigated using periodic raised dots and gratings manufactured by conventional techniques, it is more difficult to apply the same techniques to periodic micro-scale textures. Moreover, compared to the periodic textures, manufacturing randomly rough surfaces is even more difficult since the surface topography follows some probabilistic distribution and the height profile should be sampled from it. The recent advances in additive manufacturing could be helpful in investigating human tactile perception of randomly rough surfaces in a systematic manner \cite{hartcher2019surface, sahli2020tactile}.

In order to produce surfaces with features smaller than micro-scale, micro/nano-scale surface coating techniques are necessary. Skedung et al. \cite{skedung2013feeling} used such techniques to produce wrinkled surfaces with wavelengths ranging from 300 nm to 90 $\mu$m and amplitudes between 7 nm and 4.5 $\mu$m. They then conducted psychophysical studies involving similarity scaling on those surfaces. The results of the study show that the lowest amplitude of the periodic wrinkles distinguished by humans is approximately 10 nm. If this value is taken as the limit of our tactile perception (though it may not directly apply to "non-periodic" surfaces), the surfaces having roughness amplitudes below this limit can be accepted as extremely smooth. 

In tribology literature, it is known that adhesive forces significantly affect the friction between a soft object such as the human finger and an extremely smooth surface, especially at low normal contact forces (F$_n$) \cite{derler2009friction}. The friction force acting on a finger sliding on a smooth surface can be written as F$_t$ = F$_{adh}$= $\tau^*$A$_{\rm real}$, where $\tau^*$ is the interfacial shear stress and A$_{\rm real}$ is the real area of contact, which varies nonlinearly with the normal force applied by the finger on the surface. The real contact area is difficult to measure or estimate since human finger pad has surface roughness at different length scales and each asperity makes adhesive contacts down to the nanoscale and supports the adhesive shear load proportional to its own contact area, contributing to the tangential force, shearing those contacts \cite{ayyildiz2018contact}. The adhesive contacts between finger and an extremely smooth surface are mainly formed due to van der Waals, electrostatic, and hydrogen bonding forces. Derler et al. \cite{derler2009friction} measured the CoF between finger and smooth and rough glass surfaces under dry and wet conditions and concluded that adhesion significantly affected the frictional interactions with smooth (rough) surfaces under dry (wet) conditions.

Compared to the earlier studies on macro and micro textures, the number of studies investigating the human tactile perception of extremely smooth surfaces having a roughness of a few nanometers is highly limited. Moreover, to this time, very few studies investigated the effect of coating type (material) on tactile perception of such surfaces. Gueorguiev et al. \cite{gueorguiev2016touch} conducted 2AFC experiments with glass and polymethyl methacrylate (PMMA) plastic plates having similar roughness magnitudes at nanoscale and observed that the human participants with dry fingers could successfully discriminate them though they resulted in similar CoF. They attributed this result to the difference in molecular structures of glass and PMMA. Carpenter et al. \cite{carpenter2018human} showed that human participants can differentiate Silicon surfaces that differ only by a single layer of molecules. Their results demonstrate that surface chemistry plays an important role in tactile perception.
Skedung et al. \cite{skedung2018feeling} prepared a stimuli set consisting of ten glass surfaces with different coatings and measured their water contact angle, contact angle hysteresis, and surface free energy. They also measured the CoF between human finger and the samples. The surfaces were evaluated in terms of perceived similarities by ten female participants who were able to distinguish between the surfaces varying in coating material. The MDS analysis revealed that the primary perceptual dimension correlates with surface free energy, but both CoF and surface energy contribute to this dimension depending on whether the coating is organic or inorganic. Ajovalasit et al. \cite{ajovalasit2019human} prepared twenty smooth aluminum surfaces consisted of four uncoated substrates and sixteen coated substrates (four different types of coatings applied on each of the four uncoated substrates) to conduct tactile perception studies with human participants. A total of forty participants rated the surfaces based on their perceived slipperiness, roughness, and glossiness. They also measured surface roughness, CoF, surface free energy, and surface glossiness of the coated surfaces. They concluded that the coating type had significant effects on perceived slipperiness and roughness, while both the coating type and manufacturing process had significant effects on glossiness.

%/////////////////////////////////////////////////////////////////////////////////////////////////////////////////////////////////////////////////////

\chapter{Frequency-Dependent Behavior of Electrostatic Forces}\label{chapter:FrequencyDependentElectroadhesion}
\subsubsection{Summary}
In this chapter\footnote{This chapter is based on an article \cite{aliabbasi2022frequency}}, we focus on the frequency-dependent behavior of electrostatic forces between human finger and touchscreen under electroadhesion. The number of studies investigating this topic is only a few \cite{meyer2013fingertip,vezzoli2014electrovibration, vardar2016effect, forsbach2021rigorous}. Furthermore, in all the earlier studies, relatively simple circuit models were considered, which are limited in capturing the true (experimentally-observed) behavior of electrostatic force, especially at high frequencies. We argue that modeling charge transfer between the interfaces of SiO$_2$-air and air-SC as a function of frequency is critical to capture this behavior. At low frequencies, induced charges in opposite polarity accumulate at these interfaces and some of them leak to the surface of the touchscreen. As the frequency increases, the polarity changes frequently and the charges are not able to accumulate at the interfaces and the leakage decreases. In addition, the permittivity of the SC drops with increasing frequency, which adversely affects the electric field and hence the electrostatic force.

We argue that considering charge transfer is important for developing a proper electrical model to investigate electroadhesion using the fundamental laws of electric field. To our knowledge, such in-depth investigation has not been performed for a wide range of stimulation frequencies in the literature. The model proposed in this chapter fits the experimental data better than the ones available in the literature. In particular, we show that charge leakage is the main factor for the reduction in electrostatic forces at lower frequencies, while the frequency-dependent nature of the SC is the one at higher frequencies.
\section{Analytical Modeling}
\begin{figure}[!b]
\centering
\includegraphics[width=0.9\linewidth]{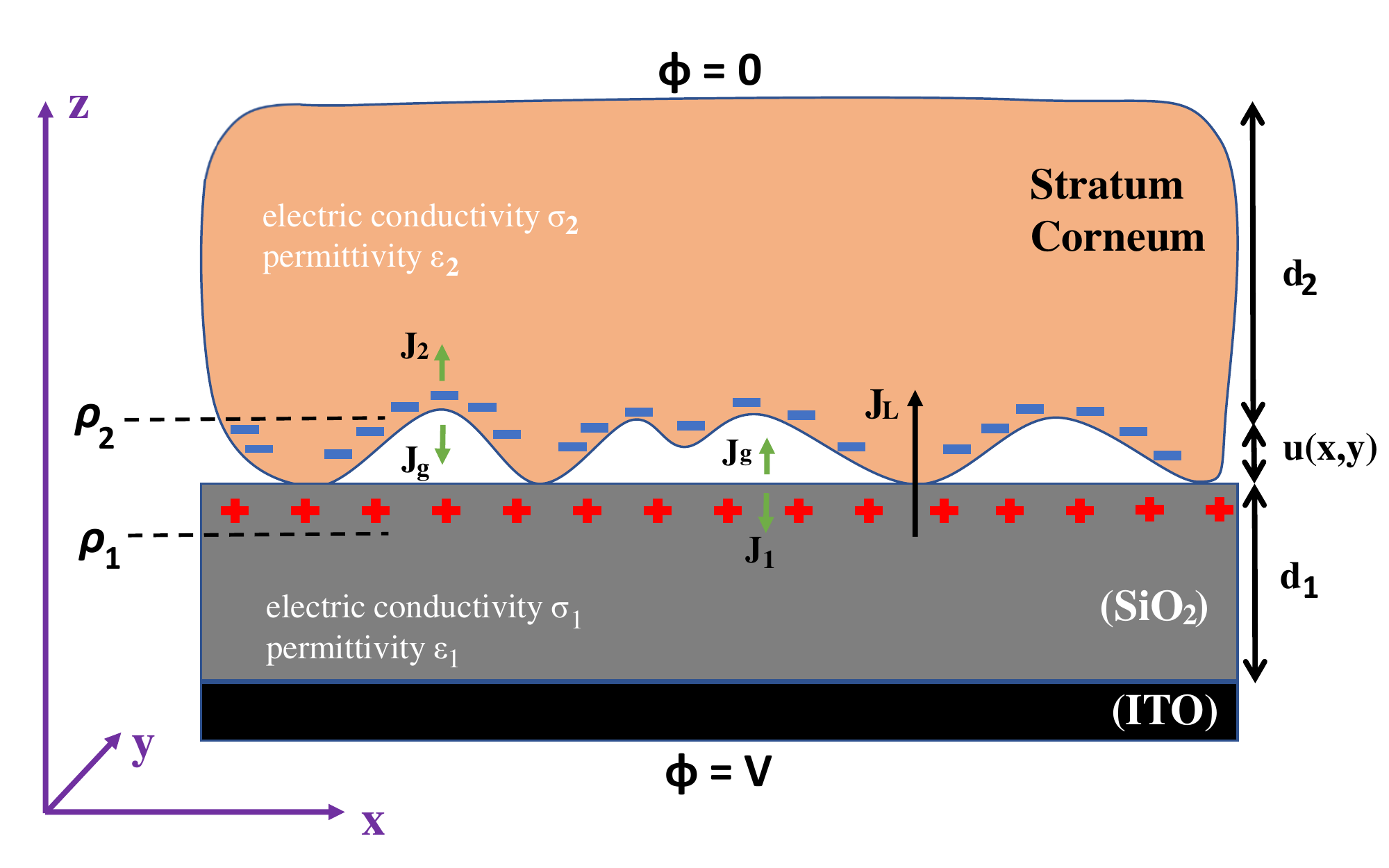}
\caption{A cross-sectional representation of the capacitive touchscreen in contact with a human finger. The touchscreen is composed of a conductive (ITO) layer beneath an insulator layer of SiO$_2$. Only the outermost layer of the human finger (SC) is displayed in the figure, which has a finite conductivity.}
\label{fig:ch3_Schematic}
\end{figure}

Fig. \ref{fig:ch3_Schematic} represents a cross-section of human finger on a touchscreen. As shown in this figure, the SC layer of finger is in full contact with the SiO$_2$ layer of touchscreen in some small regions only due to the multi-scale roughness of fingerpad surface. These contact regions form the real contact area. Since the SC is not a perfect insulator, a small number of electrical charges moves from the SC layer to the surface of the dielectric (in other words, some amount of leakage current density (J$_L$) flows from SiO$_2$ to the SC layer).

\subsection{Electrical Model}\label{sec:ch3_ElectricalModel}
We have developed an electrical model to investigate the charge transfer between human finger and touchscreen under electroadhesion. In our model, a voltage source is attached to the conductive ITO layer and the finger is electrically grounded. As a result, negative electric charges (electrons) travel from SiO$_2$ towards the voltage source and positive charges remain inside the material. On the other hand, electrons move from the voltage source towards the SC layer and negative charges accumulate inside this layer (see Fig. \ref{fig:ch3_Schematic}). As long as the air gap acts like a dielectric layer between the SiO$_2$ and SC, the positive charges accumulated in SiO$_2$ and the negative ones in the SC produce an external electric field. This electric field creates an electrostatic attraction force on the finger (F$_e$). The amplitude of the electrostatic force is proportional to the square of the input voltage amplitude and hence, the number of accumulated charges at the interfaces of SiO$_2$-air and air-SC.

In our model, the subscripts 1, g, and 2 denote SiO$_2$, air gap, and the SC, respectively. Since the electrical properties of the SC vary with stimulation frequency \cite{yamamoto1976dielectric}, the complex dielectric function for this layer can be written as \cite{chen2004electrical,persson2021general}:
\begin{equation}\label{eq1}
    \varepsilon_2=\varepsilon_2^{\prime}-j\varepsilon_2^{\dprime}=\varepsilon_{2}^\prime-j\frac{\sigma_{2}}{\omega\varepsilon_0}
\end{equation}
where, $\varepsilon_2^{\prime}$ and $\varepsilon_2^{\dprime}$ are the real and imaginary parts of the permittivity function, respectively, $\omega$ is the stimulation frequency, $\varepsilon_0$ is the permittivity of free space ($8.854\times10^{-12}\;F.m^{-1}$), and $\sigma_{2}$ is the conductivity of the SC layer. The complex permittivity of the SC is a function of frequency and the values of $\varepsilon_{2}^\prime$ and $\sigma_{2}$ in Eq. \ref{eq1} are adopted from \cite{yamamoto1976dielectric}.

The surface charge densities at the interfaces of SiO$_2$-air and air-SC are represented by $\rho_1$ and $\rho_2$, respectively. Defining J as the current density, the following expression is valid for the SiO$_2$-air interface as (see Appendix \ref{appendix:charge continuity} for details) \cite{nakamura2017modeling,zhang1999modeling}:
\begin{equation}\label{eq4}
    J_1+J_g=-\frac{\partial \rho_1}{\partial t}
\end{equation}
and similarly for the interface of air-SC one can write:
\begin{equation}\label{eq5}
    J_g+J_2=-\frac{\partial \rho_2}{\partial t}
\end{equation}
Referring to the definition of electric flux density, one can write $D_{n_1}=\varepsilon_1 E_1$, $D_{n_g}=\varepsilon_g E_g$, and $D_{n_2}=\varepsilon_2 E_2$, where D$_{n}$ is the normal component of the electric flux density, and E is the electric field. Based on the Gauss law, the relationship between the flux densities and the charge densities for each interface can be expressed as \cite{cheng1989field}:
\begin{equation}\label{eq6}
    \rho_1=D_{n_g}-D_{n_1}=\varepsilon_g E_g-\varepsilon_1 E_1
\end{equation}
\begin{equation}\label{eq7}
    \rho_2=D_{n_2}-D_{n_g}=\varepsilon_2 E_2-\varepsilon_g E_g
\end{equation}

In addition, the current densities for each layer can be written as $J_1=\sigma_1 E_1$, $J_g=\sigma_g E_g$, and $J_2=\sigma_2 E_2$. Note that in the equation for the first interface (Eq. \ref{eq4}), the direction of J$_1$ is opposite to the direction of the electric field E$_1$ and in the equation for the second interface (Eq. \ref{eq5}), the direction of J$_g$ is opposite to the direction of the electric field E$_g$. So, a negative sign must be considered in front of the current density J$_1$ in Eq. \ref{eq4} and the current density J$_g$ in Eq. \ref{eq5}. Hence, Eqs. \ref{eq4} and \ref{eq5} can be re-written as:
\begin{equation}\label{eq8}
    -\sigma_1 E_1+\sigma_g E_g=\frac{\partial}{\partial t}(\varepsilon_1 E_1-\varepsilon_g E_g)
\end{equation}
\begin{equation}\label{eq9}
    -\sigma_g E_g+\sigma_2 E_2=\frac{\partial}{\partial t}(\varepsilon_g E_g-\varepsilon_2 E_2)
\end{equation}
The boundary condition is expressed as:
\begin{equation}\label{eq10}
    V=E_1d_1+E_g u+E_2d_2
\end{equation}
where, d$_1$ and d$_2$ are the thicknesses of SiO$_2$ and SC layers, respectively and V is the excitation voltage applied to the ITO layer of the touchscreen. The interfacial separation between the SiO$_2$ and SC layers is denoted by u and it is a function of position in xy plane \cite{persson2018dependency}. The electric fields E$_1$, E$_g$, and E$_2$ are obtained using Eqs. \ref{eq6}, \ref{eq7}, and \ref{eq10} as:
\begin{equation}\label{eq11}
    E_1=\frac{V-\left(\frac{u}{\varepsilon_g}+\frac{d_2}{\varepsilon_2}\right)\rho_1-\frac{d_2}{\varepsilon_2}\rho_2}{d_1+\varepsilon_1\left(\frac{u}{\varepsilon_g}+\frac{d_2}{\varepsilon_2}\right)}
\end{equation}
\begin{equation}\label{eq12}
    E_g=\frac{V+\frac{d_1}{\varepsilon_1}\rho_1-\frac{d_2}{\varepsilon_2}\rho_2}{u+\varepsilon_g\left(\frac{d_1}{\varepsilon_1}+\frac{d_2}{\varepsilon_2}\right)}
\end{equation}
\begin{equation}\label{eq13}
    E_2=\frac{V+\frac{d_1}{\varepsilon_1}\rho_1+\left(\frac{d_1}{\varepsilon_1}+\frac{u}{\varepsilon_g}\right)\rho_2}{d_2+\varepsilon_2\left(\frac{d_1}{\varepsilon_1}+\frac{u}{\varepsilon_g}\right)}
\end{equation}
Now that the electric fields are calculated, we need to calculate the electric charge densities at both interfaces. Eqs. \ref{eq4} and \ref{eq5} can be rewritten as:
\begin{equation}\label{eq14}
    -\frac{\partial \rho_1}{\partial t}=-\sigma_1 E_1+\sigma_g E_g
\end{equation}
\begin{equation}\label{eq15}
    -\frac{\partial \rho_2}{\partial t}=-\sigma_g E_g+\sigma_2 E_2
\end{equation}
Substituting the electric fields from Eqs. \ref{eq11}, \ref{eq12}, and \ref{eq13} into Eqs. \ref{eq14} and \ref{eq15} and rearranging the parameters of both equations with respect to the charge densities and the applied voltage give:
\begin{equation}\label{eq16}
    \begin{aligned}
        -\frac{\partial \rho_1}{\partial t}=\left[\frac{\sigma_1\left(\frac{u}{\varepsilon_g}+\frac{d_2}{\varepsilon_2}\right)}{d_1+\varepsilon_1\left(\frac{u}{\varepsilon_g}+\frac{d_2}{\varepsilon_2}\right)}+\frac{\sigma_g \frac{d_1}{\varepsilon_1}}{u+\varepsilon_g\left(\frac{d_1}{\varepsilon_1}+\frac{d_2}{\varepsilon_2}\right)}\right]\rho_1 \\
        +\left[\frac{\sigma_1 \frac{d_2}{\varepsilon_2}}{d_1+\varepsilon_1\left(\frac{u}{\varepsilon_g}+\frac{d_2}{\varepsilon_2}\right)}-\frac{\sigma_g \frac{d_2}{\varepsilon_2}}{u+\varepsilon_g\left(\frac{d_1}{\varepsilon_1}+\frac{d_2}{\varepsilon_2}\right)}\right]\rho_2\\
        +\left[-\frac{\sigma_1}{d_1+\varepsilon_1\left(\frac{u}{\varepsilon_g}+\frac{d_2}{\varepsilon_2}\right)}+\frac{\sigma_g}{u+\varepsilon_g\left(\frac{d_1}{\varepsilon_1}+\frac{d_2}{\varepsilon_2}\right)}\right]V
    \end{aligned}
\end{equation}
\begin{equation}\label{eq17}
    \begin{aligned}
        -\frac{\partial \rho_2}{\partial t}=\left[\frac{\sigma_2\frac{d_1}{\varepsilon_1}}{d_2+\varepsilon_2\left(\frac{d_1}{\varepsilon_1}+\frac{u}{\varepsilon_g}\right)}-\frac{\sigma_g\frac{d_1}{\varepsilon_1}}{u+\varepsilon_g\left(\frac{d_1}{\varepsilon_1}+\frac{d_2}{\varepsilon_2}\right)}\right]\rho_1\\
        +\left[\frac{\sigma_2\left(\frac{d_1}{\varepsilon_1}+\frac{u}{\varepsilon_g}\right)}{d_2+\varepsilon_2\left(\frac{d_1}{\varepsilon_1}+\frac{u}{\varepsilon_g}\right)}+\frac{\sigma_g\frac{d_2}{\varepsilon_2}}{u+\varepsilon_g\left(\frac{d_1}{\varepsilon_1}+\frac{d_2}{\varepsilon_2}\right)}\right]\rho_2\\
        +\left[\frac{\sigma_2}{d_2+\varepsilon_2\left(\frac{d_1}{\varepsilon_1}+\frac{u}{\varepsilon_g}\right)}-\frac{\sigma_g}{u+\varepsilon_g\left(\frac{d_1}{\varepsilon_1}+\frac{d_2}{\varepsilon_2}\right)}\right]V
    \end{aligned}
\end{equation}

If we define a, b, c, d, e, and f as the equivalent coefficients, the above equations can be simplified to:
\begin{equation}\label{eq18}
    -\frac{\partial \rho_1}{\partial t}=a\rho_1+b\rho_2+c V
\end{equation}
\begin{equation}\label{eq19}
    -\frac{\partial \rho_2}{\partial t}=d\rho_1+e\rho_2+f V
\end{equation}
These coupled equations are transferred to the Laplace domain to solve for charge densities as:
\begin{equation}\label{eq22}
    P_1(s)=\frac{-c(s+e)+f b}{(s+e)(s+a)-d b}V(s)
\end{equation}
\begin{equation}\label{eq23}
    P_2(s)=\frac{-f(s+a)+c d}{(s+e)(s+a)-db}V(s)
\end{equation}
Note that initially all the materials are electrically neutral and hence, the initial conditions for both charge densities are zero. Therefore,
\begin{equation}\label{eqLap1}
    \rho_1 (t)=\mathcal{L}^{-1} \{P_1(s)\}
\end{equation}
\begin{equation}\label{eqLap2}
    \rho_2 (t)=\mathcal{L}^{-1} \{P_2(s)\}
\end{equation}
The time domain solutions for $\rho_1(t)$ and $\rho_2(t)$ are given in Appendix \ref{appendix:charge densities}.

The electrical relaxation time of a material is defined by the ratio of its permittivity to conductivity as $\tau=\varepsilon/\sigma$. According to the Maxwell-Wagner effect \cite{bhushan2012encyclopedia}, if two materials in contact have different relaxation times, charges can accumulate at the interface and current flows from one material to another when there is a potential difference between them. In fact, this current is called the leakage current. Let us discuss the current leakage in steady-state for the regions where SiO$_2$ and the SC are in full contact (Fig. \ref{fig:ch3_Schematic}). In this case, $\nabla. \vec{J_L}=0$, where J$_L$ is the density of leakage current from the SC to the surface of the touchscreen (SiO$_2$). The differential form of the Gauss law is defined as:
\begin{equation}\label{eq24-1}
    \nabla. \vec{D}=\rho_L
\end{equation}
Considering the electric flux density, the following equation is obtained:
\begin{equation}\label{eq24}
    \nabla. \vec{D}=\nabla. \varepsilon \vec{E}=\nabla. \varepsilon \frac{\vec{J_L}}{\sigma}=\nabla. \tau \vec{J_L}
\end{equation}
Substituting Eq. \ref{eq24} into Eq. \ref{eq24-1}, the density of leakage current is equal to:
\begin{equation}\label{eq26}
    J_L=\frac{\rho_L}{\tau_2-\tau_1}
\end{equation}
where, $\tau_1$ and $\tau_2$ are the relaxation times of SiO$_2$ and the SC, respectively, and $\rho_L$ is the steady-state value of leakage charge density, which is obtained from the steady-state values of interface charge densities as:
\begin{equation}\label{eq27}
    \rho_L=\rho_2-\rho_1
\end{equation}
The leakage decreases the electric field at the air gap. The reduction in the electric field at the air gap due to the leakage from the SC to the surface of the touchscreen can be expressed as:
\begin{equation}\label{electric field leakage}
    E_L=\frac{J_L}{\alpha u}
\end{equation}
where, $\alpha$ is the electric contact conductivity and calculated using the equations in Appendix \ref{appendix:electrical contact conductivity}. It is a function of stimulation frequency and the values used for its calculation are tabulated in Table. \ref{table:ch3_Parameters}. Defining $h_0=d_1/\varepsilon_1 + d_2/\varepsilon_2$ as the effective thickness, and also referring to Eq. \ref{eq12}, the total electric field at the air gap can now be written as:
\begin{equation}\label{eq elec field}
    \begin{aligned}
    \begin{split}
        E_{tot}={}& E_g - E_L\\
        &=\frac{V}{u+\varepsilon_g h_0}+\frac{d_1}{\varepsilon_1}\frac{\rho_1}{u+\varepsilon_g h_0}-\frac{d_2}{\varepsilon_2}\frac{\rho_2}{u+\varepsilon_g h_0}-\frac{J_L}{\alpha u}
        \end{split}
    \end{aligned}
\end{equation}
The first term of this equation is exactly the same electric field equation obtained by Persson \cite{persson2018dependency} for the case of no charge leakage while the second and the third terms are added in this study to take into account the accumulation of charges at the interfaces and the fourth term is added to include the leakage effect.

\subsection{Mechanistic Contact Model}\label{sec:ch3_MechanisticContactModel}

An important issue in investigating the contact interactions between human finger and touchscreen under electroadhesion is the air gap, which is nonuniform due to the relative motion between them and multi-scale roughness of the fingerpad. Persson's contact theory is utilized in this study to model variability in air gap, which takes into account the multi-scale nature of contacting surfaces \cite{persson2001theory,persson2006contact}. We use a probability function for the variable air gap as suggested in Persson's contact theory \cite{persson2018dependency, sirin2019electroadhesion} and integrate the electric field estimated in Eq. \ref{eq elec field} into the Maxwell stress tensor to estimate the electrostatic forces as a function of frequency.

The zz-component of the Maxwell stress tensor is utilized to calculate the normal electrostatic pressure as:
\begin{equation}\label{eq sigma}
    \sigma_{zz} = \ddfrac{1}{2}\varepsilon_0 E_{tot}^2
\end{equation}
Defining the probability distribution of interfacial separations as P(p,u) \cite{almqvist2011interfacial}, the average stress over the surface roughness is written as:
\begin{equation}\label{eq ensemble sigma}
    \langle \sigma_{zz} \rangle = \ddfrac{1}{2}\varepsilon_0 \int du\, P(p,u) E_{tot}^2
\end{equation}
where, $\langle...\rangle$ denotes ensemble averaging. Note that E$_{tot}$ is the difference between E$_g$ and E$_L$ (Eq. \ref{eq elec field}) and the limits of the integral for E$_g$ are from $0$ to $\infty$ while the limits for E$_L$ are from a$_c$ to $\infty$, where a$_c$= 10 $nm$ is a cut-off distance taken from \cite{persson2018dependency}. Based on Persson's contact theory \cite{almqvist2011interfacial,yang2008contact,persson2007relation}, if the nominal contact pressure applied by the finger on touchscreen (p$_0$) is not too high and not too low, the electrostatic pressure p$_e=\langle \sigma_{zz} \rangle$ adds on the external load and makes the total loading pressure p = p$_0+$p$_e$. Finally, electrostatic force can be calculated by multiplying the electrostatic pressure with the real area of contact as:
\begin{equation}\label{eq electrostatic force theory}
    F_e = p_e A_{real}
\end{equation}
More discussion on Persson's contact mechanics theory is given in Appendix \ref{appendix:persson theory}.

\begin{figure}[!b]
\centering
\includegraphics[width=.8\linewidth]{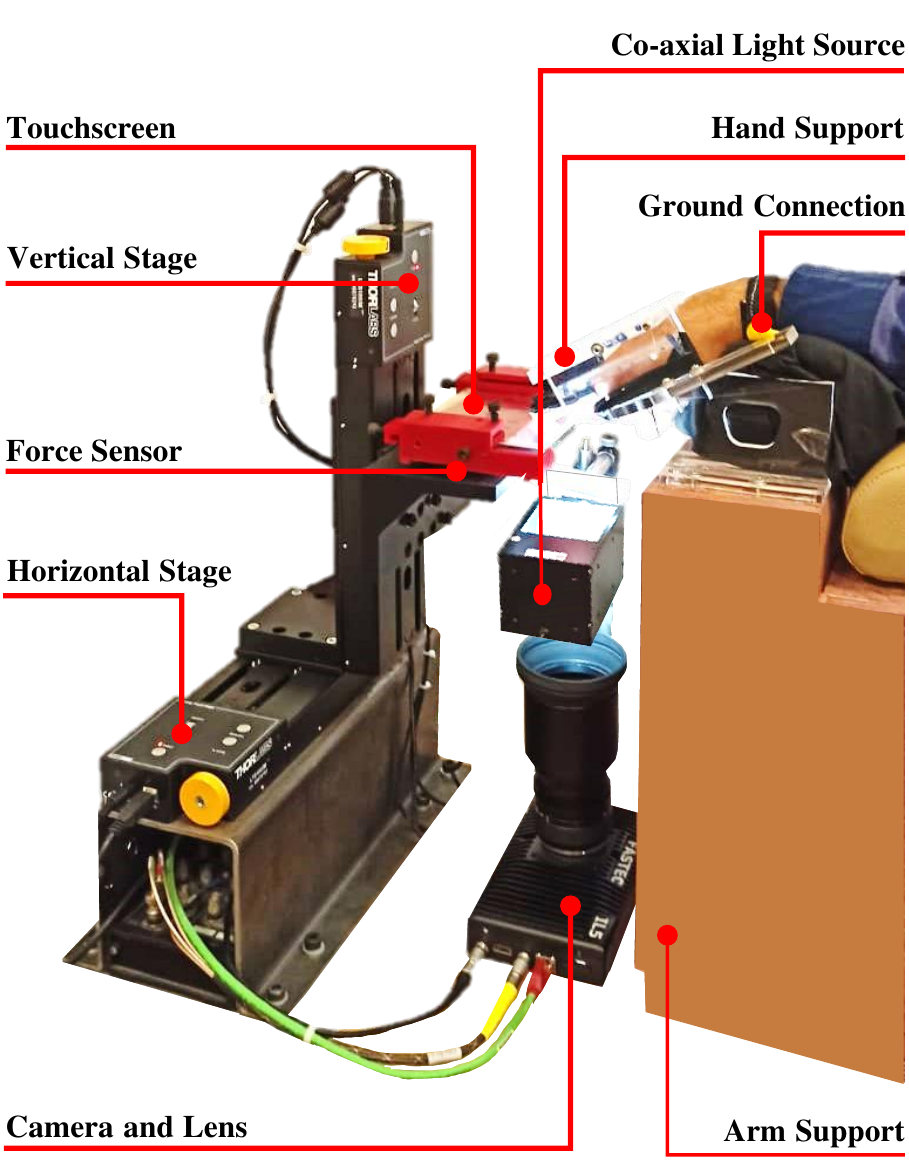}
\caption{The experimental set-up used in our study to measure the electrostatic forces acting on human finger.}
\label{fig:ch3_FrictionSetup}
\end{figure}

\section{Experimental Methods}\label{sec:ch3_ExperimentalMethods}
The experimental set-up used in this study (Fig. \ref{fig:ch3_FrictionSetup}) is the one designed and developed by Ozdamar et al. \cite{ozdamar2020step}. The major components of this set-up include a force transducer (Mini40-SI-80-4, ATI Inc.) placed under a capacitive touchscreen (SCT3250, 3M Inc.) to measure the normal and tangential forces acting on finger, two linear translational stages (LTS150, Thorlabs Inc.) to move the touchscreen with respect to finger in normal and tangential directions, a high-speed camera (IL5H, Fastec Imaging Inc.) to capture the images of fingerpad, a co-axial light source (C50C, Contrastech Inc.), a waveform generator (33220A, Agilent Inc.), and a piezo driver/amplifier (PZD700A M/S, Trek Inc.) to apply the desired voltage signals to the ITO layer of the touchscreen. The normal and tangential forces acting on finger were acquired at 2.5 kHz using a DAQ card (PCIe-6034E, National Instruments Inc.).

The goal of the experiment was to measure the CoF between finger and touchscreen under electroadhesion for different stimulation frequencies (Fig. \ref{fig:ch3_FrictionResults}). The amplitude of the voltage signal applied to the touchscreen was kept constant at 75 volts, but its frequency was varied from 1 Hz to 1 MHz (1 Hz, 10 Hz, 50 Hz, 100 Hz, 250 Hz, 500 Hz, 1 kHz, 10 kHz, 100 kHz, and 1 MHz).

Data was collected from a 28 years old male participant. Before the experiment, the surface of touchscreen was cleaned carefully with alcohol and the participant washed his hands with soap and water and dried them at room temperature. The index finger of the participant was placed inside the hand support of the set-up to keep it stationary while the touchscreen under his finger was moved in tangential direction with a constant velocity of 20 mm/s. Using a PID controller, the normal force applied by his finger on the touchscreen was kept constant at 0.5 N. The participant was asked to stay stable during the experiment and a wrist band was utilized to make him electrically grounded. A consent form, approved by the Ethical Committee for Human Participants of Koc University, was read and signed by the participant before the experiment. The study conformed to the principles of the Declaration of Helsinki and the experiment was performed in accordance with relevant guidelines and regulations.

\section{Results}\label{sec:ch3_Results}
\subsection{Experimental Results}\label{experimental results}
\begin{figure*}[!t]
\centering
\includegraphics[width=1\linewidth]{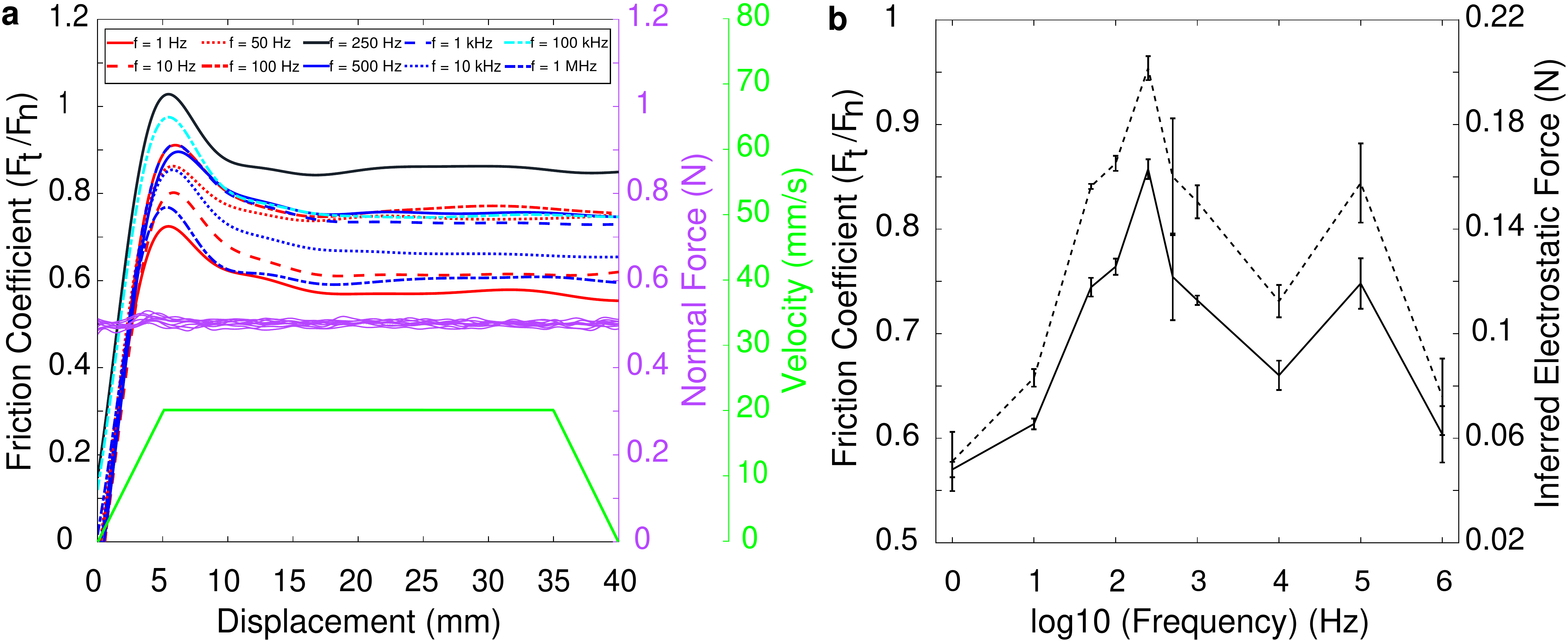}
\caption{Experimental results; (a) Coefficient of friction (CoF), normal force, and the velocity profile of the horizontal stage as a function of displacement (b) steady-state values of CoF (solid) and electrostatic attraction force (dashed) as a function of stimulation frequency.}
\centering
\label{fig:ch3_FrictionResults}
\end{figure*}

The experiment was performed in 3 separate sessions on 3 different days. Data (normal and tangential forces) were collected 3 times (i.e. 3 trials) for each frequency in each session. The CoF was obtained by dividing the recorded tangential force to the normal force. The CoF curve reported in Fig. \ref{fig:ch3_FrictionResults}a are the mean values of 9 trials (3 trials$/$session $\times$ 3 sessions) recorded for each stimulation frequency. The mean value of normal force for each stimulation frequency and velocity profile of the horizontal stage as a function of displacement are also presented in the same figure. The steady-state region for all CoF curves was taken as the interval from 15 mm to 35 mm. The steady-state value of CoF for each stimulation frequency was calculated by averaging the instantaneous values in this region, (see Fig. \ref{fig:ch3_FrictionResults}b, the deviation bars in the figure are the standard errors of means). As shown in Fig. \ref{fig:ch3_FrictionResults}b, CoF increases with increasing stimulation frequency until 250 Hz and then decreases. In addition, it has a relatively high value at 100 kHz.

The electrostatic force acting on the finger can be inferred from the experimental CoF data as suggested in \cite{basdogan2020modeling}:
\begin{equation}\label{electrostatic experiment}
    F_e=\left(1-\frac{\mu ^{OFF}}{\mu ^{ON}}\right)F_n
\end{equation}
where, $\mu^{OFF}$ and $\mu^{ON}$ are the measured CoF when electroadhesion is off and on, respectively. The mean value of $\mu^{OFF}$ (average of 9 trials: 3 trials$/$session $\times$ 3 sessions) was measured as 0.512 with a standard error of mean of 0.006. Fig. \ref{fig:ch3_FrictionResults}b presents the dependency of the electrostatic force on the stimulation frequency for the normal force of 0.5 N.

Furthermore, for the same normal force, we measured the apparent contact area of the fingertip of the participant using the high-speed camera as 100 mm$^2$. The measurement procedure is available in Ref. \cite{ozdamar2020step}.

\begin{table*}[!b]
\renewcommand{\arraystretch}{1.3}
\caption{The list of parameters and their values used in the model}
\label{table:ch3_Parameters}
\centering
\begin{tabular}{|l|l|l|c|}
\hline
\textbf{Parameter}&\textbf{Definition}&\textbf{Value}&\textbf{Unit}\\ \hline\hline
$\varepsilon_1$&Relative permittivity of SiO$_2$&3.9&\\ \hline
d$_1$, d$_2$&Thicknesses of SiO$_2$ and SC&1, 200&$\mu$m\\ \hline
$\sigma_1$, $\sigma_2$& Conductivities of SiO$_2$ and air&10$^{-13}$, 10$^{-14}$&1$/\Omega$m\\ \hline
Y$_1$, Y$_2$&Elastic moduli of SiO$_2$ and SC&70$\times10^{3}$, 10&MPa\\ \hline
$\nu_1$, $\nu_2$&Poisson's ratios of SiO$_2$ and SC&0.15, 0.5&\\ \hline
$h_{rms}$&RMS roughness amplitude of fingerpad&22&$\mu$m\\ \hline
$H$&Hurst exponent&0.86& \\ \hline
$q_L$&The shortest cut-off wavevector&9$\times 10^2$&1/m\\\hline
$q_0$&The long-distance roll-off wavevector&8$\times 10^3$&1/m\\ \hline
$q_1$&The short-distance cut-off wavevector&1$\times 10^{10}$&1/m\\ \hline
$p_0$&Applied normal pressure&5&kPa\\ \hline
$A_0$&Apparent contact area&100&mm$^2$\\ \hline
\end{tabular}
\end{table*}

\subsection{Modeling Results}\label{sec:ch3_ModelingResults}
\begin{figure}[!t]
\centering
\includegraphics[width=1\linewidth]{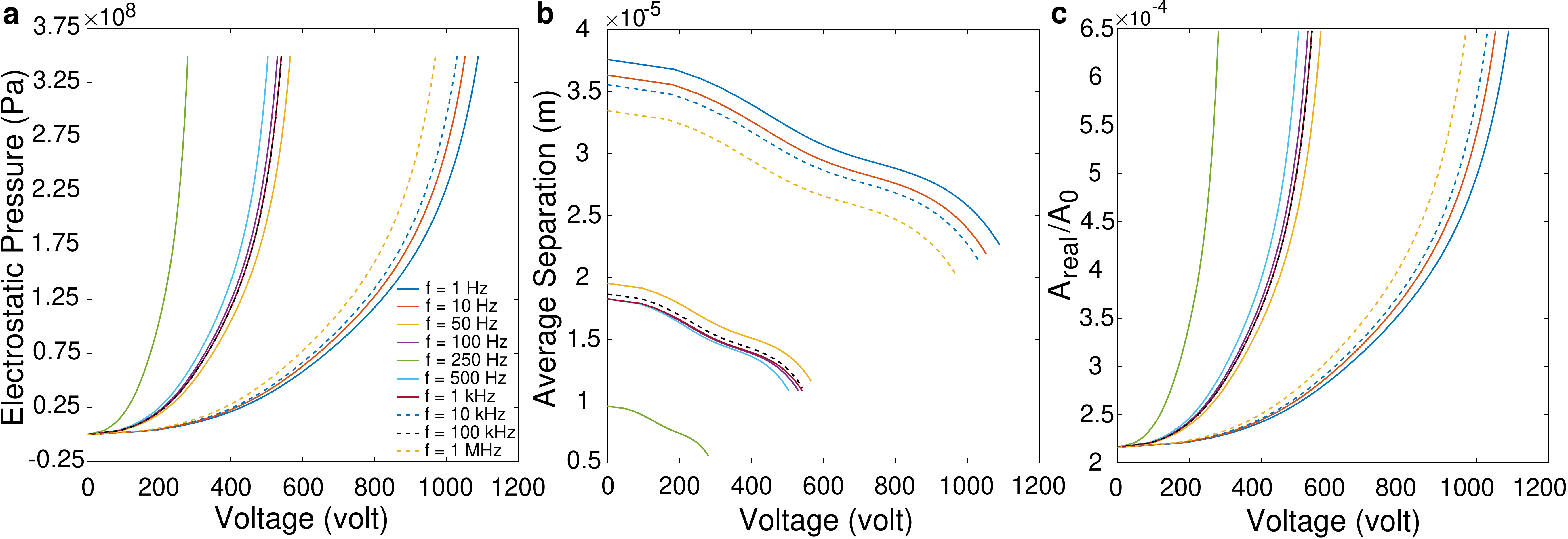}
\caption{(a) The electrostatic pressure, (b) the average interfacial separation, and (c) the contact area ratio with respect to the applied voltage amplitude.}
\label{fig:ch3_PressureGapArea}
\end{figure}

The values of parameters used in the proposed model are tabulated in Table. \ref{table:ch3_Parameters}. These values were taken from the related references in the literature \cite{basdogan2020modeling,sirin2019electroadhesion}. Since the air gap between the SiO$_2$ and SC layers is varying, the electrostatic pressure, the average separation between them, and the contact area ratio are calculated for different voltage amplitudes and presented in Fig. \ref{fig:ch3_PressureGapArea}. As shown in the figure, the values of those parameters change as a function of stimulation frequency for a constant AC voltage amplitude. They can simply be obtained by drawing a vertical line from x-axis and intersecting the curves in Fig. \ref{fig:ch3_PressureGapArea}a, b, and c for all frequencies.  

\begin{figure*}[!b]
\centering
\includegraphics[width=1\linewidth]{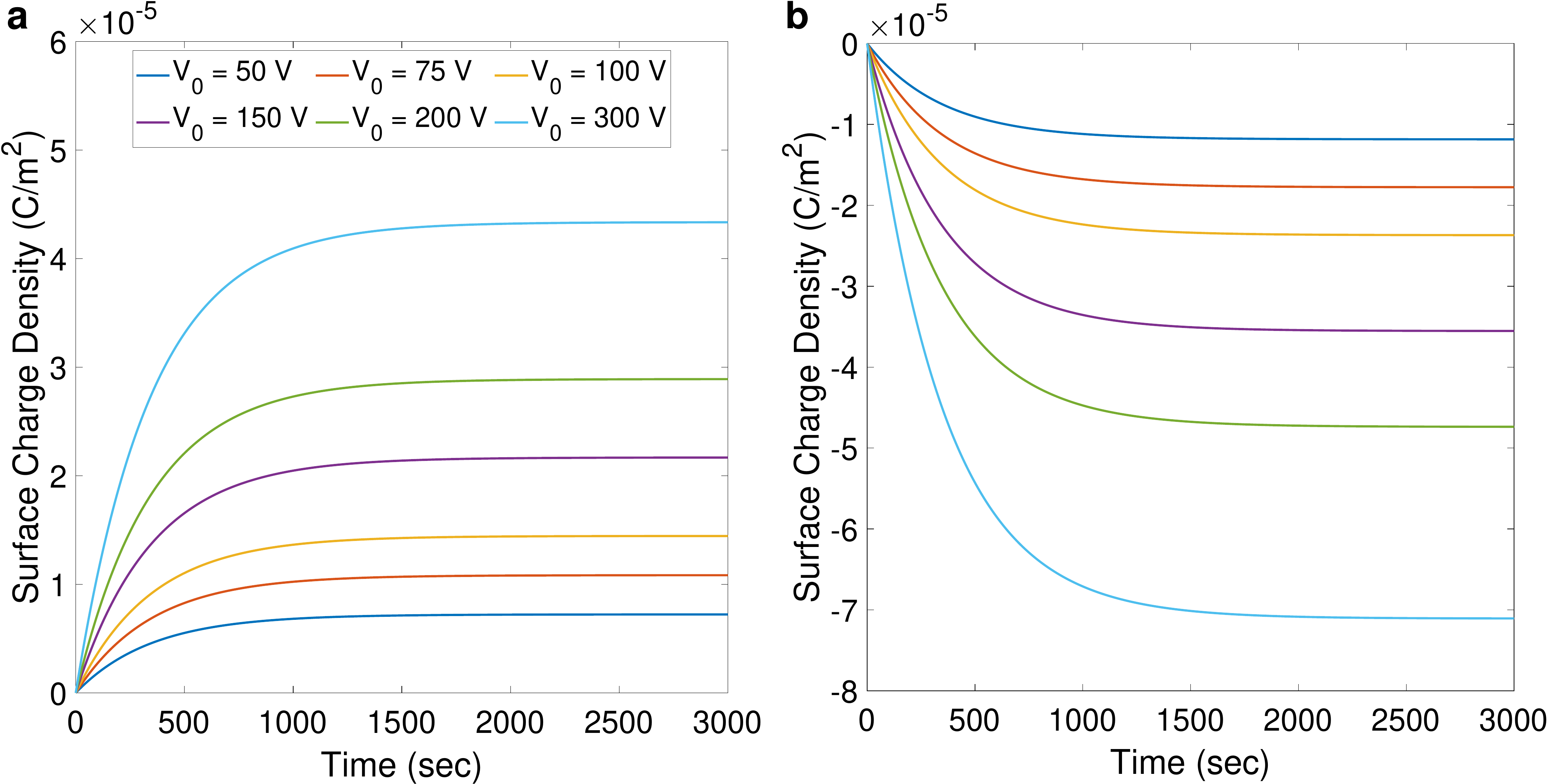}
\caption{Accumulated charges at the interfaces of (a) SiO$_2$-air and (b) air-SC as a function of time for different DC input voltages.}
\centering
\label{fig:ch3_DCCharges}
\end{figure*}

\begin{figure}[!b]
\centering
\includegraphics[width=0.6\linewidth]{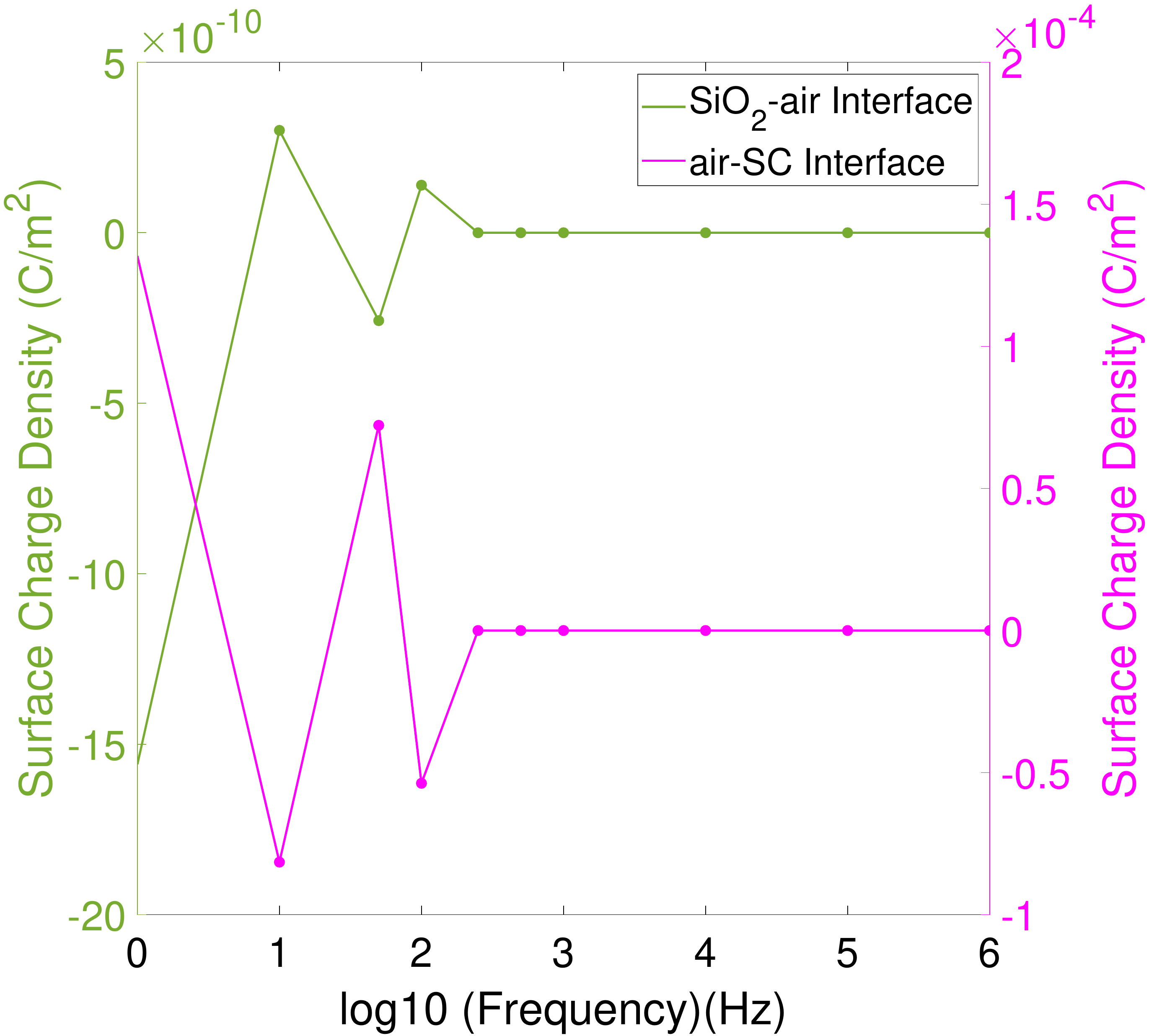}
\caption{Accumulated charges at the interfaces of SiO$_2$-air and air-SC at steady-state with respect to stimulation frequencies.}
\label{fig:ch3_ACCharges}
\end{figure}

For a DC input voltage ($V(s)=V_0/s$), the accumulated charges at both interfaces were calculated directly using Eqs. \ref{eqLap1} and \ref{eqLap2}. The amount of charge accumulated at the interfaces of SiO$_2$-air and air-SC were plotted as a function of time for different voltage amplitudes in Fig. \ref{fig:ch3_DCCharges}a and Fig. \ref{fig:ch3_DCCharges}b, respectively. For the results reported in Fig. \ref{fig:ch3_DCCharges} the amplitude of the voltage signal for the simulations was selected as 50, 75, 100, 150, 200, and 300 volts. Using the proposed model, we also investigated the behavior of interface charges for different stimulation frequencies for an AC input voltage of 75 volts (Fig. \ref{fig:ch3_ACCharges}). Dark green and magenta colors are used to distinguish the SiO$_2$-air and air-SC charge densities, respectively. Note that the electrical permittivity and resistivity of the SC change with the stimulation frequency and Eq. \ref{eq1} accounts for this dependency. The values of the variable air gap for calculating surface charges under DC and AC voltages are taken from Fig. \ref{fig:ch3_PressureGapArea}b.

The charge densities at the interfaces depend on the stimulation frequency. Naturally, the density of the leaked charge, which is the difference between the charge densities at the interfaces (Eq. \ref{eq27}), also depends on the frequency. Since the electrostatic force is a function of the electric field, which is calculated by using the charge densities, its magnitude varies with the stimulation frequency. Furthermore, the electrical properties of the SC change with frequency, which also affects the magnitude of electrostatic force.

\begin{figure*}[!b]
\centering
\includegraphics[width=1\linewidth]{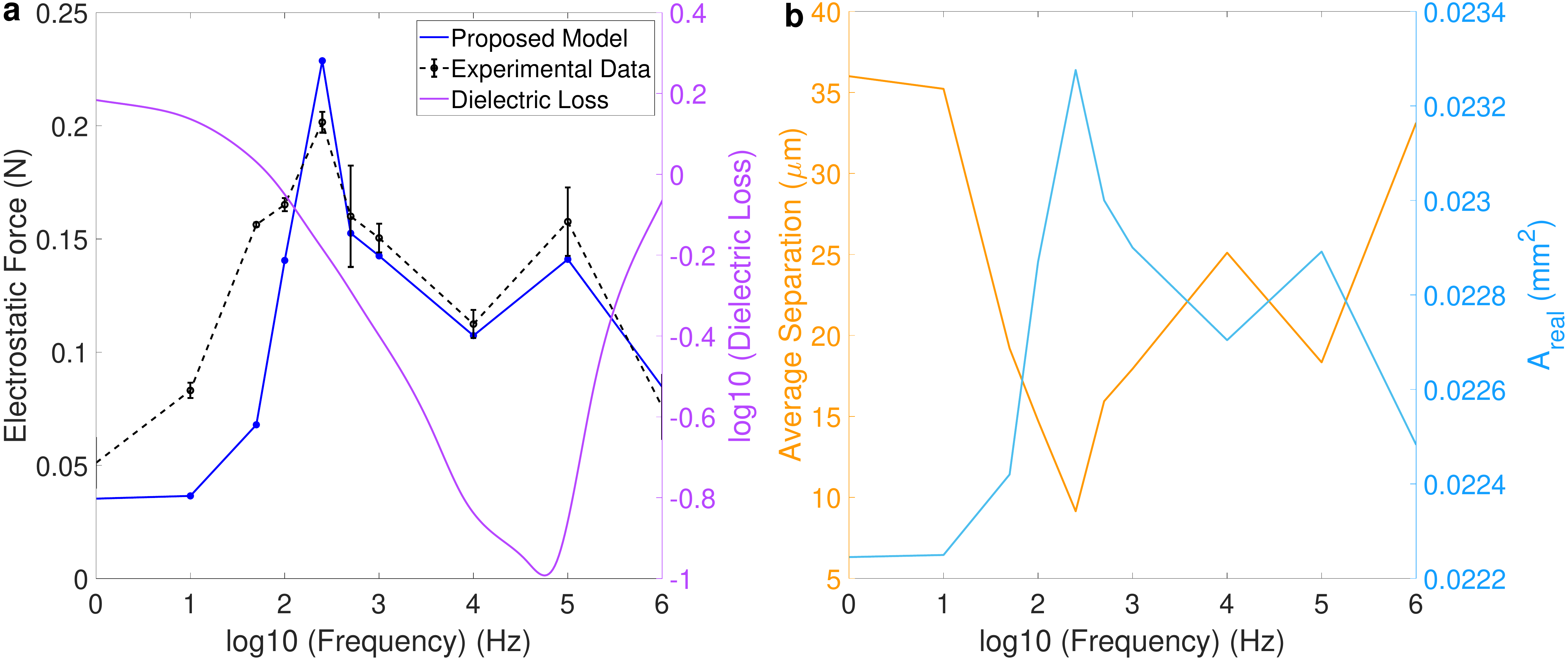}
\caption{(a) Electrostatic forces inferred from the experimental data (dashed black) and the one estimated by the model (dark blue), and the electric loss for the SC (purple), and (b) average separation (orange) and real contact area (light blue) as a function of the stimulation frequency.}
\centering
\label{fig:ch3_ElectrostaticForce}
\end{figure*}

For an AC voltage amplitude of 75 volts, the change in electrostatic force as a function of stimulation frequency is shown in Fig. \ref{fig:ch3_ElectrostaticForce}a (see the dark blue-colored curve). This curve shows the electrostatic force response, in which both the effect of charge leakage and the frequency-dependent electrical properties of the SC were taken into account. In Fig. \ref{fig:ch3_ElectrostaticForce}b, we report the change in average separation distance (orange) and the real contact area (light blue) as a function of stimulation frequency. All results reported in Fig. \ref{fig:ch3_ElectrostaticForce} are based on the results given in Fig. \ref{fig:ch3_PressureGapArea}.

\begin{figure*}[!t]
\centering
\includegraphics[width=1\linewidth]{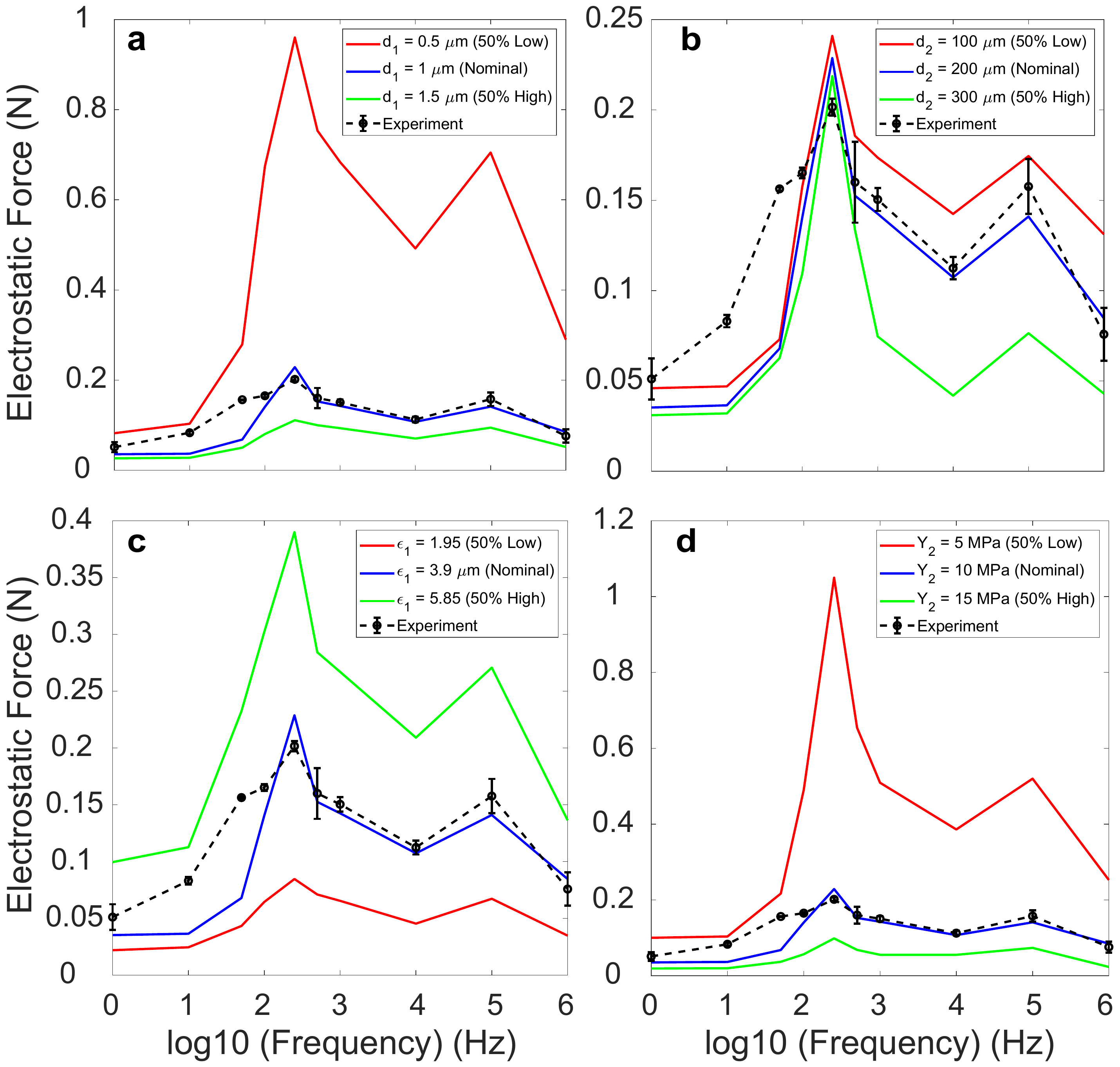}
\caption{Dependency of the electrostatic force on (a) thickness of the insulator layer in touchscreen, (b) thickness of the SC, (c) permittivity of the insulator layer in touchscreen, and (d) elastic modulus of the SC.}
\centering
\label{fig:ch3_Comparison}
\end{figure*}

\begin{figure*}[!t]
\centering
\includegraphics[width=1\linewidth]{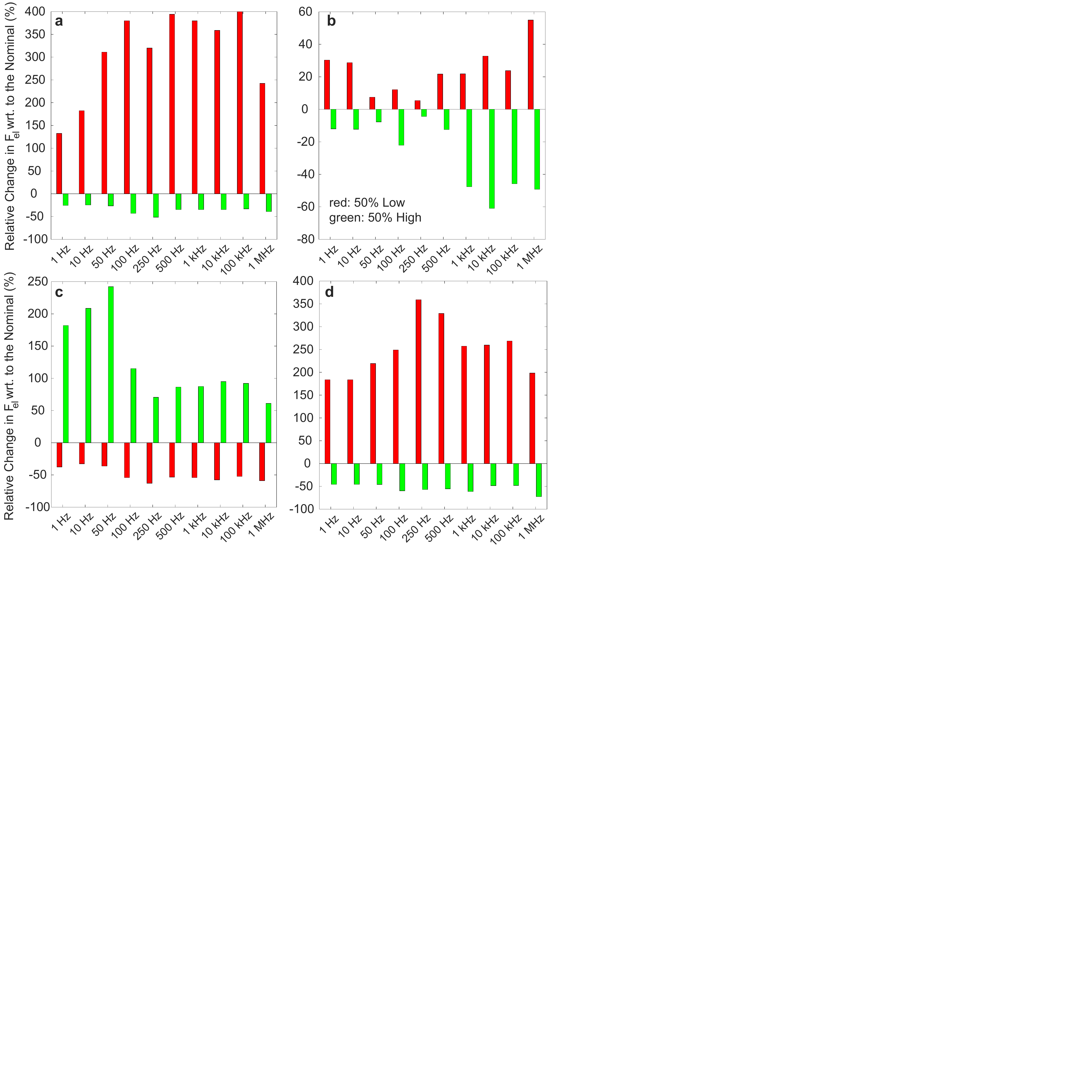}
\caption{The relative percentage change in electrostatic force for $50\%$ decrease and $50\%$ increase of (e) thickness of the insulator layer in touchscreen, (f) thickness of the SC, (g) permittivity of the insulator layer in touchscreen, and (h) Young's modulus of the SC with respect to the optimal case.}
\centering
\label{fig:ch3_ComparisonBar}
\end{figure*}

Since our proposed model estimates the magnitude of electrostatic force as a function of frequency well, we can now investigate the influence of some parameters on the force response. Fig. \ref{fig:ch3_Comparison} shows the results of this investigation for different thicknesses of the insulator layer of touchscreen (a) and the SC layer (b), the permittivity of the insulator layer of touchscreen (c), and the Young's modulus of the SC layer (d). In all plots, the blue-colored curve is the closest estimation of the model to the experimental data. The red and green curves present the model outcome for $50\%$ lower and $50\%$ higher parameter values with respect to the nominal ones, respectively. Corresponding to each plot in Fig. \ref{fig:ch3_Comparison}, the percent change in electrostatic force is reported in Fig. \ref{fig:ch3_ComparisonBar}a, b, c, and d for the frequencies of stimulation.

\section{Discussion}\label{sec:ch3_Discussion}
The lumped parameter models proposed in the earlier studies \cite{meyer2013fingertip,vezzoli2014electrovibration,vardar2016effect,vardar2017effect} are based on simple electrical components like capacitors and resistors and are not able to capture the true frequency-dependent behavior of electrostatic forces as observed in our experiment (Fig. \ref{fig:ch3_FrictionResults}b). The electrostatic force response estimated in the earlier modeling studies starts from zero and increases with frequency until it saturates at some value but does not show a decaying behavior as in Fig. \ref{fig:ch3_FrictionResults}b. Moreover, it is known that the electrostatic force still exists at DC voltages, though it has a low magnitude. Hence, a proper model should not return a zero value for electrostatic force at zero frequency. Besides, the electroadhesion occurs due to the exchange of electrical charges at the interfaces of contacting surfaces and hence, fundamental laws of electric field should be used instead of lumped circuit models to investigate the frequency-dependent behavior of electrostatic force between finger and a touchscreen under electroadhesion.

We initially investigated the accumulation of charges at the interfaces for a DC input voltage. As anticipated, positive charges accumulate at the interface of SiO$_2$-air since the conductive ITO layer is attached to the positive port of DC voltage source (Fig. \ref{fig:ch3_DCCharges}a). Negative charges travel from SiO$_2$ to the power source and hence the remaining ones are positive. Since the finger is electrically grounded, negative charges gather at the SC and the ratio of negative to positive charges increases in time. Because electrical relaxation times of SiO$_2$ and SC are different, negative charges drift from the SC to the surface of SiO$_2$ and gather there, resulting in a decrease in the electric field at the air gap.

Fig. \ref{fig:ch3_ACCharges} shows that the accumulation of charges at the interfaces highly depends on the stimulation frequency. Compared to the SiO$_2$-air interface, the amount of charge accumulated at the air-SC interface is higher since the SC has higher conductivity than SiO$_2$. For both interfaces, increasing the stimulation frequency reduces the accumulation of charges, making the total count zero eventually.

In Fig. \ref{fig:ch3_ElectrostaticForce}a, we present the electrostatic forces estimated by our model (dark blue-colored curve) for the stimulation frequencies ranging from 1 to 10$^6$ Hz. We can virtually divide this range into two regions with respect to the peak value of 250 Hz. In the first region (below 250 Hz), charge leakage is significant as the stimulation frequency is low. Fig. \ref{fig:ch3_ACCharges} also supports this claim since most of the charge transfer occurs at low frequencies. Sirin et al. \cite{sirin2019electroadhesion} also argued that the SC acts like a conductor below 30 Hz and the accumulated charges leaks from its inner layer to its outer surface. As reported in Fig. \ref{fig:ch3_ACCharges}, there is no charge transfer at either interface in the second region (above 250 Hz). Our proposed model (dark blue-colored curve) displays a response similar to the one observed in our experiments (dashed black-colored curve) for both regions.

We present the average separation between the finger and touchscreen (orange-colored curve) and the real area of contact (light blue-colored curve) in Fig. \ref{fig:ch3_ElectrostaticForce}b. The ranges of values obtained for the average separation and the real contact area are in line with the values reported in our earlier modeling studies \cite{sirin2019electroadhesion,ayyildiz2018contact}. The real contact area has a similar trend as electrostatic force, where it is initially low, but increases to a maximum value at 250 Hz, and then decreases as the stimulation frequency is increased. On the other hand, the average separation distance follows a trend opposite to that of the electrostatic force.

We also performed a limit analysis (similar to the one performed by Forsbach and He{\ss} \cite{forsbach2021rigorous}) to investigate the behavior of our model at high frequencies. At high frequencies, only the first term of the electric field equation (Eq. \ref{eq elec field}) is effective. Substituting the complex permittivity function of the SC layer from Eq. \ref{eq1} into Eq. \ref{eq elec field} gives:
\begin{equation*}
    E_{tot}\approx\ddfrac{V}{u+\varepsilon_g \left(\ddfrac{d_1}{\varepsilon_1}+\ddfrac{d_2}{\varepsilon^{\prime}_2-j\ddfrac{\sigma_2}{\omega\varepsilon_0}}\right)}
\end{equation*}
At high frequencies ($\omega \xrightarrow[]{} \infty$), the above term is reduced to:
\begin{equation*}
    \lim_{\omega \to \infty} E_{tot}\approx\ddfrac{V}{u+\varepsilon_g \left(\ddfrac{d_1}{\varepsilon_1}+\ddfrac{d_2}{\varepsilon^{\prime}_2}\right)}
\end{equation*}
where, the conductivity term disappears and only the permittivity terms are left. Since the real part of the SC's permittivity decreases with increasing frequency \cite{yamamoto1976dielectric}, the electric field decreases with frequency as well. Therefore, our limit analysis shows that the electrostatic force should decrease with increasing frequency.

A peak value of the electrostatic force was observed around 250 Hz in both the experimental and modeling results, which was not reported in the earlier studies. On the other hand, we observed a relatively high value for the electrostatic force at 100 kHz frequency in our experimental and modeling results (Fig. \ref{fig:ch3_ElectrostaticForce}). Since the electrical properties of the SC change with the stimulation frequency, we hypothesized that the high value of the electrostatic force can be related to the behavior of its complex permittivity function. We know that the real ($\varepsilon^{\prime}$) and imaginary ($\varepsilon^{\dprime}$) parts of the complex permittivity function (Eq. \ref{eq1}) account for the storage and loss of electrical energy, respectively and the dielectric loss is defined by the loss tangent as \cite{kao2004dielectric}:
\begin{equation*}
    \tan(\delta)=\frac{\varepsilon^{\dprime}}{\varepsilon^{\prime}}
\end{equation*}
The purple-colored curve in Fig. \ref{fig:ch3_ElectrostaticForce}a presents the dielectric loss for the SC. As shown in the figure, the curve has the lowest value at approximately 100 kHz, which results in a relatively high electrostatic force. Putting all together, we conclude that the dominant factor affecting the frequency-dependent behavior of the electrostatic forces at frequencies below 250 Hz is the charge leakage, while it is the electrical properties of the SC at frequencies above 250 Hz.

Finally, we investigated the influence of the model parameters on the electrostatic force response. This investigation revealed that decreasing the thickness of the insulator layers and the Young's modulus of the SC resulted in increase in electrostatic force. Even a relatively small reduction in these parameters results in a comparably large increase in the force response for the frequency range investigated in this study.

On the other hand, increasing the permittivity of the insulator layer in the touchscreen also results in an increase in electrostatic force. Furthermore, Fig. \ref{fig:ch3_Comparison}c and Fig. \ref{fig:ch3_ComparisonBar}c show that increasing the permittivity of the insulator strongly influences the electrostatic force response at low frequencies. Increasing permittivity decreases charge leakage and hence increases the electrostatic attraction force. Similarly, the charge leakage from SC is higher at low frequencies and hence the change in thickness of SC has a greater effect on the electrostatic force at higher frequencies compared with lower frequencies (see \ref{fig:ch3_Comparison}b and Fig. \ref{fig:ch3_ComparisonBar}b).

%//////////////////////////////////////////////////////////////////////////////////////////////////////////////////////////////////////////////////////////////////////////////////

\chapter{Experimental Estimation of Electrostatic Forces Using Electrical Impedance Measurements}\label{chapter:ExperimentalElectrostaticForcesFromImpedance}
\subsubsection{Summary}
\begin{figure*}[t!]
\centering
\includegraphics[width=1\columnwidth]{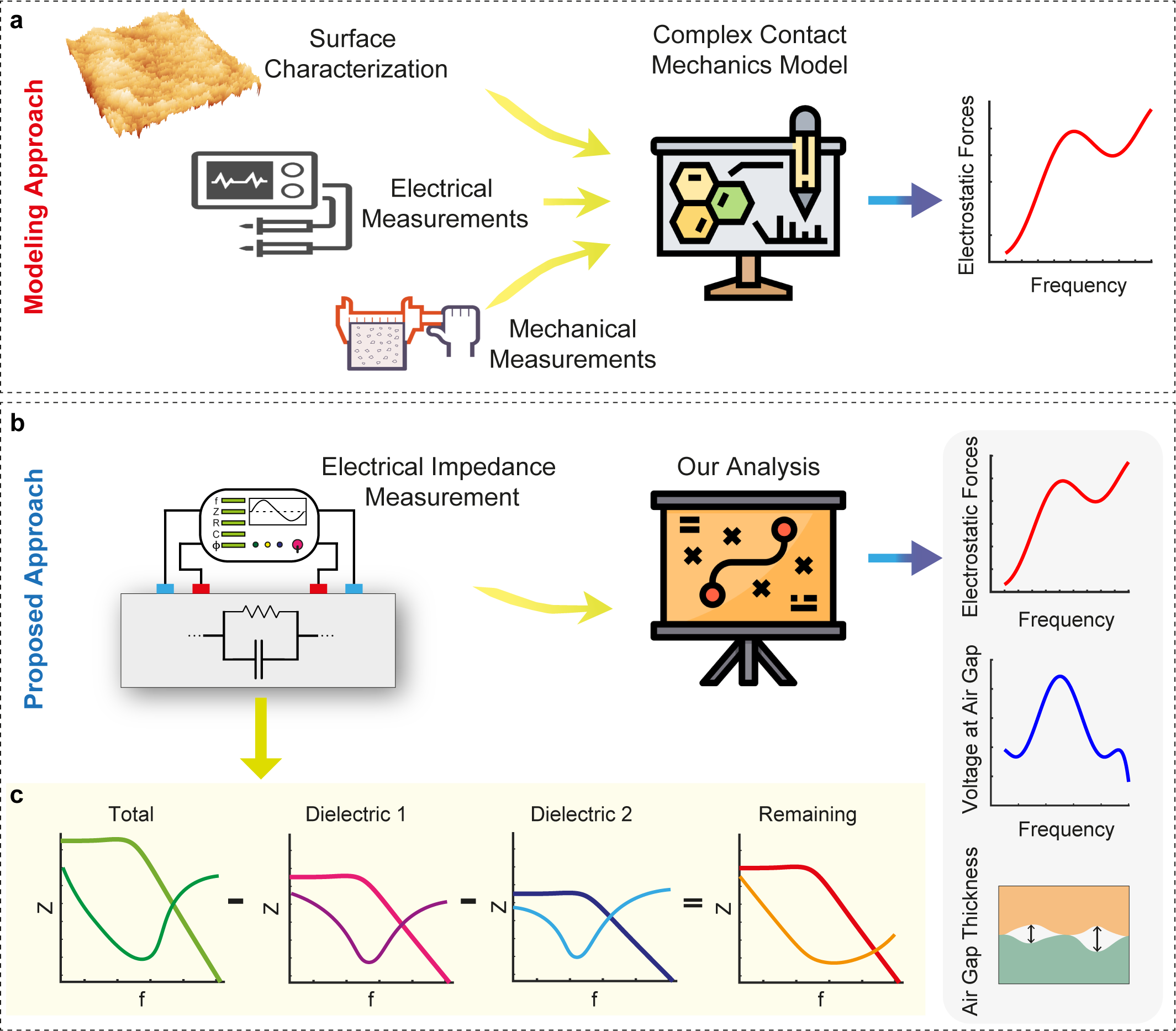}
\caption{a) Modeling approaches for calculating electrostatic forces require surface characterization and electrical and mechanical measurements. b) Our proposed approach for inferring the electrostatic forces, along with the air gap thickness, is based on electrical impedance measurements. c) In our approach, the impedances of dielectrics 1 and 2 are subtracted from the total impedance of the coupled system to obtain the remaining impedance, which enables us to infer the air gap thickness first and then the electrostatic forces.}
\label{fig:ch4_Approach}
\end{figure*}

Despite the advantages of electroadhesion, there is a lack of comprehensive understanding of the physics behind it, which involves electrical and mechanical interactions between the contacting surfaces. Moreover, our knowledge of the parameters influencing the magnitude of electrostatic forces that attract these surfaces to each other is highly limited. Designing controlled experiments to investigate the effects of these parameters on electroadhesion is not trivial due to the complex nature of the contact problem. Hence, there is insufficient experimental data on this topic in the literature. Furthermore, it is difficult to develop mathematical models to estimate electrostatic forces since the thickness of the air gap between the contacting surfaces is not uniform across the contact area and changes with the input voltage and frequency, mechanical and electrical properties of the contacting objects and their surface topography, initial contact forces, environmental factors, etc. (see Fig. \ref{fig:ch4_Approach}a). In addition, the polarization and depolarization of dielectrics in response to changes in the electric field as a function of frequency further complicates the modeling problem. Finally, the charge leakage through the contacting surfaces due to the differences in electrical properties of contacting surfaces is not trivial to measure or model, despite its noticeable effects on the electrostatic forces \cite{aliabbasi2022frequency}. To the best of our knowledge, there is no study in the literature clearly explaining the reasons behind this leakage phenomenon.

In this chapter\footnote{This chapter is based on articles \cite{aliabbasi2024experimental, aliabbasi2023effect}}, we present a systematic experimental approach to infer i) the average thickness of the air gap, ii) the voltage at the air gap, and iii) the electrostatic forces between the contacting dielectrics based on the measurement of electrical impedances. The overview of the proposed approach is presented in Fig. \ref{fig:ch4_Approach}b. Typically, the total impedance of the interface equals the summation of the impedances of dielectrics 1 and 2 and what we call the remaining impedance (see Fig. \ref{fig:ch4_Approach}c). The remaining term in this study represents the impedances of the air gap and the electrode polarization. Shultz et al. \cite{shultz2018electrical} considered the air gap impedance as the only remaining impedance and ignored the electrode polarization impedance. This undesired impedance, due to what is known as EDL \cite{kuang1998low}, explains the charge leakage phenomenon reported in our earlier study \cite{aliabbasi2022frequency}, and also the difference in the strength of the electric fields when DC versus AC voltage is applied to one of the dielectrics. Finally, we show that removing the electrode polarization impedance from the remaining impedance leads to the ``true" impedance of the air gap and enables us to infer the average thickness of the air gap and the magnitude of electrostatic forces.

\section{Extended Physics of Electroadhesion}\label{sec:ch4_ExtendedPhysicsElectroadhesion}
\begin{figure*}[t!]
\centering
\includegraphics[width=1\columnwidth]{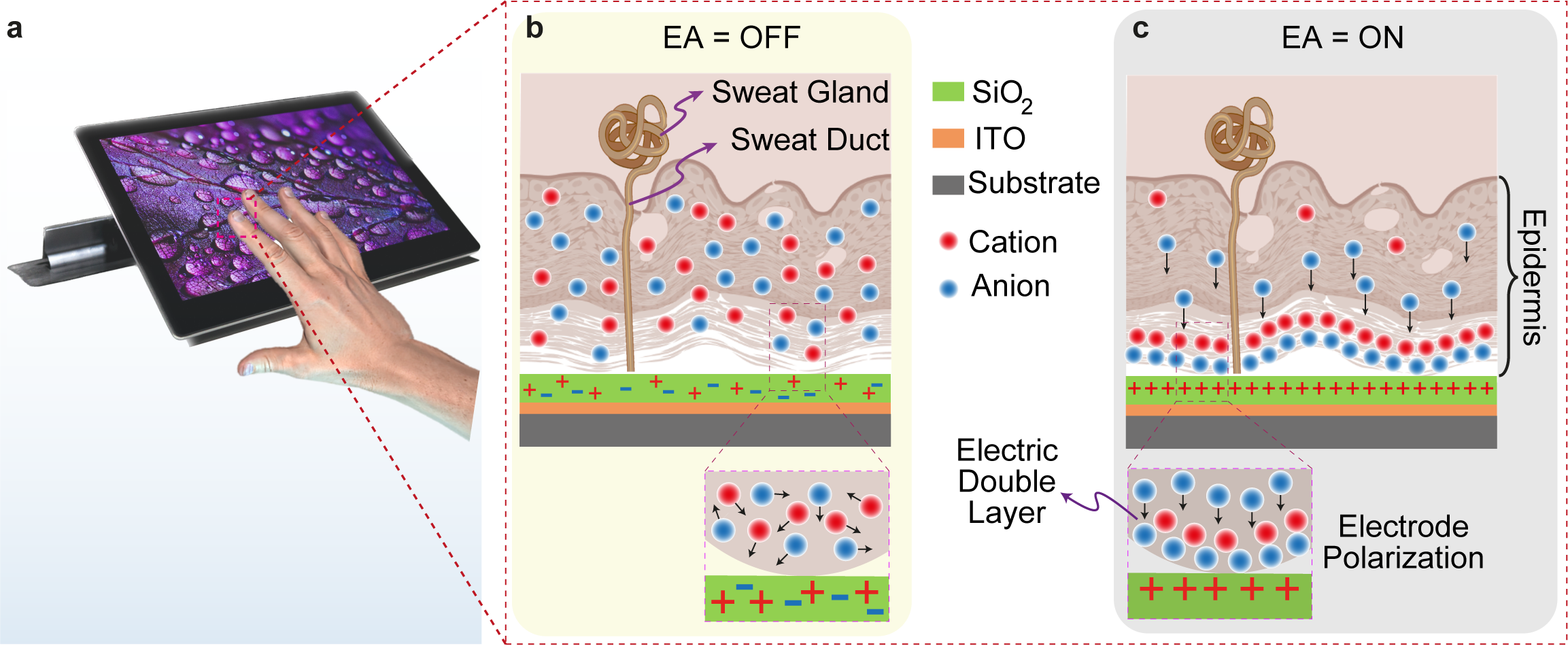}
\caption{a) A human finger in contact with a touchscreen, and b) an enlarged schematic cross-section of the skin-touchscreen interface when electroadhesion is OFF. The outermost layer of skin (SC), which acts as an electrical insulator for the finger, is in contact with the insulator layer of the touchscreen (SiO$_2$) resting on top of a conductor layer (ITO). The electrical charges and ions are distributed randomly in both dielectric layers. c) An enlarged schematic cross-section of the skin-touchscreen interface when voltage is applied to generate electrostatic forces between the finger and the touchscreen. The magnitude of these forces is influenced by the electrical impedances of the air gap and the electrode polarization. The electrode polarization occurs due to the EDL at low frequencies.}
\label{fig:ch4_Introduction}
\end{figure*}

Fig. \ref{fig:ch4_Introduction}a presents a human finger in contact with the surface of a touchscreen. The electric charges and ions in the touchscreen and the finger are distributed randomly when there is no voltage applied to the conductive layer of the touchscreen (see Fig. \ref{fig:ch4_Introduction}b). However, when voltage is applied, an electrostatic attraction force builds up between the human finger and the touchscreen due to the parallel plate capacitor principle (see Fig. \ref{fig:ch4_Introduction}c). This principle requires two oppositely charged conductive surfaces, separated by dielectric (insulating) material. The SC, which serves as the finger's insulator layer, is primarily made up of dead cells. This layer, similar to all dielectric layers, can partially prevent the passage of electrical charges. The soft tissue under SC is conductive and this is why the capacitive touchscreens in our model can detect finger position in the first place. Similarly, the tactile display (i.e., surface capacitive touchscreen) comprises a thin insulator layer (SiO$_2$) on top of a conductive layer of ITO.

When human finger contacts the voltage-induced touchscreen, the ions in the finger tissue are pulled toward the touchscreen's surface and form the first layer on the inner finger surface. This layer comprises ions of opposite charge to that of the touchscreen while the second layer contains loosely anchored ions of the same charge. The free ions in the finger with opposite charges to those of the touchscreen are attracted to the touchscreen, pushing the ones in the first layer out and causing leakage of electrons from the finger to the surface of the touchscreen. At low frequencies below approximately 30 Hz, the free ions have sufficient time to push the ones in the first layer out. Hence, more charges leak and consequently, the strength of the electric field at the interface is reduced.

\section{Materials and Methods}\label{sec:ch4_MaterialsMethods}
In order to estimate the impedance of the air gap between the finger and touchscreen, the electrical impedances of human skin and touchscreen were measured separately and subtracted from the total sliding impedance in a recent study \cite{shultz2018electrical}. Hence, the remaining impedance was treated as the air gap impedance, considered in series with the skin and touchscreen impedances. However, we demonstrate that the remaining term represents the impedance of not just the gap between the finger and the touchscreen but also the electrode polarization \cite{grimnes2015bioimpedance}. 

\subsection{Stimuli}\label{sec:ch4_Stimuli}
\begin{figure*}[b!]
\centering
\includegraphics[width=0.5\columnwidth]{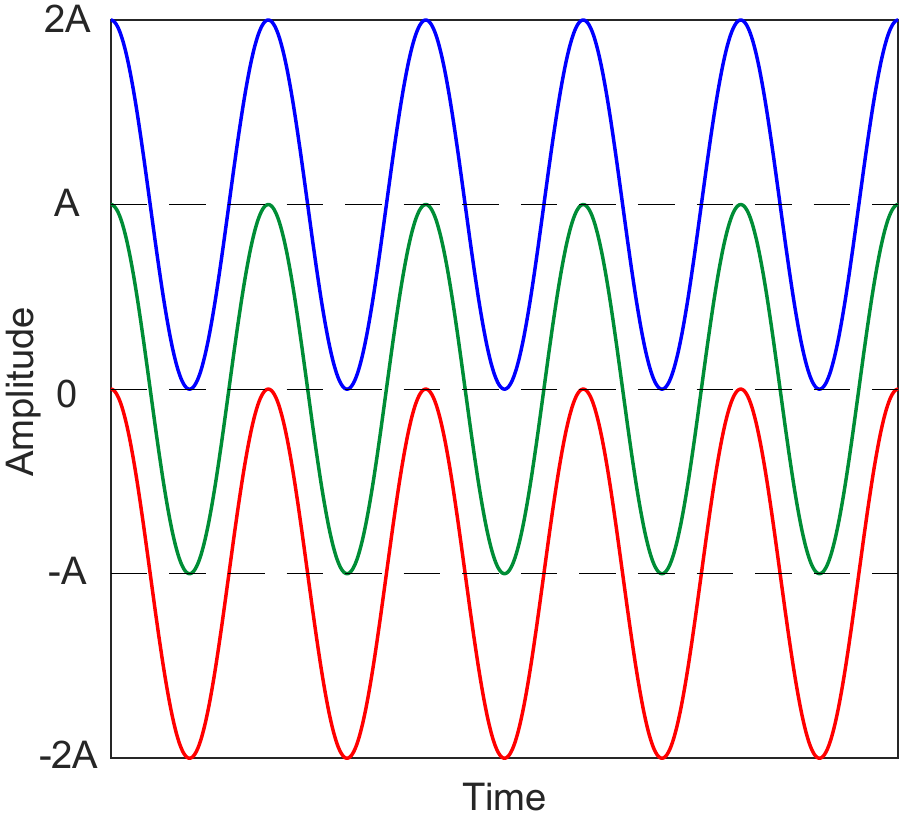}
\caption{The three stimuli (the input voltage signal applied to the ITO layer of the touchscreen) used in this study; green, blue, and red colors represent the sinusoidal voltage signals with no, positive, and negative DC offsets, respectively.}
\label{fig:ch4_Stimuli}
\end{figure*}

We performed our impedance measurement under AC stimulation using the input signals shown in Fig. \ref{fig:ch4_Stimuli}. Throughout the paper, we report the results of the impedance measurements performed with sinusoidal voltage signals having positive and negative DC offsets with blue and red-colored curves, respectively, and no DC offset with green-colored curves.

\subsection{Participants}\label{sec:ch4_Participants}
One healthy right-handed participant (30 years old) was selected to take part in this study. A consent form was read and signed by the participant before the experiment, which was approved by the Ethical Committee for Human Participants of Koc University. The study conformed to the principles of the Declaration of Helsinki, and the experiment was performed following relevant guidelines and regulations.

Since emotions can alter the psychological sweating mechanism, the participant was asked to inform the experimenter regarding his emotional feelings and any pains or aches in his body before starting the experiments. Stress, anxiety, discomfort, excitement, and sleeplessness could be the factors affecting any electrical measurements involving the human skin. Before each experimental trial, the participant washed his hands with soap, rinsed with water, and dried them with a clean towel. He also waited in the experimentation room for five minutes without touching anything before each trial. This protocol was strictly followed for all trials to ensure that the sebum or sweat from the finger's surface was removed and the skin had enough time to reach its normal hydration level.

\subsection{Apparatus for Measuring Electrical Impedance}\label{sec:ch4_ApparatusMeasuringElectricalImpedances}
In all electrical impedance measurements performed by an impedance analyzer (MFIA 5 MHz, Zurich Instruments Inc.), the four-electrode method was chosen. Since electrical impedance measurements are prone to noise, the impedance analyzer was calibrated before each measurement session.

\subsubsection{Electrochemical Bioimpedance of the Skin}
\begin{figure*}[t!]
\centering
\includegraphics[width=1\columnwidth]{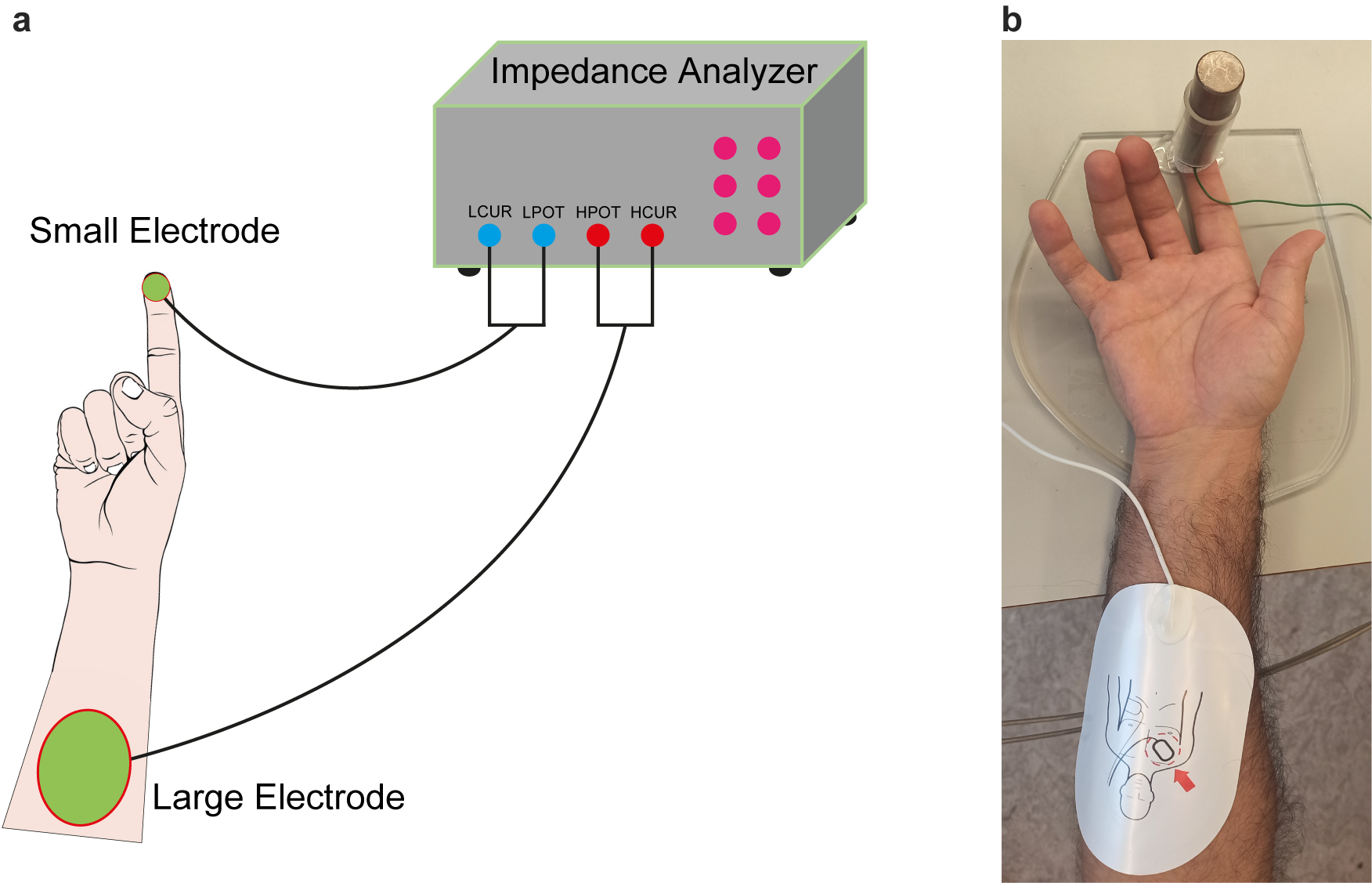}
\caption{Set-up for measuring the electrical impedance of skin: a) the four-electrode impedance measurement method was chosen in our measurements, where a small electrode (hydrogel/metal) was attached to the tip of the participant’s index finger and a larger electrode with approximately ten times larger area was attached to the ventral forearm of the same hand. b) A weight of 100 grams, equivalent to 1 N normal force, was placed on top of the small electrode during the impedance measurements of the skin.}
\label{fig:ch4_SetupSkin}
\end{figure*}

The electrochemical bioimpedance of the skin was measured using two types of electrodes: hydrogel and metal electrodes. The measurements were performed in three separate sessions on three different days and the data were collected in ten consecutive trials (10 trials$/$session $\times$ 3 sessions). Fig. \ref{fig:ch4_SetupSkin}a presents the experimental apparatus for the bioimpedance measurements. A small hydrogel electrode (1050NPSM Neonatal Pre Wired Small Cloth ECG Electrodes, Cardinal Health Inc.) was attached to the participant's right-hand index finger. A larger electrode (HeartStart FR2 Defibrillator Electrode Pads, Philips Medical Systems Inc.) was attached to the ventral forearm of the same hand. This electrode has a contact area approximately ten times larger than the small electrode to minimize its contribution to bioimpedance measurements. During the measurements, the applied current always entered the skin through one electrode and exited it from the other depending on the polarity of the signal. We replaced the hydrogel electrode with a metal one and repeated the impedance measurements. The hydrogel electrode has a solid gel layer that reduces the undesired electrode polarization effects. A weight of 100 grams, equivalent to a normal force of 1 N, was placed on top of the small electrode at the fingertip and kept vertically aligned using a custom-made circular tube (see Fig. \ref{fig:ch4_SetupSkin}b).

\begin{figure*}[b!]
\centering
\includegraphics[width=1\columnwidth]{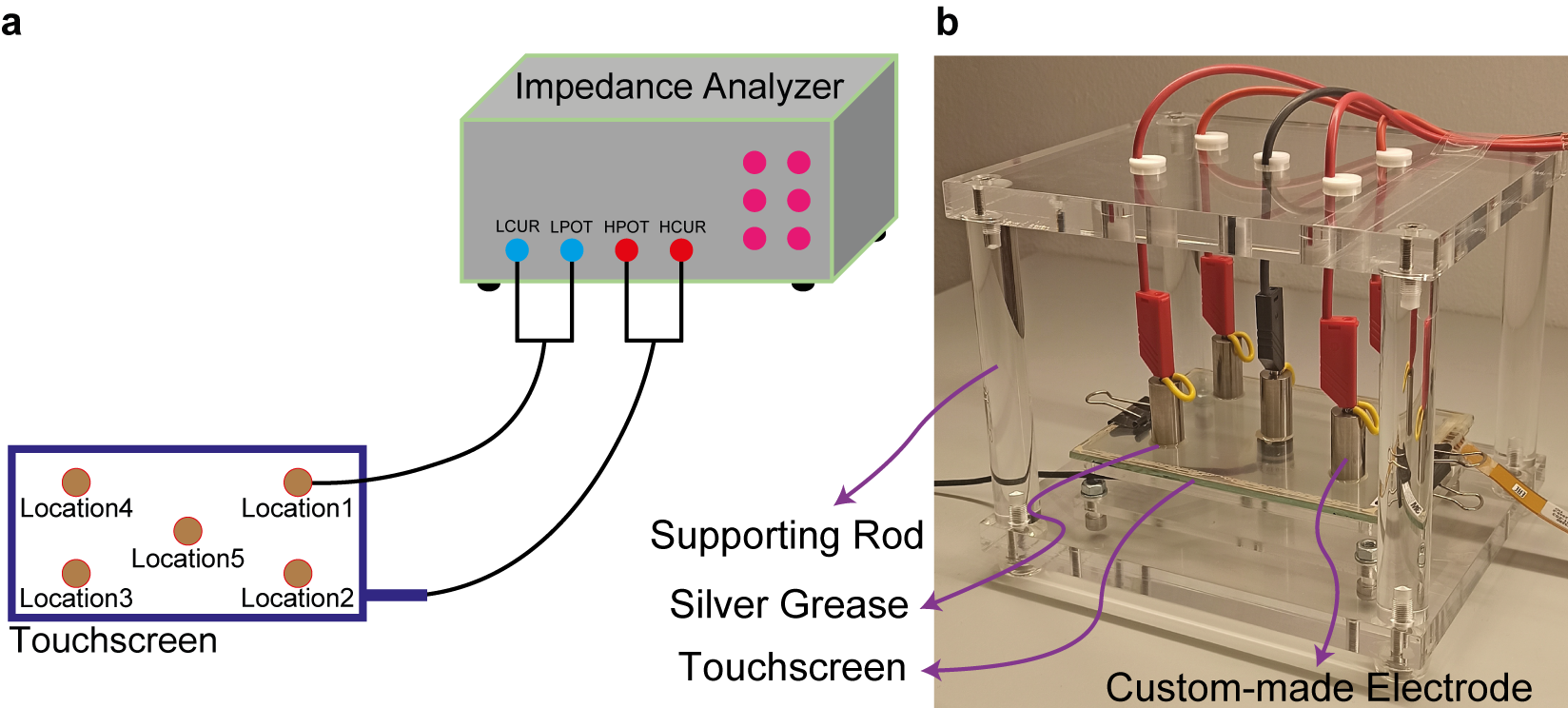}
\caption{Set-up for measuring the electrical impedance of touchscreen: a) the four-electrode impedance measurement method was chosen. One terminal of the impedance analyzer was connected to the conductive layer of the touchscreen and the other terminal was connected to the touchscreen’s surface using a custom-made electrode. Data were collected from five different locations on the surface of the touchscreen. The electrical connection between the touchscreen’s surface and each electrode was established by a thin layer of silver grease. b) A solid structure was designed and manufactured from plexiglass to hold the touchscreen and the cables of electrodes stationary to reduce the effects of any external disturbance.}
\label{fig:ch4_SetupTouchscreen}
\end{figure*}

\subsubsection{Electrical Impedance of the Touchscreen}
The electrical impedance of the touchscreen was measured using a set of five custom-made metal electrodes (see Fig. \ref{fig:ch4_SetupTouchscreen}a). The electrodes were attached to the touchscreen's surface at five different locations using a very thin layer of silver grease (8463A, MG Chemicals). The data were collected in ten consecutive trials from each location (10 trials$/$location $\times$ 5 locations). The electrode locations were selected to cover the whole surface of the touchscreen. In a pilot study, we observed that the impedance measurements were sensitive to the thickness of the silver grease between the electrodes and the touchscreen. Efforts were made to perform the measurements with the thinnest possible layer of silver grease, but this approach resulted in relatively large variations in the impedance measurements. A support structure was also manufactured to keep the connection cables stable and minimize the electrical noise (see Fig. \ref{fig:ch4_SetupTouchscreen}b).

\begin{figure*}[b!]
\centering
\includegraphics[width=1\columnwidth]{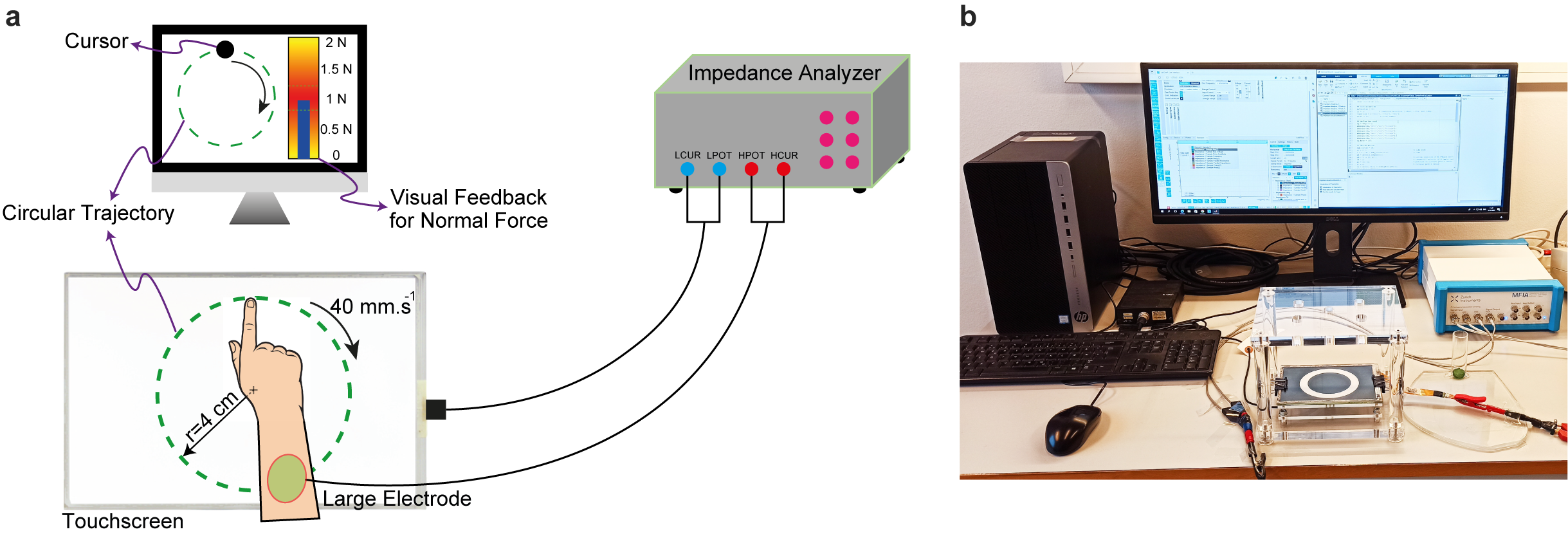}
\caption{Set-up for measuring the electrical impedance of total electrical impedance while the participant’s finger was sliding on the touchscreen. a) The four-electrode impedance measurement method was chosen. One terminal of the impedance analyzer was connected to the conductive layer of the touchscreen, while the other terminal was connected to the ventral forearm of the participant through a large electrode. The normal force applied by the index finger of the participant was measured by a force sensor placed beneath the touchscreen. During the experiment, the participant was instructed to maintain a normal force of approximately 1 N. A user interface was designed to provide visual feedback to the participant through a computer monitor to help him keep the normal force within the range of 0.8-1.2 N. A circular trajectory was displayed on the computer monitor with a cursor moving at a velocity of 40 mm/s. The participant was asked to follow this trajectory by moving his finger on the touchscreen at the same speed. b) To enable more accurate tracking of the cursor, the same circular trajectory was printed on paper and placed beneath the touchscreen.}
\label{fig:ch4_SetupSliding}
\end{figure*}

\subsubsection{Total Electrical Impedance of Sliding Finger on Touchscreen}
The total electrical impedance of the finger was measured while the participant's finger was sliding on the surface of the touchscreen (i.e., sliding condition). The measurements were performed in three separate sessions on three different days and the data were collected in ten consecutive trials (10 trials$/$session $\times$ 3 sessions). A user interface was developed in Matlab (The MathWorks Inc.) to provide visual feedback to the participant while his finger made circular movements on the surface of the touchscreen (see Fig. \ref{fig:ch4_SetupSliding}a and b). As shown in Fig. \ref{fig:ch4_SetupSliding}a, the black-colored cursor displayed on the computer monitor circled on the dashed-green trajectory (r$=4$ cm) with a constant velocity of 40 mm/s. The participant was asked to follow the cursor's motion by moving his finger on the touchscreen at the desired speed. A circular trajectory of the same size as the dashed-green trajectory was printed on paper and placed beneath the touchscreen to help the participant follow it more easily. The blue bar on the computer monitor displayed the magnitude of normal force applied by the participant's finger to the touchscreen in real time. Two horizontal dashed-green lines at 0.8 N and 1.2 N showed the lower and upper limits of the normal force to assist the participant control his normal force close to 1 N. The normal force was measured using a force transducer (Nano 17, ATI Industrial Automation Inc.) placed beneath the touchscreen and acquired by a data acquisition card (PCIe-6034E, National Instruments Inc.) at 100 Hz. The participant was trained before the experiment to get familiar with the procedure. He could successfully maintain the normal force close to the desired value while moving his finger on the circle ($1\pm 0.15$ N).

\subsubsection{Total Electrical Impedance of Stationary Finger on Touchscreen}
\begin{figure*}[t!]
\centering
\includegraphics[width=1\columnwidth]{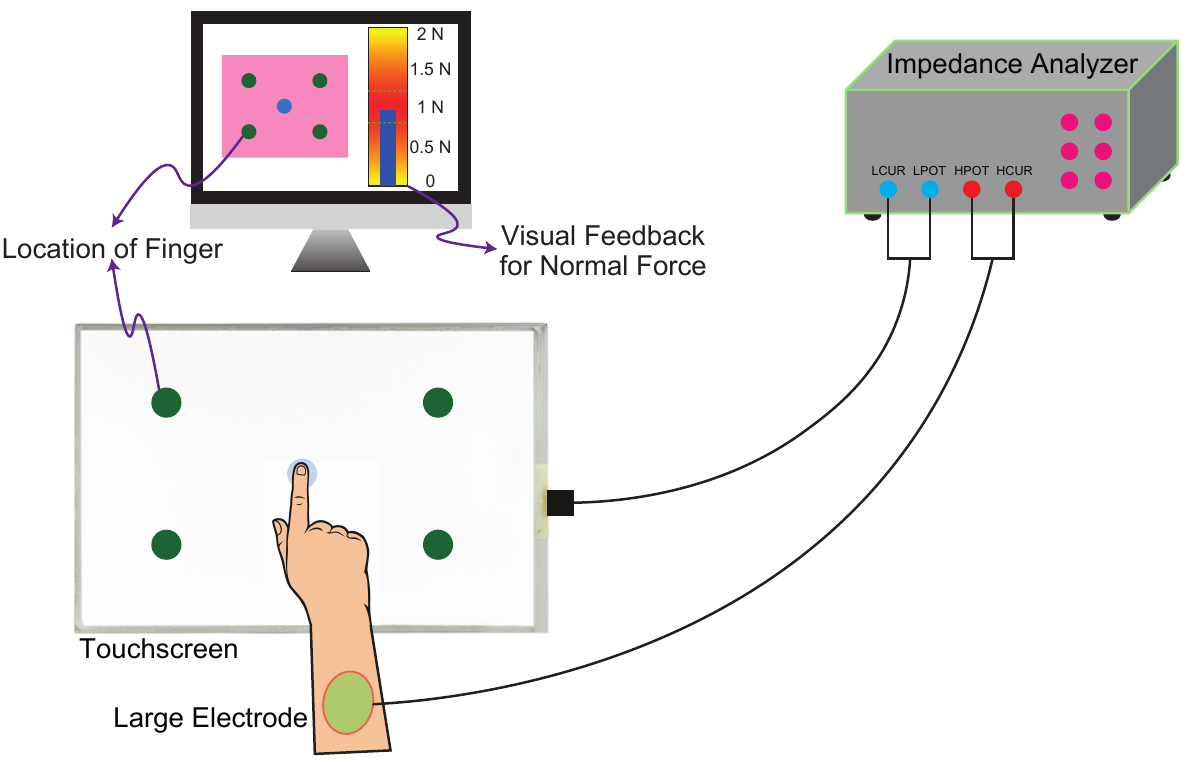}
\caption{Set-up for measuring the total electrical impedance while the participant’s finger was stationary on the touchscreen. a) The four-electrode impedance measurement method was chosen. One terminal of the impedance analyzer was connected to the conductive layer of the touchscreen and another one was connected to the ventral forearm of the participant’s hand using a large electrode. The normal force applied by the participant during the experiment was measured through a force sensor placed beneath the touchscreen. The participant was asked to maintain the normal force at approximately 1 N throughout the experiment. A user interface was designed to provide visual feedback to the participant through a computer monitor to help him keep the normal force within the range of 0.8-1.2 N. Data were collected from five different locations on the
touchscreen. All locations were initially displayed with green-colored circles on the computer monitor. Each measurement location was signaled one by one to the participant by changing its color to blue.}\label{fig:ch4_SetupStationary}
\end{figure*}

We measured the total electrical impedance of the finger while it was stationary on the touchscreen (see Fig. \ref{fig:ch4_SetupStationary}). The measurements were performed in three separate sessions on three different days and the data were collected in ten consecutive trials (10 trials$/$session $\times$ 3 sessions). The data were collected from five different sites on the touchscreen, and the participant was instructed to change the location of his finger after every two trials. The participant was asked to keep his finger stable during the measurements while trying to maintain the normal force close to the desired value of 1 N.

\subsection{Calculation of the Remaining Impedance}\label{sec:ch4_CalculationRemainingImpedance}
\begin{figure*}[t!]
\centering
\includegraphics[width=0.5\columnwidth]{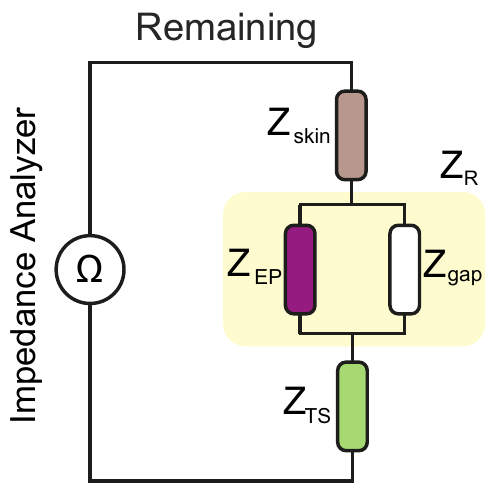}
\caption{The schematic circuit model for the finger-touchscreen interactions: the remaining impedance is calculated by subtracting the measured impedances of skin and touchscreen from the total sliding impedance.}\label{fig:ch4_SetupRemaining}
\end{figure*}

Since electrical impedances of skin, touchscreen, and the remaining are assumed to be in series based on the model shown in Fig. \ref{fig:ch4_SetupRemaining}, the skin (Z$_{Skin}$) and touchscreen (Z$_{TS}$) impedances were subtracted from the total sliding impedance (Z$_{Total(Sliding)}$) to obtain the remaining impedance using the following equation:
\begin{equation}\label{eq:ch4_Subtraction}
    Z_{R} = Z_{Total(Sliding)}-Z_{Skin}-Z_{TS}
\end{equation}
It is essential to insert real and imaginary parts of the measured impedances properly into this equation. Otherwise, it is not possible to calculate the phase angle of the remaining impedance or its resistance/capacitance.

\section{Results and Discussion}\label{sec:ch4_ResultsDiscussion}

\subsection{Electrical Impedance Measurements}
Our electrical impedance measurements are divided into four groups:\\
i) electrochemical bioimpedance of human skin\\
ii) electrical impedance of touchscreen\\
iii) total electrical impedance of sliding finger on touchscreen\\
iv) total electrical impedance of stationary finger on touchscreen.

\begin{figure*}[t!]
\centering
\includegraphics[width=1\columnwidth]{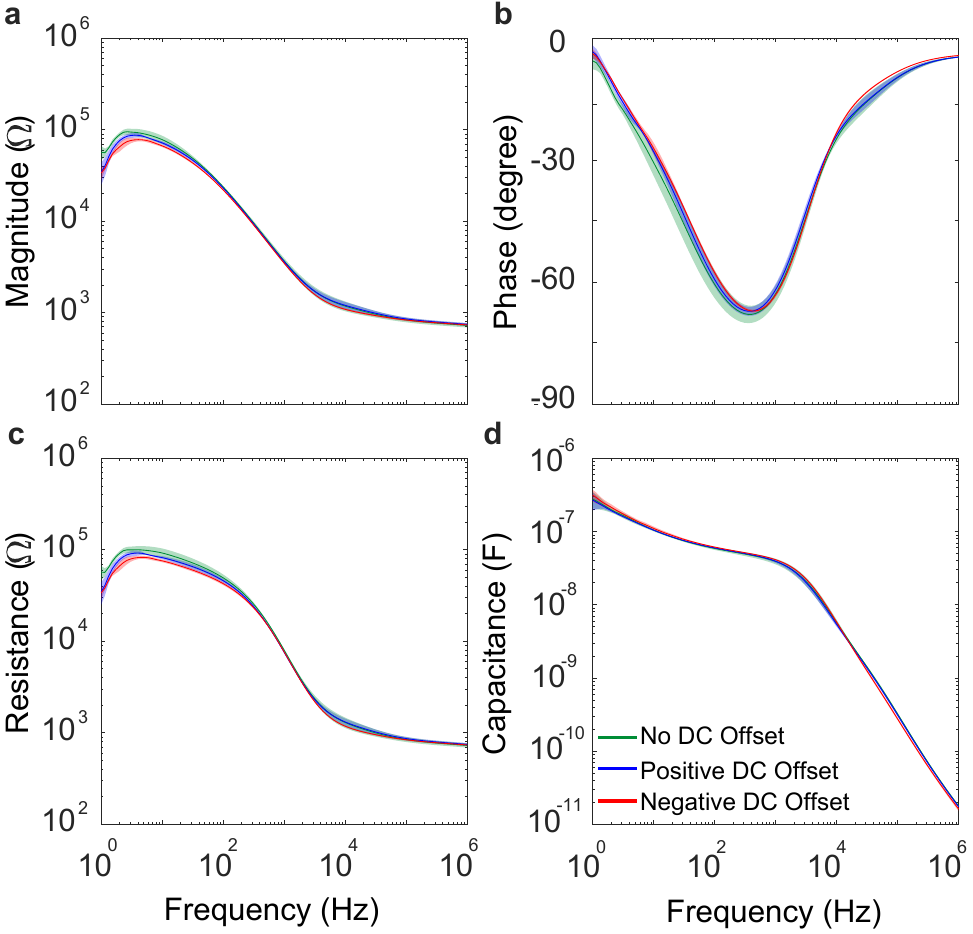}
\caption{Electrical impedance measurements of skin as a function of frequency. The solid curves in the plots represent the mean values of impedance a) magnitude, b) phase, c) resistance, and d) capacitance using hydrogel electrode. The shaded regions around the solid curves represent the standard error of means.}
\label{fig:ch4_ResultsSkinHydrogel}
\end{figure*}

\begin{figure*}[t!]
\centering
\includegraphics[width=1\columnwidth]{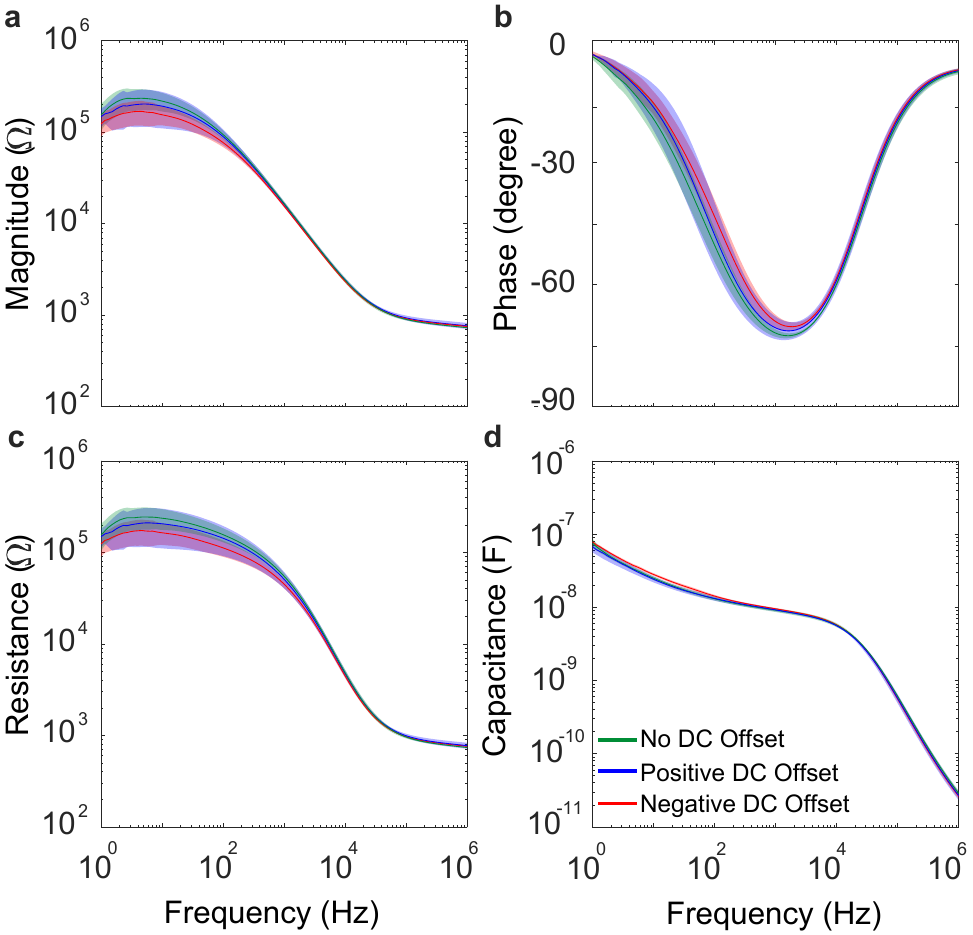}
\caption{Electrical impedance measurements of skin as a function of frequency. The solid curves in the plots represent the mean values of impedance a) magnitude, b) phase, c) resistance, and d) capacitance using metal electrode. The shaded regions around the solid curves represent the standard error of means.}
\label{fig:ch4_ResultsSkinMetal}
\end{figure*}

The average magnitude and phase of the impedance, resistance, and capacitance as a function of frequency for the skin with their standard error of means are shown in Fig. \ref{fig:ch4_ResultsSkinHydrogel}a, b, c, and d and Fig. \ref{fig:ch4_ResultsSkinMetal}a, b, c, and d using hydrogel and metal electrodes, respectively. The impedance magnitude of skin, similar to its resistance, was found to be higher when measured with metal electrodes compared to hydrogel electrodes due to the effect of electrode polarization impedance. Moreover, the electrode polarization impedance caused phase lag in the impedance response. These results align with the earlier measurements of the skin bioimpedance \cite{grimnes2015bioimpedance}. The capacitance data shows that the dispersion in the skin occurs with the metal electrode at approximately one decade of frequency later than that of the hydrogel electrode. Various polarization mechanisms, which are responsible for different types of dispersions, occur in human skin, but it is not easy to observe all of them since a broad range of frequencies needs to be scanned. In biological materials, $\alpha$, $\beta$, and $\gamma$ dispersions are typically observed, which are caused by ionic processes, charging up of the membrane or orientation of permanent dipoles, and orientational relaxation, respectively \cite{schwan1991dielectric}. In our measurements, we do not observe $\alpha$ and $\gamma$ dispersions since they occur at very low and high frequencies, respectively. We observed $\beta$ dispersion around 1 kHz for the measurements performed with the hydrogel electrode and around 10 kHz for the measurements performed with the metal electrode. More information on polarization and dispersion is provided in Appendix \ref{appendix:polarization and dispersion}.

Relatively large variations were observed in the impedance magnitude of skin at low frequencies, which is attributed to the influence of electrodermal activity. In fact, measurements of skin bioimpedance at low frequencies are easily influenced by even small movements of the measurement site, environmental factors, and emotional arousal \cite{boucsein2012electrodermal}. The effect of electro-osmosis was also observed in our measurements. The voltage signal with negative DC offset (red-colored curve) resulted in an impedance slightly lower than that of the positive DC offset (blue-colored curve) due to the increase in skin conductance caused by electro-osmosis \cite{grimnes1983skin}. While both electrodermal activity and electro-osmosis occur at low frequencies, the dominance of electrodermal activity makes the electro-osmosis phenomenon less evident in our results. See Appendix \ref{appendix:electro-osmosis and electrodermal} for more information about electro-osmotic and electrodermal activity of human skin.

\begin{figure*}[t!]
\centering
\includegraphics[width=1\columnwidth]{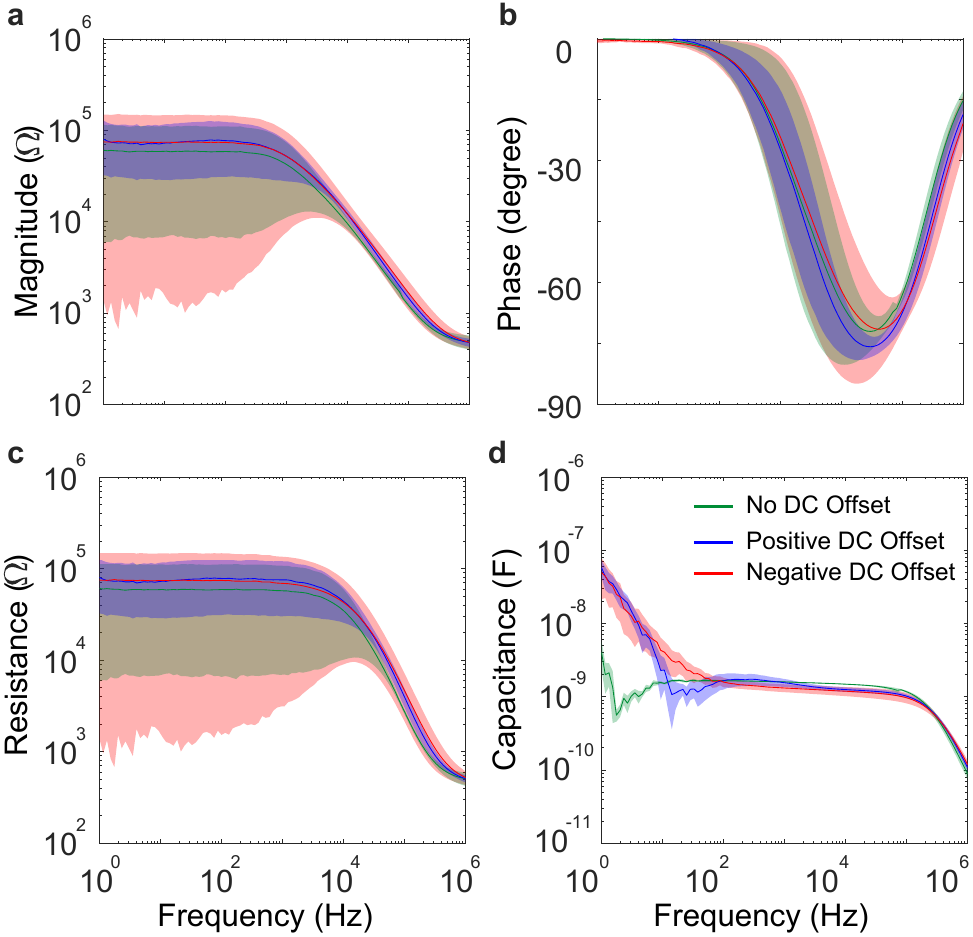}
\caption{Electrical impedance measurements of touchscreen as a function of frequency. The solid curves in the plots represent the mean values of impedance a) magnitude, b) phase, c) resistance, and d) capacitance. The shaded regions around the solid curves represent the standard error of means.}
\label{fig:ch4_ResultsTouchscreen}
\end{figure*}

\begin{figure*}[t!]
\centering
\includegraphics[width=1\columnwidth]{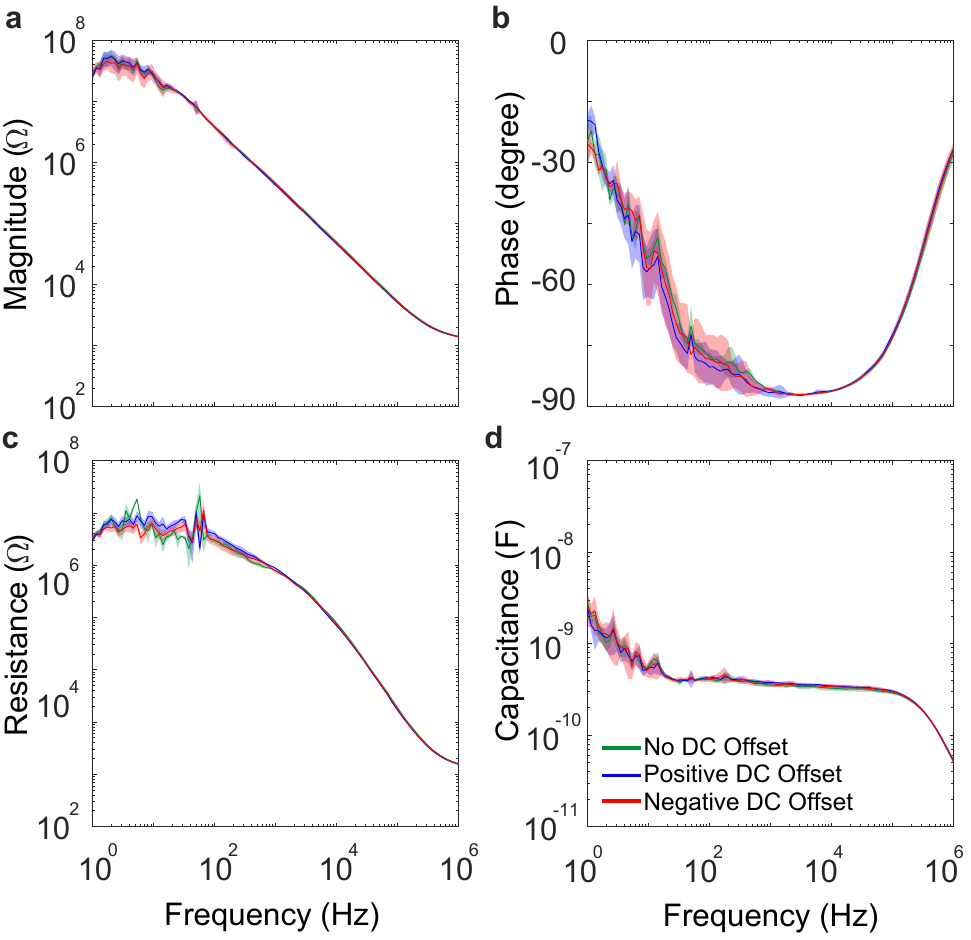}
\caption{Total electrical impedance measurements while the participant’s finger was sliding on the touchscreen as a function of frequency. The solid curves in the plots represent the mean values of impedance a) magnitude, b) phase, c) resistance, and d) capacitance. The shaded regions around the solid curves represent the standard error of means.}
\label{fig:ch4_ResultsSliding}
\end{figure*}

\begin{figure*}[t!]
\centering
\includegraphics[width=1\columnwidth]{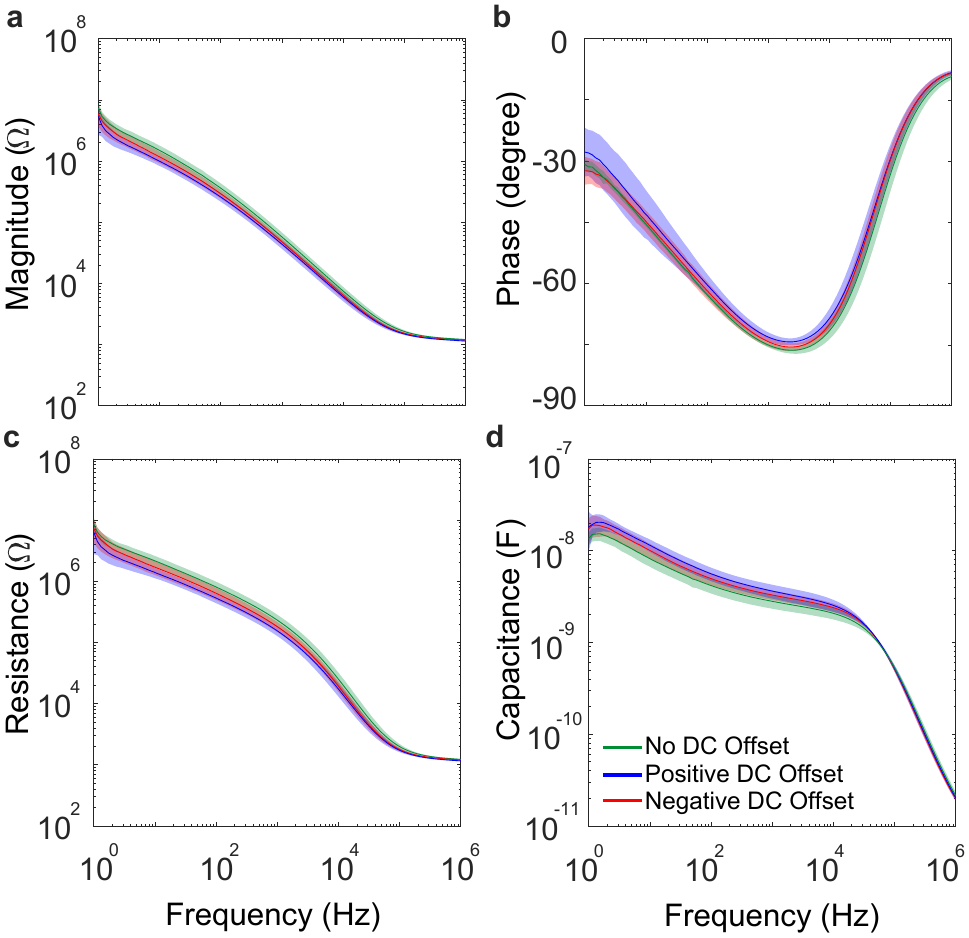}
\caption{Total electrical impedance measurements while the participant’s finger was stationary on the touchscreen as a function of frequency. The solid curves in the plots represent the mean values of impedance a) magnitude, b) phase, c) resistance, and d) capacitance. The shaded regions around the solid curves represent the standard error of means.}
\label{fig:ch4_ResultsStationary}
\end{figure*}

The average results of the impedance measurements performed on the touchscreen at five different locations on its surface, along with their standard error of means, are presented as a function of frequency in Fig. \ref{fig:ch4_ResultsTouchscreen}a, b, c, and d. The results suggest that the impedance of the touchscreen does not change significantly with the polarity of the applied voltage signal. Similar to the skin bioimpedance measurements, a dispersion was observed at approximately 10 kHz, which could be related to the electrical properties of the insulating layer of the touchscreen (i.e., SiO$_2$).

Fig. \ref{fig:ch4_ResultsSliding}a-h and Fig. \ref{fig:ch4_ResultsStationary}a-h show the average and standard error of means of the measured total electrical impedance as a function of frequency when the finger was sliding and stationary on the touchscreen, respectively. The magnitude of the total impedance measured under the sliding condition is approximately an order of magnitude higher than that of the stationary condition. In stationary condition, sweat accumulates in the air gap between the finger and the touchscreen and shorts out the air gap impedance due to its high conductivity \cite{shultz2018electrical}. However, the finger leaves sweat behind on the touchscreen's surface as it slides, resulting in a higher impedance with respect to the stationary condition. In addition, we also argue that the sweat ducts, together with the skin tissue, contribute to the conduction of current and when the sweat pores of the skin make contact with the surface of touchscreen, sweat acts as a lubricant to increase the conduction through the sweat ducts in the stationary condition. Hence, in stationary condition, not only does the air gap impedance short out, but the conduction of the skin also increases due to the higher conductivity of the sweat ducts than the soft tissue itself. We believe that these two mechanisms cause a reduction in the total impedance under the stationary condition compared to the sliding one, though more evidence is required to support or reject them. Finally, the dispersion in the stationary condition occurred approximately one decade of frequency sooner than that of the sliding condition, which suggests that some polarization mechanisms under stationary condition failed earlier, weakening the capacitive nature of the contact interface.

The results on phase angle also support the argument that the contact interface shows more capacitive behavior under the sliding condition, as its phase angle levels off just shy of 90 degree (Fig. \ref{fig:ch4_ResultsSliding}b). The phase plot can be divided into three regions with more resistive behavior at low and high frequencies and more capacitive behavior in the middle. The phase angle for the total stationary impedance follows a similar trend to that of the total sliding impedance, with 20 degree lower values in the middle region. The phase angles of the skin and touchscreen are not only lower than that of the total but also their curves are shifted in phase with respect to that of the total. Their capacitive regions are narrower, and the phase angles for the skin and touchscreen reach their maximum values at approximately 250 Hz and 100 kHz, respectively. Hence, they show the most capacitive behavior at those frequencies.

\begin{figure*}[t!]
\centering
\includegraphics[width=1\columnwidth]{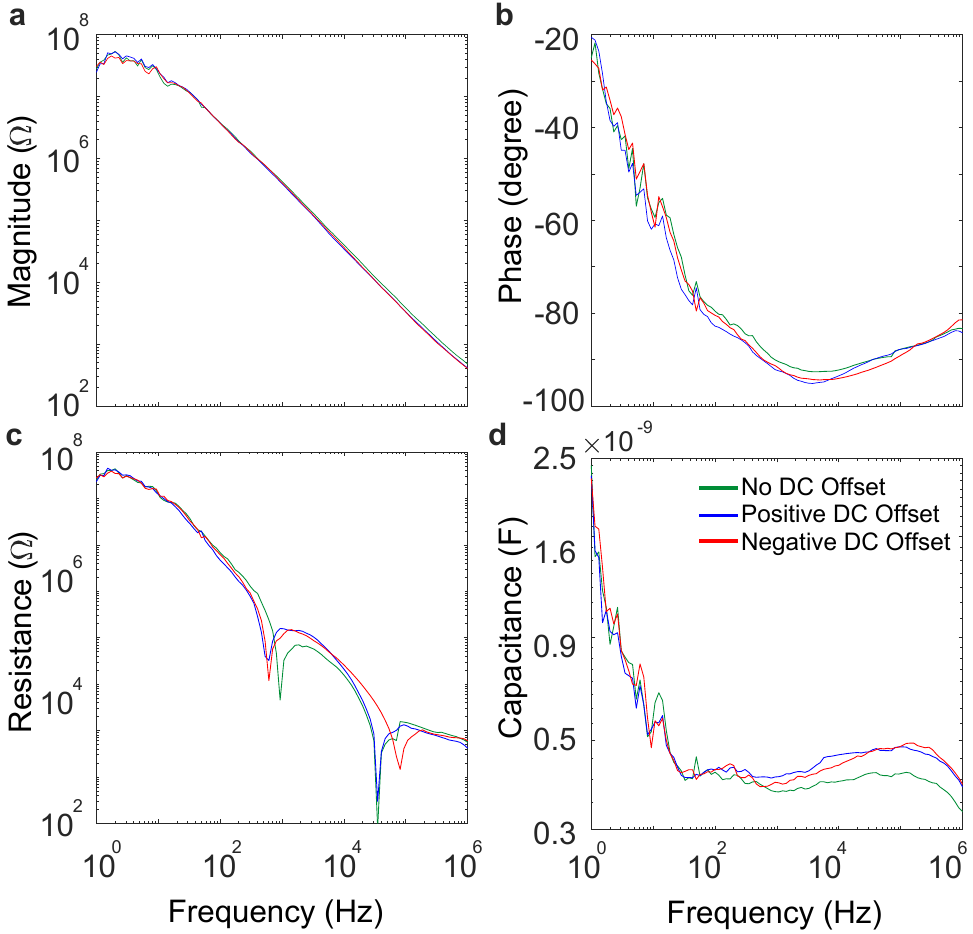}
\caption{The remaining impedance as a function of frequency. The solid curves in the plots represent the mean values of impedance a) magnitude, b) phase, c) resistance, and d) capacitance. The shaded regions around the solid curves represent the standard error of means.}
\label{fig:ch4_ResultsRemaining}
\end{figure*}

The results for the remaining impedance, which was calculated by subtracting the skin and touchscreen impedances from the total sliding impedance, are presented in Fig. \ref{fig:ch4_ResultsRemaining}a, b, c, and d. Moreover, the remaining admittance (inverse of impedance) magnitudes are plotted in Fig. \ref{fig:ch4_FurtherAnalysis}a. A one-decade-per-decade line (dotted black) fits the admittance curve well for frequencies larger than 30 Hz. This indicates that the nature of the remaining impedance after this frequency is almost purely capacitive due to the air gap. The phase angle of the remaining impedance also shows that the impedance is more resistive at low frequencies and becomes capacitive at higher frequencies. Hence, the remaining impedance can be modeled by a resistance (R$_{EP}$) and a capacitance (C$_{EP}$) in parallel for the electrode polarization and another capacitance parallel to those for the air gap (C$_{gap}$). We observe that the circuit for the electrode polarization is effective at frequencies lower than 30 Hz while the air gap capacitance dominates the impedance otherwise. In other words, the electrode polarization impedance vanishes after approximately one decade of frequency \cite{schwan1968electrode, schwan1992linear}.

\begin{figure*}[t!]
\centering
\includegraphics[width=0.4\columnwidth]{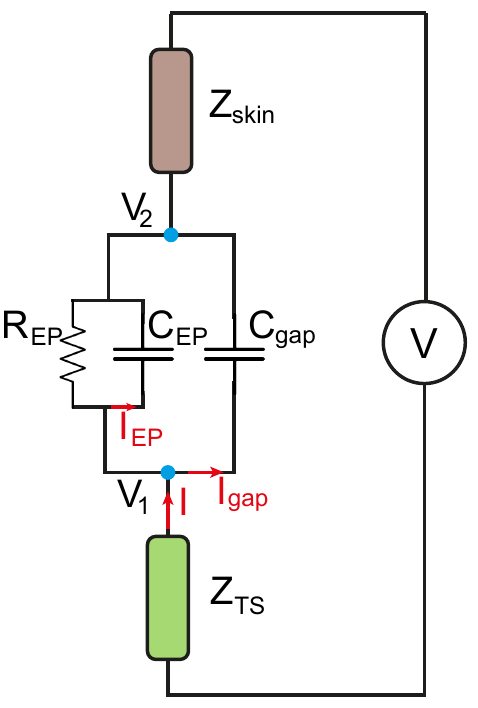}
\caption{Schematic representation of the circuit model proposed for the finger in contact with a touchscreen under electroadhesion.}
\label{fig:ch4_Circuit}
\end{figure*}

\begin{figure*}[t!]
\centering
\includegraphics[width=0.95\linewidth]{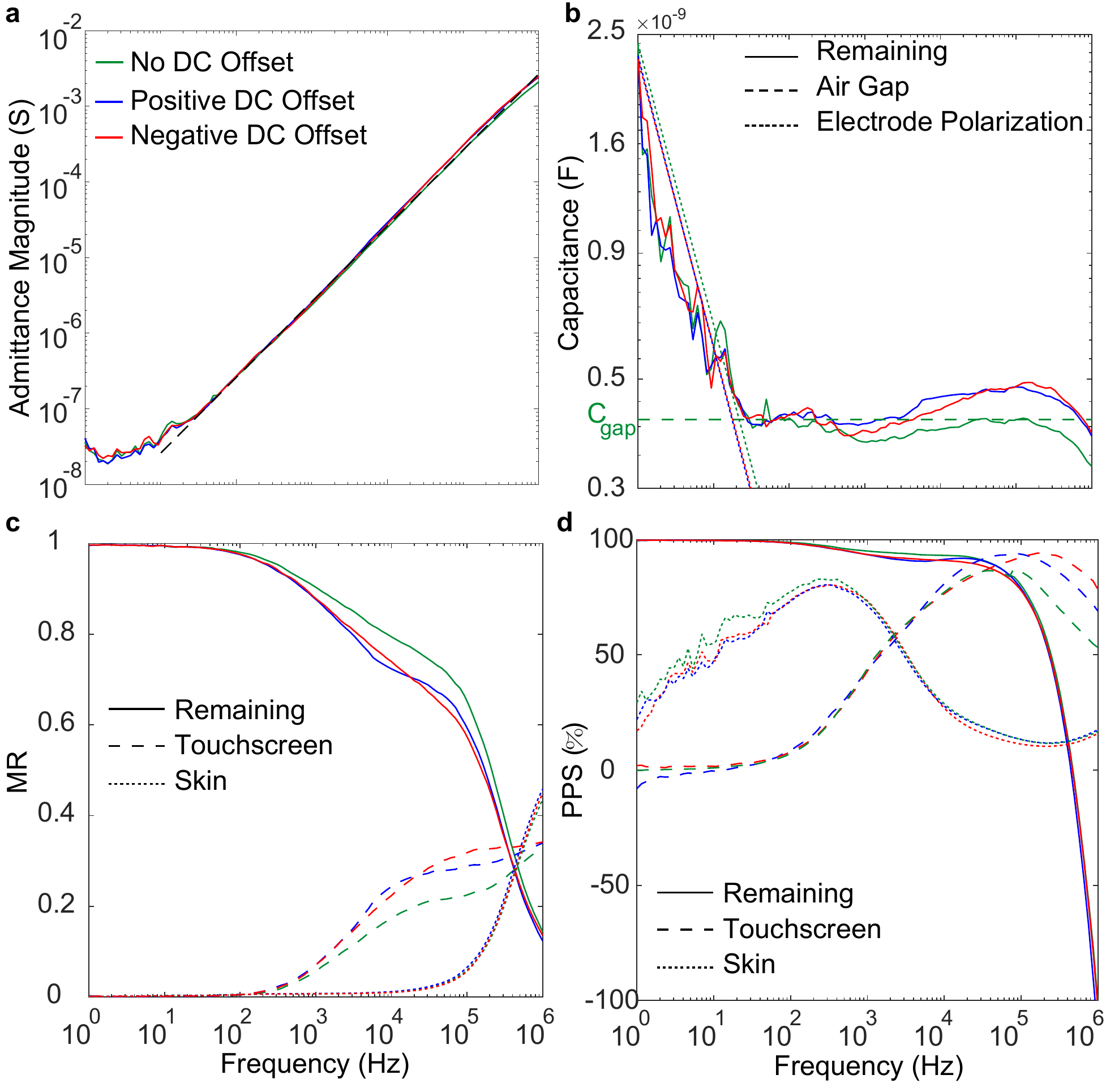}
\caption{a) The change in remaining admittance as a function of frequency. The dashed line in black color fits the admittance curve well for frequencies higher than approximately 30 Hz, indicating purely capacitive behavior at those frequencies. b) The change in capacitance as a function of frequency. The solid, dashed, and dash-dotted curves in the plot show the capacitances for the remaining term, air gap, and electrode polarization, respectively. The electrode polarization capacitance dominates the remaining capacitance at low frequencies and vanishes afterward. c) Magnitude ratio (MR) of each impedance component to the total sliding impedance as a function of frequency. The solid, dashed, and dash-dotted curves in the plot show the MR of the remaining term, touchscreen, and skin, respectively. d) Percent phase synchronization (PPS) of each impedance component to that of the total sliding impedance. The solid, dashed, and dash-dotted curves in the plot show PPS of the remaining term, touchscreen, and skin, respectively.}
\label{fig:ch4_FurtherAnalysis}
\end{figure*}

 Fig. \ref{fig:ch4_Circuit} shows our proposed schematic circuit model taking into account the effect of electrode polarization. The reason behind electrode polarization is the EDL, which forms at the contact interface of the finger surface \cite{kuang1998low}. The EDL also explains that the electrical behavior of the finger-touchscreen interface is more conductive at low frequencies due to the charge leakage, while more capacitive at high frequencies since the charge leakage diminishes. This clarifies why the strength of the electric field is weaker under DC stimulation compared to AC stimulation. In reference to Fig. \ref{fig:ch4_Circuit}, the capacitances C$_{EP}$ and C$_{gap}$ short out as the frequency of stimulation approaches zero (i.e., DC condition) and all the current is channeled to the resistance R$_{EP}$, which results in higher charge leakage compared to that of the AC condition.

\subsection{Estimation of Air Gap Thickness} \label{sec:ch4_EstimationGapThickness}
In order to estimate the air gap capacitance from the experimental data, we refer to the remaining admittance (Fig. \ref{fig:ch4_FurtherAnalysis}a), which is composed of remaining conductance and susceptance. Above 30 Hz, the conductance term almost vanishes and we can write the remaining admittance as:
\begin{equation}
    |Y_R|=B_R/{\sin\Phi_R}
\end{equation}
where, B$_R$ and $\Phi_R$ represent the remaining susceptance and phase angle, respectively. At high frequencies, the phase angle is roughly 90 degrees, and hence the remaining admittance equals the remaining susceptance as:
\begin{equation}
    |Y_R|=B_R=\omega C_R
\end{equation}
where, $\omega$ and C$_R$ are the frequency of stimulation and the remaining capacitance, respectively. Hence, at frequencies higher than 30 Hz, the remaining capacitance is solely equal to the air gap capacitance and its value can be calculated by:
\begin{equation}
    C_R=C_{gap}={|Y_R|}/{\omega}
\end{equation}
A constant value of C$_{gap}$ = 413 pF was obtained using this equation and the experimental data of remaining admittance. The capacitances of the air gap and the electrode polarization are plotted as a function frequency in Fig. \ref{fig:ch4_FurtherAnalysis}b. In this figure, the estimated air gap capacitance (413 pF) is represented by a horizontal dashed line in green color. 

On the other hand, the capacitance of the air gap is equal to:
\begin{equation}
    C_{gap} = \varepsilon_0 \varepsilon_{gap} {A_{app}}/{u}
\end{equation}
where, $\varepsilon_{gap}$ is the permittivity of the air (approximately 1.00059) and $A_{app}=\;130$ mm$^2$ is the apparent contact area of the participant's index finger, measured by a camera and using the set-up presented in Section \ref{sec:ch3_ExperimentalMethods}. Substituting these values into the above equation returns the average thickness of the air gap as $u$ = 2.78 $\mu$m.

\subsection{Inferring Electrostatic Forces from Electrical Impedance Measurements}\label{sec:ch4_InferringElectrostaticForcesFromImpedance}
The earlier approaches in the literature for estimating the electrostatic forces \cite{meyer2013fingertip, vezzoli2014electrovibration, vardar2016effect, forsbach2021rigorous} utilized circuit models without paying attention to the leakage of charges from the finger to the surface of the touchscreen. In Chapter \ref{chapter:FrequencyDependentElectroadhesion}, we inferred the electrostatic forces between a human finger and a voltage-induced touchscreen using an electro-mechanical model that relies on fundamental laws of electric fields. We claimed that considering simple electrical elements such as resistors and capacitors are inadequate to model the true experimentally-observed behavior of electrostatic forces changing as a function of frequency. This claim is justified by our current experimental results, where both skin and touchscreen impedances (see Fig. \ref{fig:ch4_ResultsSkinHydrogel}, \ref{fig:ch4_ResultsSkinMetal}, and \ref{fig:ch4_ResultsTouchscreen}) exhibit the behavior of a resistor in parallel with a constant phase element (CPE) as suggested earlier for modeling skin bioimpedance \cite{grimnes2015bioimpedance}. In particular, the phase angle measurements for the skin and the touchscreen in our study cannot be interpreted by a simple circuit model of a resistance in parallel with a capacitance. Hence, we use a black-box approach and directly utilize the experimental data of impedance measurements for the skin and touchscreen rather than constructing individual circuit models for them. On the other hand, the behavior of the remaining impedance (see Fig. \ref{fig:ch4_ResultsRemaining}) is relatively easy to interpret, and hence we modeled its behavior using an electrical circuit.

\begin{figure*}[t!]
\centering
\includegraphics[width=0.7\columnwidth]{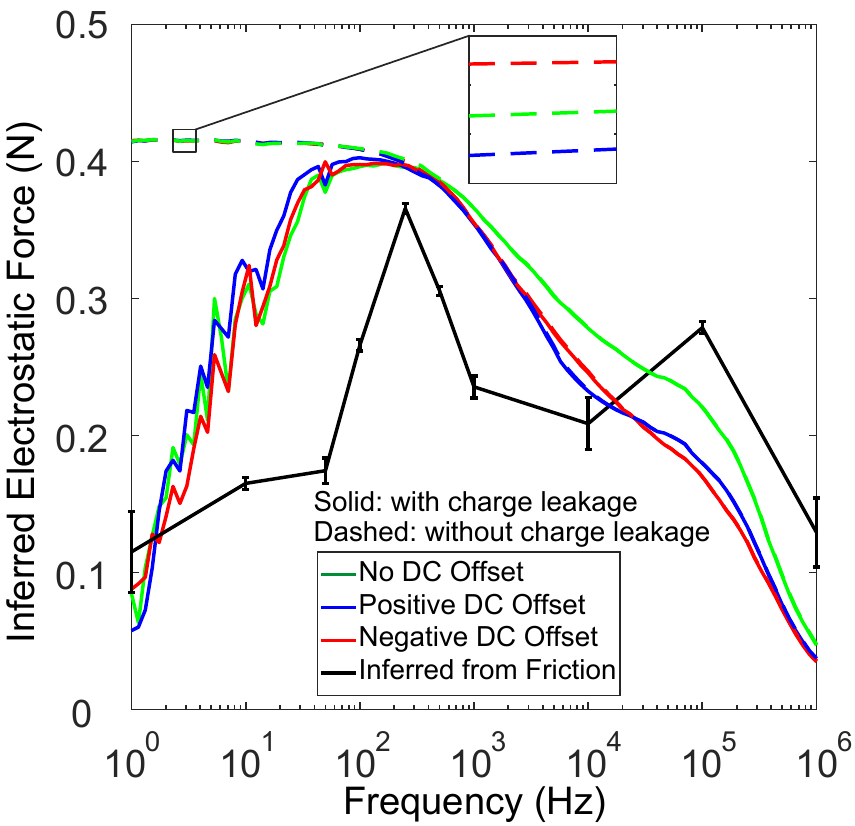}
\caption{Electrostatic forces inferred from the friction (black-colored curve) and electrical impedance (green, blue, and red-colored solid curves) measurements.}
\label{fig:ch4_ElectrostaticForce}
\end{figure*}

We use Eq. \ref{eq:ch1_MaxwellTensor} to calculate the electrostatic forces as a function of frequency. Since the average air gap was already estimated from the experimental data by the method discussed in the earlier section, the only unknown in Eq. \ref{eq:ch1_MaxwellTensor} is the voltage across the air gap, which was estimated using the circuit model given in Fig. \ref{fig:ch4_Circuit}. If an AC voltage (V) with an amplitude V$_0$ is applied to the conductive layer of the touchscreen, then the total current passing through the system is $I = {V}/{Z_{Total}}$. A potential drop occurs across the insulator layers of the touchscreen and skin. Hence, the voltages at nodes 1 and 2 in Fig. \ref{fig:ch4_Circuit} become $V_1 = V - Z_{TS} I$ and $V_2 = Z_{Skin}I$, respectively. Then, the voltage across the remaining impedance, $\Delta V$, is the difference between V$_1$ and V$_2$ ($\Delta V = V_1 - V_2$). Since the current passing through R$_{EP}$ equals the difference between the total current I and the capacitive currents for the air gap (I$_{gap}$) and the electrode polarization (I$_{EP}$), then one can write the potential difference across the air gap as:
\begin{equation}\label{eq:delta Vg}
    \Delta V_{gap} = \Delta V - R_{EP} \left[I - (I_{gap} + I_{EP}) \right]
\end{equation}
Substituting Eq. \ref{eq:delta Vg} into Maxwell's stress tensor (Eq. \ref{eq:ch1_MaxwellTensor}) gives the electrostatic forces, which are plotted in Fig. \ref{fig:ch4_ElectrostaticForce} as a function of frequency for V$_0=$ 75 volts. The magnitude of electrostatic forces increases with the stimulation frequency until reaching a peak value at 250 Hz and decreases afterward. A smaller peak also occurs at 100 kHz. The dashed curves in the figure represent the case in which the electrode polarization was not considered in the model (i.e., the charge leakage from the finger to the surface of the touchscreen was not taken into account). Under this circumstance, the magnitude of electrostatic forces is high and constant until 250 Hz and then drops. This supports our earlier findings in Chapter \ref{chapter:FrequencyDependentElectroadhesion}, suggesting that charge leakage is the dominant factor in reducing the magnitude of electrostatic forces at low frequencies. 

In order to further investigate the frequency-dependent behavior of electrostatic forces, we analyze the relative contribution of each impedance (skin, touchscreen, and remaining) to the total sliding impedance. First, the ratio of magnitudes of each impedance to the total sliding impedance was calculated by \cite{shultz2018electrical}:
\begin{equation}\label{eq:ratio}
    MR_i=\frac{Z_i}{Z_{total(Sliding)}}
\end{equation}
where, i indicates each component (skin, touchscreen, remaining). The results (Fig. \ref{fig:ch4_FurtherAnalysis}c) show that the remaining impedance contributes the most to the total impedance until 100 kHz. The touchscreen and skin start to contribute to the total impedance after 100 Hz and 10 kHz, respectively. Second, we analyze how well the phase of each impedance (skin, touchscreen, remaining) was synchronized with that of the total sliding impedance using the following metric: 
\begin{equation}\label{eq:sync}
    PPS_i=\left[1-\frac{|\Phi_i-\Phi_{Total(Sliding)}|}{|\Phi_{Total(Sliding)}|}\right] \times 100
\end{equation}
where, PPS$_i$ is the percent phase synchronization of the impedance component $i$ with respect to the phase of the total sliding impedance. As shown in Fig. \ref{fig:ch4_FurtherAnalysis}d, the phase of the remaining impedance is highly synchronized with that of the total sliding impedance until 100 kHz, where a sharp asynchronization starts afterward. The phases of the skin and touchscreen are most synchronized with the phase of the total at approximately 250 Hz and 100 kHz, respectively. Based on the MR and PPS plots and our earlier findings, we claim that the air gap impedance dominates the total sliding impedance for a broad range of stimulation frequencies (from approximately 30 Hz to 100 kHz). Moreover, the two peak amplitudes observed at 250 Hz and 100 kHz in our measurement of electrostatic forces (see Fig. \ref{fig:ch4_ElectrostaticForce}) match with the peaks observed in the PPS curves for the skin and touchscreen impedances (see Fig. \ref{fig:ch4_FurtherAnalysis}d), respectively. We argue that the synchronization of the skin phase at 250 Hz and the touchscreen phase at 100 kHz with the phase of the total impedance amplifies the strength of the electric field at those frequencies, resulting in high amplitudes in electrostatic forces.

\begin{figure*}[t!]
\centering
\includegraphics[width=0.95\columnwidth]{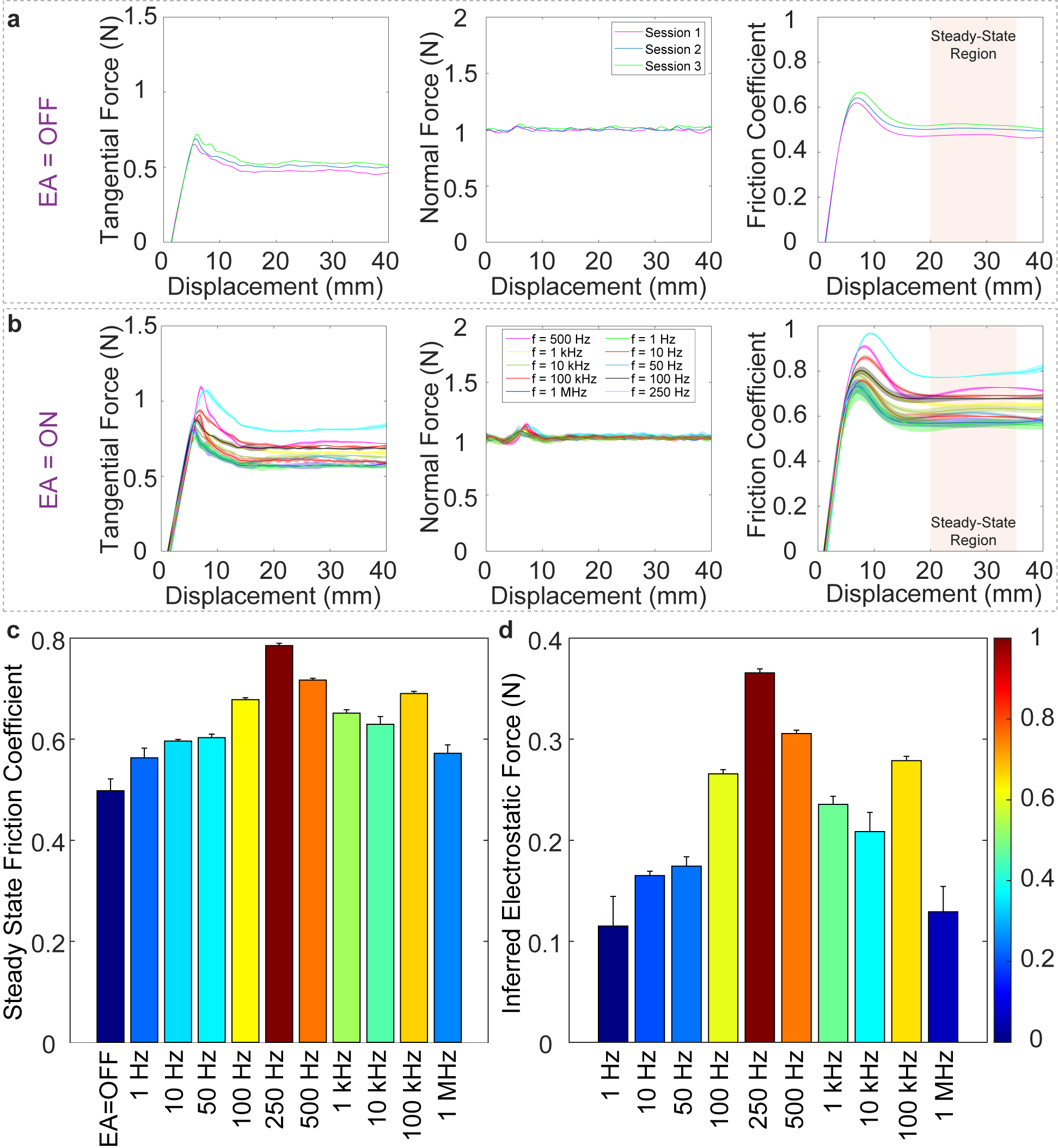}
\caption{a) The mean tangential and normal forces acting on the participant's index finger and the CoF as a function of displacement when electroadhesion was turned off. b) The mean tangential and normal forces acting on the participant's index finger and the CoF as a function of displacement when electroadhesion was turned on. Sinusoidal voltage signals at ten different frequencies with an amplitude of 75 volts were applied to the conductive layer of the touchscreen to generate electrostatic forces between finger and touchscreen. c) The mean steady-state CoF when EA = OFF and for the ten stimulation frequencies when EA = ON. d) The mean electrostatic forces as a function of stimulation frequency when EA = ON.}
\label{fig:ch4_FrictionResults}
\end{figure*}

\subsection{Inferring Electrostatic Forces from Friction Measurements}\label{sec:ch4_InferringElectrostaticForcesFromFriction}
The electrostatic forces inferred from the impedance measurements were compared with the ones inferred from friction measurements performed by the set-up shown in Fig. \ref{fig:ch3_FrictionSetup}. The average tangential and normal forces and the CoFs are presented in Fig. \ref{fig:ch4_FrictionResults}a and b as a function of relative displacement between finger and the touchscreen for EA = OFF and EA = ON conditions, respectively. The steady-state values of CoF are shown in Fig. \ref{fig:ch4_FrictionResults}c and the electrostatic forces inferred from friction measurements are presented in Fig. \ref{fig:ch4_FrictionResults}d. The electrostatic forces inferred from friction and electrical impedance measurements show good agreement, as shown in Fig. \ref{fig:ch4_ElectrostaticForce}. They both exhibit an inverted parabolic behavior where the electrostatic forces increase until approximately 250 Hz and drop afterward, with another peak at approximately 100 kHz. A similar inverted parabolic behavior was reported in \cite{friesen2021building} for the mechanical vibrations of sliding human finger under electroadhesion, measured by Laser Doppler Vibrometer for the stimulation frequencies between 10 Hz and 1 kHz.

%//////////////////////////////////////////////////////////////////////////////////////////////////////////////////////////////////////////////////////////////////////////

\chapter{Effects of Voltage Type and Finger Moisture on Electroadhesion}\label{chapter:RoleMoistureElectroadhesion}
With the current technology, generating electrostatic forces between a human finger and a touchscreen to modulate the frictional forces between them is quite straightforward. Producing a detectable tactile sensation necessitates the use of an AC voltage signal. A DC input voltage signal, even with a higher amplitude, does not appear to generate a similar sensation, but the root causes of this difference are not fully known. Moreover, finger moisture and the humidity of the environment are known to affect our tactile perception under electroadhesion \cite{sirin2019fingerpad}, but less attention is given to explaining the physics behind the change in electroadhesion intensity due to the accumulated sweat at the interface between the finger and the touchscreen. Chatterjee et al. \cite{chatterjee2023preferential} observed that the finger leaves more residue (primarily sweat and sebum) in the areas of a touchscreen where electroadhesion was active. They suggested that 1) the electrohydrodynamic deformation of sebum droplets adhere to the finger valleys, which results in the creation of extra capillary bridges and leftover droplets on the screen's surface after they break, and 2) the electric field-induced stabilization of sebum capillary bridges exist between the finger ridges and the screen, which leads to the merging and formation of larger droplets.

In this chapter\footnote{This chapter is based on an article \cite{aliabbasi2024effect}}, we investigate the human tactile detection thresholds of electroadhesion for DC and AC voltage signals applied to the touchscreen. We show that the detection threshold under the AC condition is significantly lower than that of the DC condition. Using an electrical circuit model, we explain the physical mechanism behind this discrepancy. We also highlight the adverse effect of fingertip skin moisture on the tactile perception of electroadhesion by correlating the finger moisture of the participants measured by a Corneometer with their threshold values for the AC signal. Our electrical impedance measurements show that total impedance drops drastically when sweat accumulates at the interface.

\section{Materials and Methods}
\subsection{Participants}
The tactile perception experiment was conducted with ten adult participants (four females and six males) having an average age of 29.4 years (SD: 5.9) and the electrical impedance measurements were performed with one male participant (age: 32 years old). All participants provided written informed consent to undergo the procedure, which was approved by the ethics committee of Koc University. The investigation conformed to the principles of the Declaration of Helsinki and the experiments were performed in accordance with relevant guidelines and regulations.

\subsection{Tactile Perception Experiment}

\begin{figure}[!t]
\centering
\includegraphics[width=0.9\textwidth]{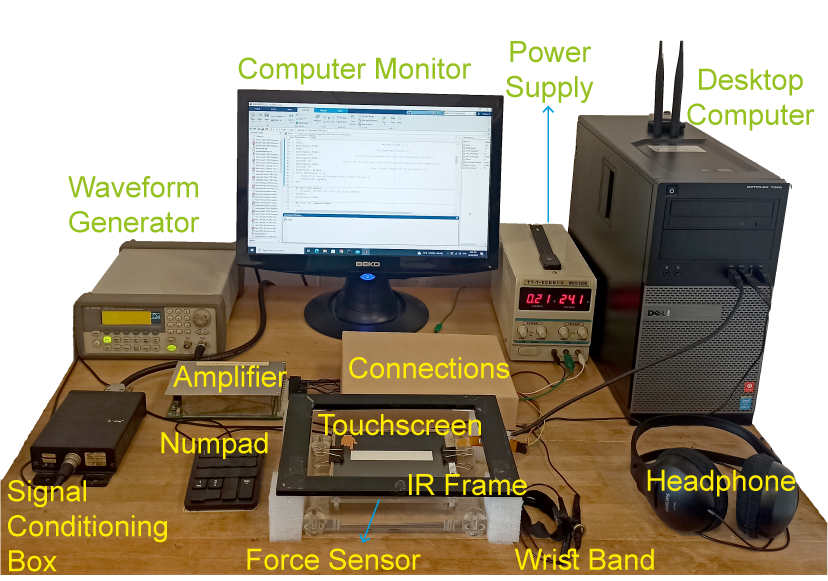}
\caption{The set-up used in our psychophysical experiments.} \label{fig:ch5_Setup}
\end{figure}

We used the set-up shown in Fig. \ref{fig:ch5_Setup} for the tactile perception experiment. In this set-up, the input voltage signals were generated by a waveform generator (33220A, Agilent Inc.) connected to a PC via a TCP/IP protocol. The signals were then amplified by a piezo amplifier (E-413, PI Inc.) and applied to the conductive layer of a capacitive touchscreen (SCT3250, 3M Inc.) for displaying tactile stimulus to the participants. The touchscreen was rigidly fixed with holders to avoid unwanted vibrations during the experiments. A DC power supply (MCH-303D, Technic Inc.) was utilized to provide 24 V DC voltage for operating the amplifier. A high-resolution force sensor (Nano 17, ATI Industrial Automation Inc.) was placed below the touchscreen to measure contact forces. These forces were acquired by a 16-bit analog data acquisition card (PCI-6034E, National Instruments Inc.) with 10 kHz sampling frequency. An IR frame (IRTOUCH Inc.) was placed above the touchscreen to detect finger position.

Before the experiments, the participants washed their hands with soap, rinsed with water, and dried them at room temperature, and the touchscreen was cleaned with alcohol. Throughout the experiments, the participants were asked to wear an elastic strap on their stationary wrist for grounding and put on headphones playing white noise to prevent their tactile perception from being affected by any external auditory cue.

To investigate human tactile detection thresholds for DC and AC voltage stimulations, we conducted the absolute detection experiment using the 2AFC paradigm as implemented by Vardar et al. \cite{vardar2016effect,vardar2017effect}. During the experiments, participants were asked to slide their index fingers on the touchscreen from left to right twice for a distance of 100 mm in each trial. The tactile stimulus was displayed only in one of the passes, which was randomized to eliminate any bias. Participants were asked to determine the pass (interval) in which they felt a tactile effect. In order to regulate their scan speed, a visual cursor moving with a speed of 20 mm/s was displayed on the computer monitor and the participants were asked to follow it with their finger. To assist the participants with controlling their applied normal forces on the touchscreen, another visual feedback displayed the real-time magnitude of the applied normal force. The normal force and scanning speed were recorded in each trial. The average normal force for all participants was 0.334 N (SD: 0.084) and 0.332 N (SD: 0.073) under DC and AC conditions, respectively. The average speed for all participants was 20.01 mm/s (SD: 1.34) and 20.06 mm/s (SD: 0.87) under DC and AC conditions, respectively. If the normal force or the scan speed of a participant was not in the desired range (0.1-0.6 N; 10-30 mm/s), the trial was repeated until a measurement within the range was obtained. Before starting the experiments, participants were given instructions and asked to complete a training session. This training session enabled participants to adjust their finger scanning speed and normal force before the actual experimentation.

\begin{figure}[!t]
\centering
\includegraphics[width=\textwidth]{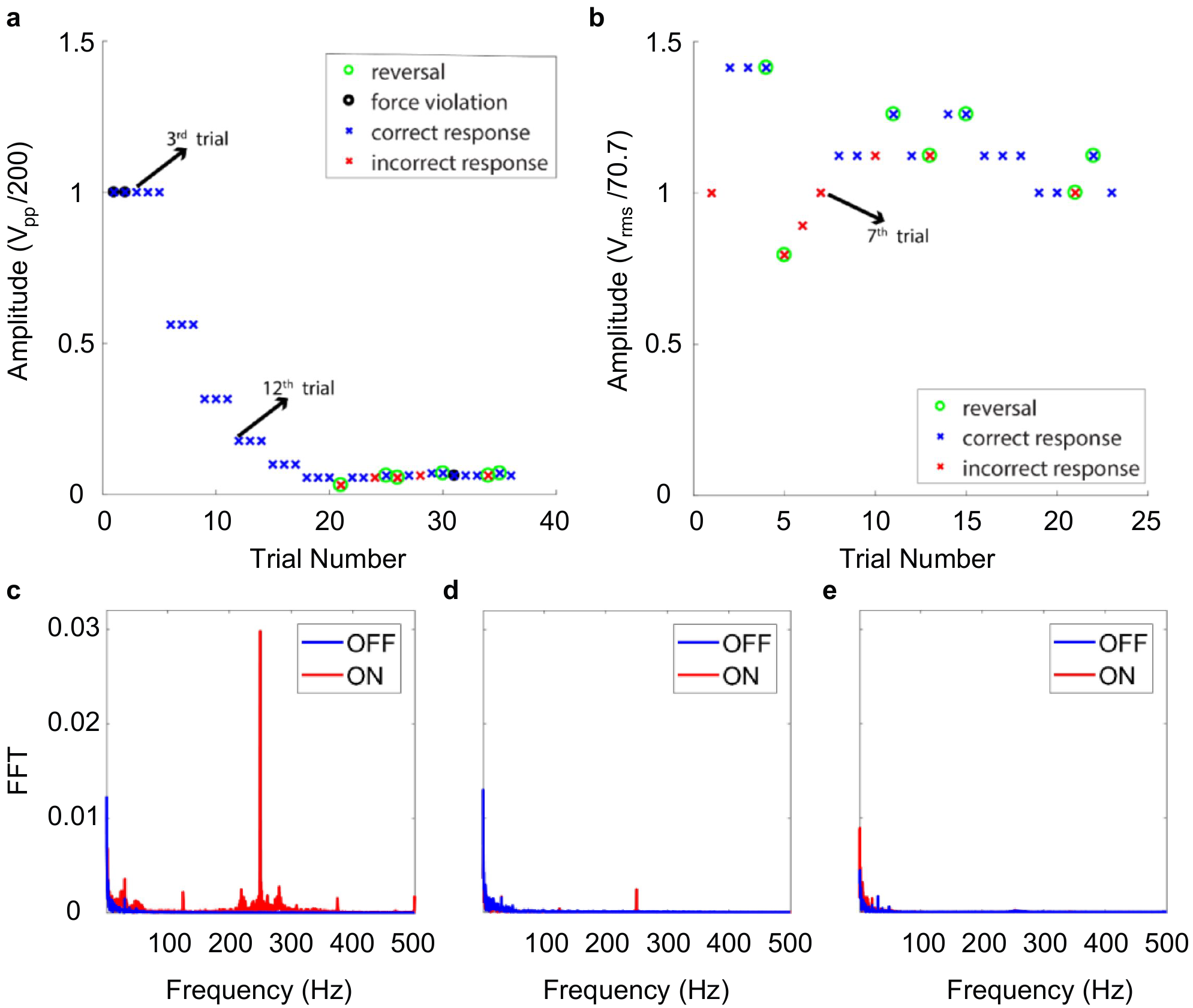}
\caption{An exemplar staircase data obtained by the threshold experiments under a) AC and b) DC conditions for participant S2. FFT analysis of the tangential force under AC condition for the c) 3\textsuperscript{rd} and d) 12\textsuperscript{th} trials and under DC condition for e) 7\textsuperscript{th} trial.}
\label{fig:ch5_FFT}
\end{figure}

The amplitude of the voltage signal applied to the touchscreen (and hence the magnitude of the tactile stimulus) was altered using the three-up/one-down adaptive staircase method as described in Vardar et al. \cite{vardar2018tactile}. In order to determine the detection thresholds for the sinusoidal AC signal at 125 Hz and the corresponding DC signal, the experiment started with a voltage amplitude of 200 V$_{pp}$ and 70.7 V$_{rms}$, respectively. It is worth noting here that the initial voltage amplitude provided sufficiently high intensity for all participants under the AC condition but not under the DC condition. However, for safety reasons, the maximum voltage was limited to 100 V$_{rms}$ in DC experiments. If a participant gave three correct responses (not necessarily consecutive), the voltage level was decreased by 5 dB. If a participant gave one incorrect response, the voltage level was increased by 5 dB. The change in response from correct to incorrect or vice versa was counted as one reversal. After one reversal, the step size was decreased to 1 dB. The experiments were stopped automatically if the reversal count was five at the ±1 dB level (Fig. \ref{fig:ch5_FFT}a and b). The threshold was calculated as the mean of the last five reversals. The moisture level of each participant was measured by a Corneometer (CM 825, Courage - Khazaka Electronic) four times just before and right after the experiment and the average of eight measurements was reported for each participant.

\subsection{Electrical Impedance Measurements}
Due to its time-consuming nature, the electrical impedance measurements were performed with one participant only. We performed the total electrical impedance measurements while the participant's finger was sliding on the surface of the touchscreen under the wet lubrication condition. In nominal condition (reported in Chapter \ref{chapter:ExperimentalElectrostaticForcesFromImpedance}), no excess liquid was added to the interface between the finger and the touchscreen. However, in wet condition, four droplets of 5 $\mu$L 0.9\% Isotonic Sodium Chloride (NaCl) were applied at four different locations on the touchscreen to imitate a very moist finger. Similar to the nominal condition in Chapter \ref{chapter:ExperimentalElectrostaticForcesFromImpedance}, the measurements for the wet condition were also performed in three separate sessions on three different days and the data were collected in ten consecutive trials. The procedures and set-ups for these measurements are explained in Section \ref{sec:ch4_ApparatusMeasuringElectricalImpedances}.

\section{Results}
\subsection{Tactile Perception Experiments}
\begin{figure}[!t]
\centering
\includegraphics[width=1\textwidth]{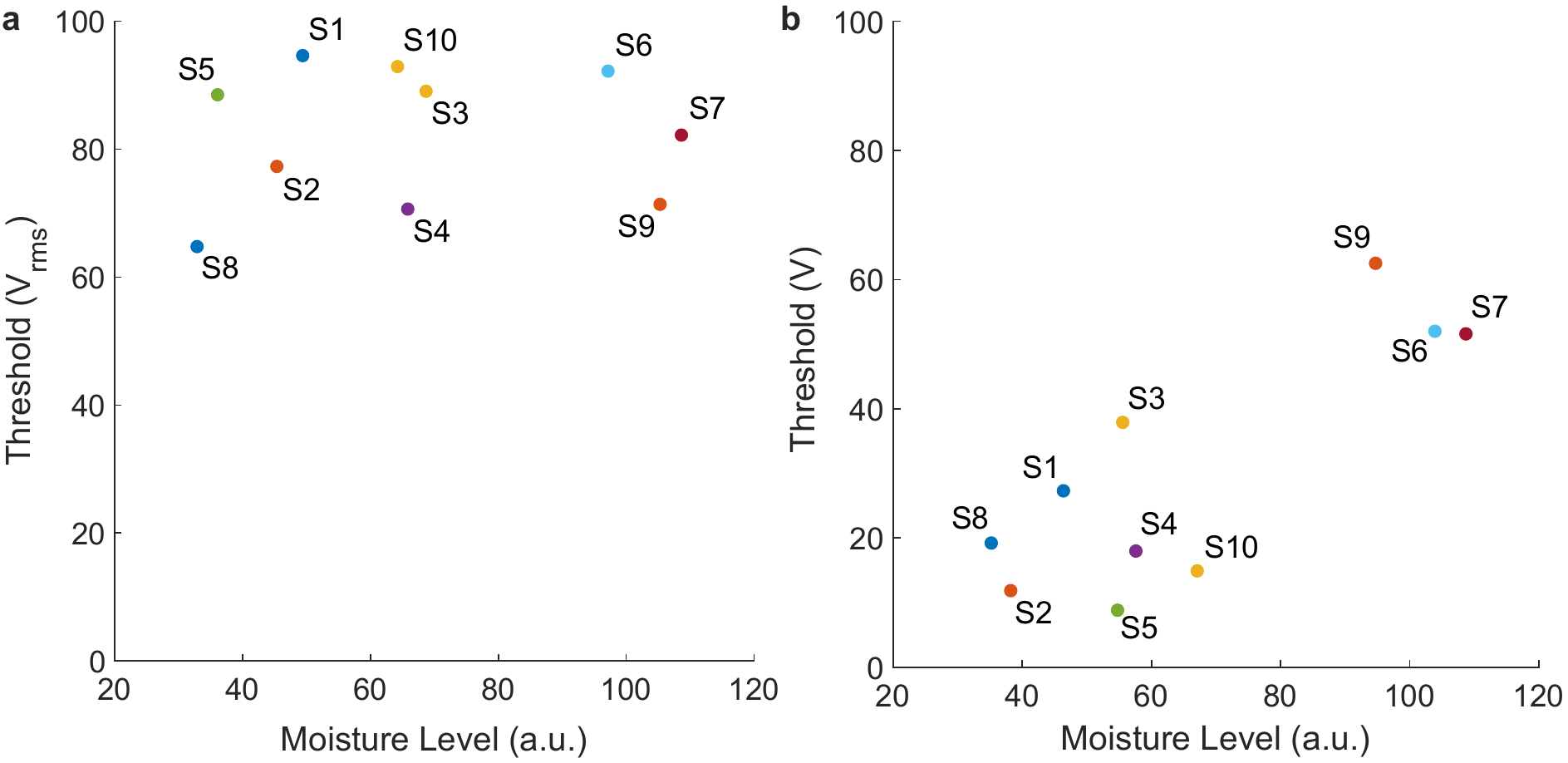}
\caption{The threshold voltage versus moisture level of each subject under a) DC and b) AC conditions. Moisture measurements are given in arbitrary units (a.u.) that range from 20 (dry skin) to 120 (very wet skin).} \label{fig:ch5_Threshold}
\end{figure}

Fig. \ref{fig:ch5_Threshold}a and b present the threshold voltages and fingertip skin moisture level of each participant under DC and AC conditions. The results showed that the threshold voltage under the AC condition was significantly lower than that of the DC condition for all participants. There was a clear convergence in threshold values for all participants under the AC condition, while there was no convergence under the DC condition (note that the experiments were stopped automatically if the reversal count was five at ±1 dB level). The threshold voltage was calculated as the mean of the last five reversals. The average threshold value obtained under AC condition (30.41 volts) is consistent with the threshold value reported in \cite{vardar2016effect} at 125 Hz stimulation frequency. We performed ANOVA with repeated measures on the moisture levels of participants using the measurement time (before and after the experiment) and the signal type (DC vs. AC) as the main factors. We did not observe a significant difference between the moisture levels before and after the experiments (F (1,9) = 0.205, p = 0.661) and DC versus AC (F (1,9) = 5.126, p = 0.051). However, the threshold voltages of the participants having a moist finger (S6, S7, S9) were higher than the other participants under AC condition (see Fig. \ref{fig:ch5_Threshold}a). The Pearson correlation showed a positive and moderate correlation between threshold voltages and moisture level (r = 0.81, p $<$ 0.01). This result is consistent with the results of our earlier study \cite{sirin2019fingerpad} and supports our claim that moisture has an adverse effect on the capacity of electroadhesion to modulate friction, which in turn affects the tactile perception of electroadhesion.

\subsection{Electrical Impedance Measurements}
\begin{figure}[!t]
\centering
\includegraphics[width=\textwidth]{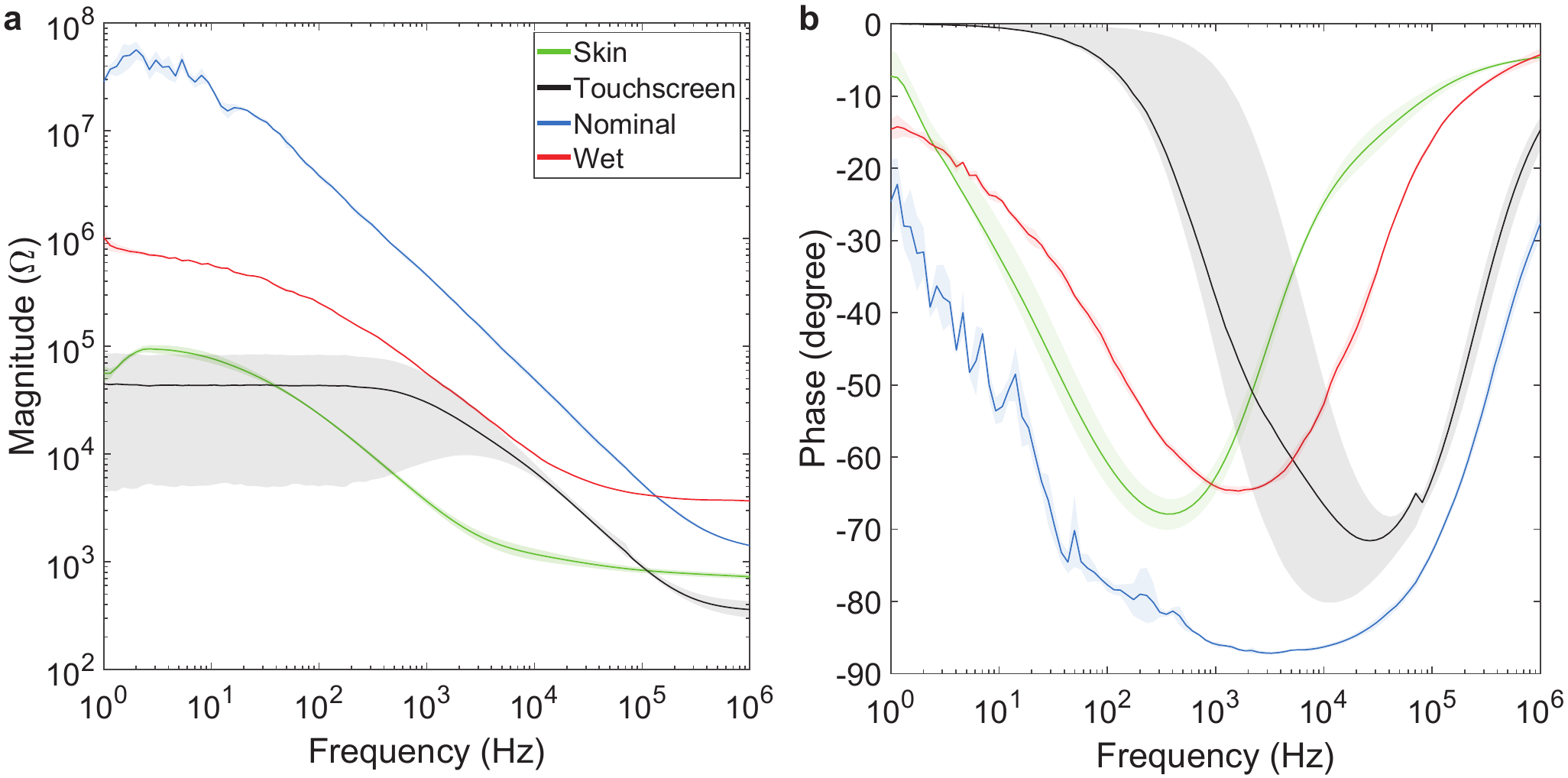}
\caption{The change in average impedance a) magnitude and b) phase as a function of frequency for finger skin (green), touchscreen (black), sliding finger in nominal condition (blue), and sliding finger in wet condition (red).}
\label{fig:ch5_ImpedanceResults}
\end{figure}

\begin{figure}[!b]
\centering
\includegraphics[width=\textwidth]{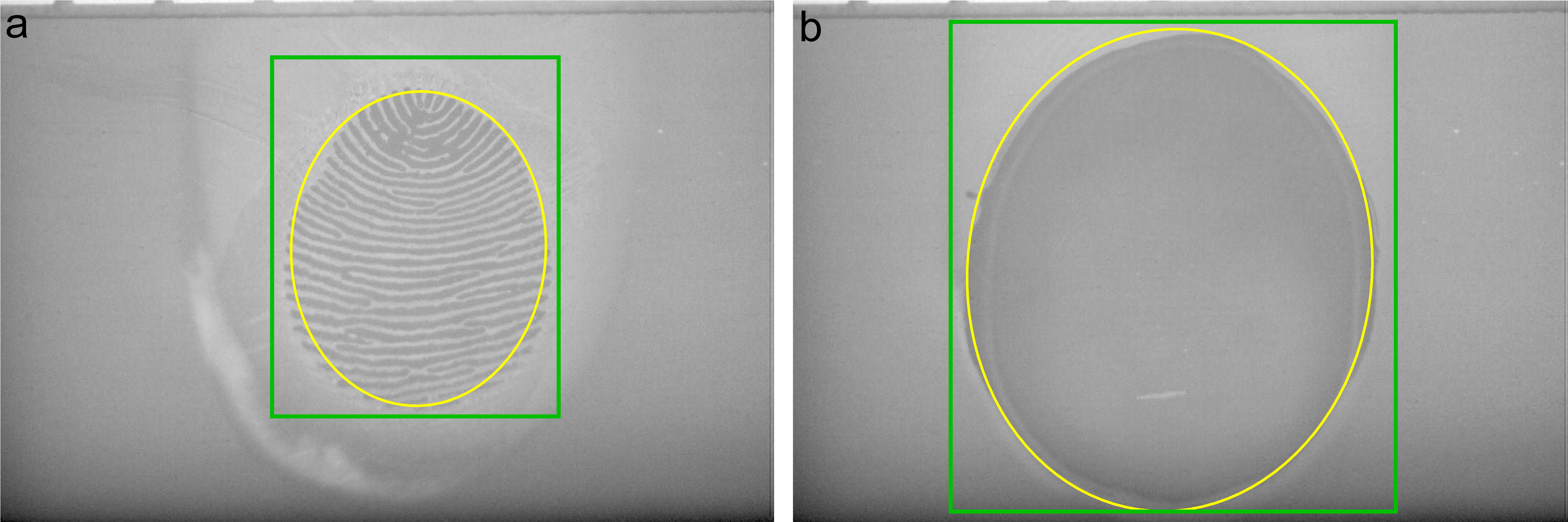}
\caption{Images of the participant’s fingerpad in contact with the surface of the touchscreen for a normal force of 1 N: a) nominal and b) wet contacts (no sliding). The green-colored rectangle and yellow-colored contour represent the region of interest for the image processing operations and fitted ellipse to the apparent contact area, respectively.}
\label{fig:ch5_FingerImages}
\end{figure}

Fig. \ref{fig:ch5_ImpedanceResults}a and b present the average magnitude and phase of the electrical impedance measurements as a function of frequency, respectively. Note that the skin, touchscreen, and nominal finger curves are already reported in Chapter \ref{chapter:ExperimentalElectrostaticForcesFromImpedance} and included in this figure for better comparison. The shaded regions around the curves represent the standard error of means. Green, black, blue, and red colored curves show the electrical impedance measurements for skin, touchscreen, sliding finger in nominal condition, and sliding finger in wet condition, respectively. As shown in Fig. \ref{fig:ch5_FingerImages}a, we measured the apparent contact area of the participant's fingertip using a high-speed camera as 130 mm$^2$ for a normal force of 1 N using the set-up presented in Section \ref{sec:ch3_ExperimentalMethods}. The apparent contact area increased to 300 mm$^2$ when the interface between the finger and the touchscreen is filled with NaCl in the wet condition (Fig. \ref{fig:ch5_FingerImages}b). Since the impedance measurements are affected by the contact area, we normalized the electrical impedance for the wet condition by multiplying its real and imaginary parts by the ratio of 300/130.

As shown in Fig. \ref{fig:ch5_ImpedanceResults}a, the summation of the skin and touchscreen impedance magnitudes is not equal to the magnitude of the total sliding impedance for both the nominal and wet conditions. Hence, another impedance must be in series with the skin and touchscreen impedances, which we named the remaining impedance in Chapter \ref{chapter:ExperimentalElectrostaticForcesFromImpedance}. We subtract the skin (Z$_{Skin}$) and touchscreen (Z$_{TS}$) impedances from the total sliding impedance (Z$_{Sliding}$) to obtain the remaining impedance as suggested in \cite{shultz2018electrical}.

\begin{figure}[!t]
\centering
\includegraphics[width=\textwidth]{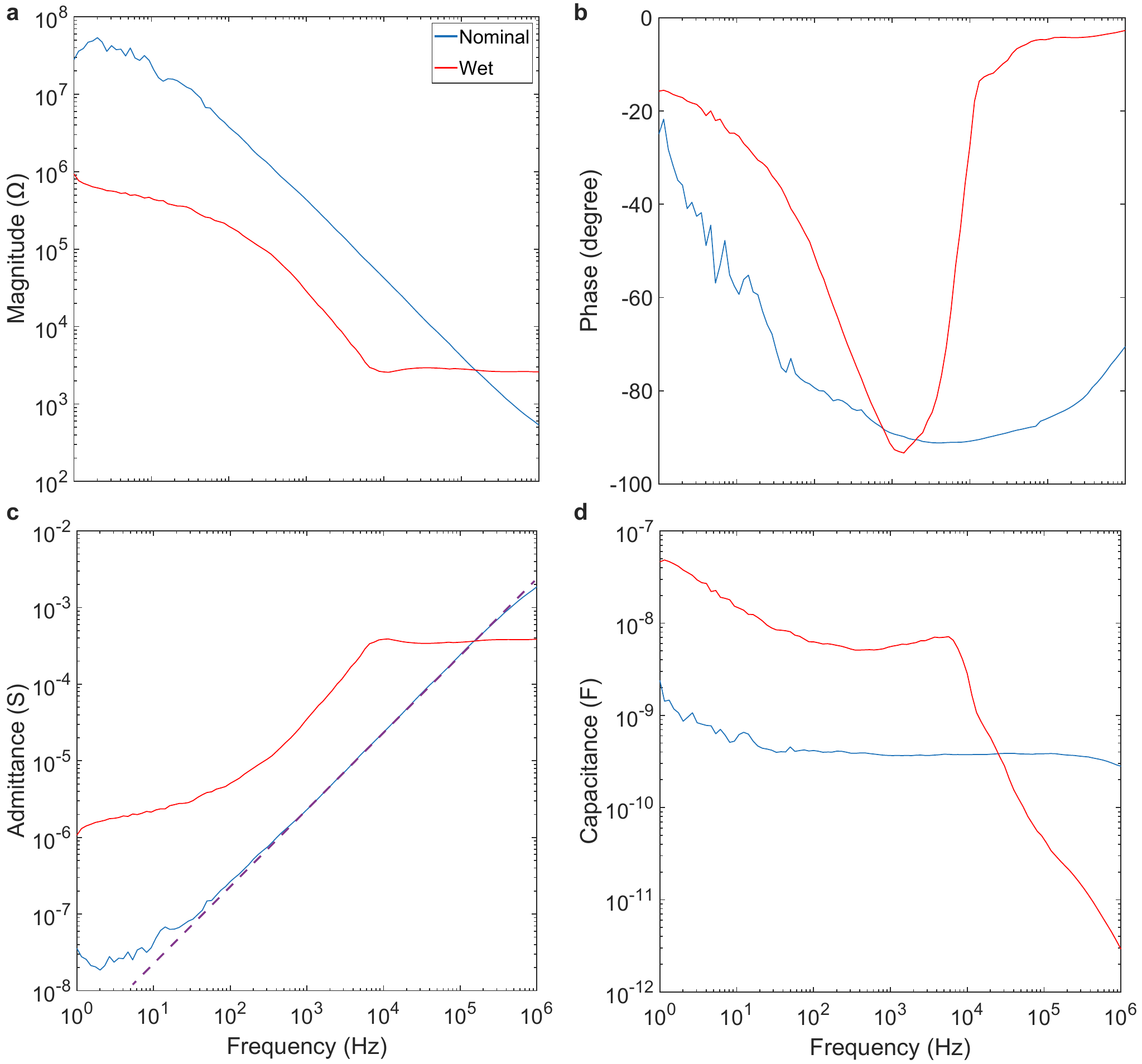}
\caption{a) The magnitude and b) phase of the remaining impedances for the sliding finger under nominal (blue) and wet (red) conditions as a function of frequency. c) The change in remaining admittances as a function of frequency for the sliding finger under nominal (blue) and wet (red) conditions as a function of frequency. For the frequencies higher than 30 Hz, the dashed-purple line fits the admittance curve of the sliding finger under nominal condition well, indicating the purely capacitive behavior of the nominal finger at those frequencies. d) The change in the capacitance of the sliding finger under nominal and wet conditions as a function of frequency.}
\label{fig:ch5_RemainingResults}
\end{figure}

The magnitude and phase of the remaining impedances for the sliding finger under the nominal (blue-colored curve) and wet (red-colored curve) conditions are presented in Fig. \ref{fig:ch5_RemainingResults}a and b, respectively, as a function of frequency. The magnitude of the remaining impedance for the nominal condition was significantly higher than that of the wet condition. In other words, the liquid at the interface of the finger and the touchscreen caused a drop in impedance magnitude of more than ten folds compared to the nominal condition. As shown in Fig. \ref{fig:ch5_RemainingResults}b, the phase angles of the remaining impedances for the nominal and wet conditions showed a resistive behavior at lower frequencies. This resistive behavior was followed by purely capacitive behavior after approximately 30 Hz for the nominal condition. However, the phase angle of the wet condition showed a capacitive behavior for a narrow range of frequencies, followed by a sharp return to the resistive behavior. Fig. \ref{fig:ch5_RemainingResults}c shows the remaining admittances for the nominal and wet conditions. As discussed in Chapter \ref{chapter:ExperimentalElectrostaticForcesFromImpedance}, a one-decade-per-decade line (dashed purple) fits well to the admittance curve of the nominal condition after 30 Hz, suggesting a constant capacitance after that frequency (Fig. \ref{fig:ch5_RemainingResults}d). Hence, the remaining impedance of the nominal condition can be modeled by a single capacitance (C$_{gap}$) at higher frequencies, representing the air gap between the finger and the touchscreen. At frequencies lower than 30 Hz, there is a parasitic capacitance due to electrode polarization \cite{schwan1968electrode}.

\section{Discussion}
We investigated the effect of input voltage signal type (DC vs. AC) on the tactile perception of electroadhesion. The earlier studies \cite{bau2010teslatouch,vardar2017effect} investigated the human tactile threshold under AC stimulation, but not correlated with the moisture level of participants. Moreover, the results of our tactile detection experiment showed that the threshold voltage under the AC condition was significantly lower than that of the DC condition for all participants (Fig. \ref{fig:ch5_Threshold} a and b). This result is consistent with our earlier findings suggesting that the Pacinian channel is mainly responsible for the tactile perception of electroadhesion \cite{vardar2017effect,vardar2016effect}] at the frequency of our stimulation (125 Hz). Since the rapidly adapting receptors are not stimulated when a DC stimulation is applied to the touchscreen (Fig. \ref{fig:ch5_FFT}e), it is not surprising that the magnitude of the perceived tactile stimulus is reduced. Our threshold experiment showed that the detection of electroadhesion stimuli not only depends on the amplitude of the voltage applied to the touchscreen but also on the human psychophysical sensitivity to tactile stimuli, as investigated by Vardar et al. \cite{vardar2017effect}. Each psychophysical channel is sensitive to different input frequencies, which partially overlap. In our threshold experiments, the frequency of the input voltage under AC condition was 125 Hz, which was primarily detected by the Pacinian channel at 250 Hz. Since the electrostatic force is proportional to the square of the input voltage, the frequency of the output force signal is twice the frequency of the input voltage applied to the touchscreen \cite{vardar2016effect}. As shown in Fig. \ref{fig:ch5_FFT}c, the FFT analysis of the tangential force under AC condition showed a peak at 250 Hz (see the 3\textsuperscript{rd} and 12\textsuperscript{th} trials). The FFT magnitude of the peak in the 3\textsuperscript{rd} trial (Fig. \ref{fig:ch5_FFT}c) is already high due to the high voltage applied to the touchscreen, while the one in the 12\textsuperscript{th} trial (Fig. \ref{fig:ch5_FFT}d) is lower in magnitude than those of some other frequencies. However, it is known that human vibrotactile perception is frequency-dependent, with a sensitivity peak around 250 Hz. If the energy contained by each frequency component is multiplied by the inverse of the normalized human sensitivity curve as suggested by Vardar et al. \cite{vardar2016effect}, then one can more easily appreciate why the small peak observed at 250 Hz in the 12\textsuperscript{th} trial affects the participant’s tactile perception significantly. This argument is also supported by the fact that the participant has successfully given a correct response under AC condition not only in the 12\textsuperscript{th} trial but also in the following eight trials (Fig. \ref{fig:ch5_FFT}a), despite the amplitudes of voltage signals applied to the touchscreen being much lower than the one, for example, applied to the touchscreen in the 7\textsuperscript{th} trial (Fig. \ref{fig:ch5_FFT}b) under DC condition.

The significant difference in threshold values of participants under DC and AC stimulations can also be interpreted using the circuit model proposed in this study (Fig. \ref{fig:ch4_Circuit}), which was developed based on the results of electrical impedance measurements. The impedance measurements show that the interface of the finger and touchscreen is more conductive at lower frequencies due to the leakage of electrical charges from the finger to the surface of the touchscreen. At higher frequencies, the charge leakage diminishes and the behavior of the interface becomes more capacitive. This understanding further clarifies the weaker electric field strength observed during DC stimulation compared to AC stimulation. As depicted in Fig. \ref{fig:ch4_Circuit}, the capacitances C$_{EP}$ and C$_{gap}$ become effectively short-circuited as the stimulation frequency tends towards zero (i.e., under DC conditions), diverting all the current flow towards the resistance R$_{EP}$, consequently leading to a higher amount of charge leakage in comparison to that observed under AC conditions.

We also investigated the effect of moisture on the tactile detection threshold voltage and found that the participants having a moist finger (S6, S7, S9) had significantly higher threshold levels than the other participants (Fig. 3). For those participants, we hypothesize that the friction force due to the moisture was already significant and a relatively small increase in the same force due to electroadhesion did not contribute much to their tactile perception. Earlier studies performed on smooth glass surfaces (without electroadhesion) showed that the coefficient of friction increases when there is a thin layer of water between the finger and the glass surface \cite{adams2007friction}. This was explained by the softening of the finger, also known as plasticization, which results in an increase in the contact area and, thus, the tangential frictional force. In addition to the finger softening due to moisture, the air gap between the finger and the surface is filled by water particles, causing a reduction in the electrostatic force. This observation was verified by our electrical impedance measurements performed with the sliding finger under the nominal and wet conditions (Fig. \ref{fig:ch5_ImpedanceResults}). In the wet condition, we intentionally added liquid (NaCl) to the interface between the finger and the touchscreen so that the interface remains lubricated throughout the frequency sweep. As shown in Fig. \ref{fig:ch5_ImpedanceResults}a, the magnitude of the electrical impedance for the wet condition dropped more than an order of magnitude compared to that of the nominal condition. This indicates that in the presence of sweat at the interface between the finger and the touchscreen, the electrical charges can move between the finger and the touchscreen more easily, reducing the magnitude of the electric field at the interface.

%//////////////////////////////////////////////////////////////////////////////////////////////////////////////////////////////////////////////////////////////////////////

\chapter{Tactile Perception of a Touchscreen's Smooth Surface}\label{chapter:TactilePerceptionTouchscreen}
\subsubsection{Summary}
In this chapter\footnote{This chapter is based on an article \cite{aliabbasi2023tactile}}, we aim to investigate the effect of the top coating layer of the touchscreen on our tactile perception with and without electroadhesion. Within the time frame of this thesis, we focus on the latter (no electroadhesion). For this purpose, we produce five touchscreens with different materials as the top coating layer and perform psychophysical experiments and physical measurements to characterize them. The relationship between tactile perception of extremely smooth surfaces coated with different materials (i.e., touchscreen) and the adhesive contact interactions between human finger and those surfaces can be due to the potential effect of surface chemistry. This topic can also be interesting to glass companies in developing standards for the classification of smooth surfaces with different coatings since the roughness and stickiness of such surfaces are both important for aesthetics and functionality of an end product. Stickiness, for example, may contribute not just to the affective attributes of a glass product, such as pleasantness, but also to its functional attributes, such as scratch resistance. Our study shows that human finger can function as a tactile sensor to detect the differences in surface chemistry and, hence, successfully differentiate the surfaces with different coatings.

\section{Materials and Methods}\label{sec:ch6_MaterialsMethods}

\subsection{Sample Preparation}\label{sec:ch6_SamplePreparation}
Multi-layer thin film stacks were deposited on soda-lime float glasses in an inline horizontal coater using the magnetron sputtering technique. Each individual thin film layer was deposited in desired thicknesses by adjusting the related process parameters such as carrier speed, power, temperature, and process gases. A total of five substrate surfaces (S1-S5) were produced for testing in this study (see Table \ref{table:ch6_CoatingLayers}). Note that only the material type at the top coating layer and the layer below the top one are reported in the table, though there are other layers below. The layer thicknesses are in the order of nanometers, but we are not allowed to provide the details here since they are the trade secrets of the manufacturing company. With respect to the type of material at the top layer, the samples are grouped into three (G1, G2, G3): Titanium-based (S1, S2), Silicon-based (S3, S4), and Zirconium-based (S5). The difference between the samples S1 and S2 is the coating material utilized in the layer below the top layer. We aimed to see if the coatings in lower layers have any significant effect on the tactile perception of the top layer.

\begin{table}[!b]
\renewcommand{\arraystretch}{1.3}
\caption{The five sample surfaces used in our study.}
\label{table:ch6_CoatingLayers}
\centering
\begin{tabular}{|l|l|c|}
\hline
\textbf{Sample} & \textbf{Top layer} & \textbf{Layer below the top} \\ \hline\hline
S1 &  Titanium Oxide (TiOx) & Silicon Oxynitride (SiOxNy) \\ \hline
S2 & Titanium Oxide (TiOx) & Silicon Nitride (SiN) \\ \hline
S3 & Silicon Oxynitride (SiOxNy) &  Silicon Nitride (SiN) \\ \hline
S4 &  Silicon Nitride (SiN) & Nickel Chromium (NiCr)  \\ \hline
S5 &  Zirconium Oxide (ZrOx) & Silicon Nitride (SiN) \\
\hline
\end{tabular}
\end{table}

\subsection{Tactile Perception Experiments}\label{sec:ch6_TactilePerception}
\subsubsection{Participants}\label{sec:ch6_Participants}

Eight healthy participants (three females, five males; mean age = 24.88, SD = 3.14) were selected to take part in this study. A consent form was read and signed by the participants before the experiment, which was approved by the Ethical Committee for Human Participants of Koc University. The study conformed to the principles of the Declaration of Helsinki, and the experiment was performed following relevant guidelines and regulations.

\subsubsection{Apparatus}\label{sec:ch6_TactilePerceptionApparatus}

\begin{figure*}[t!]
\centering
\includegraphics[width=0.9\linewidth]{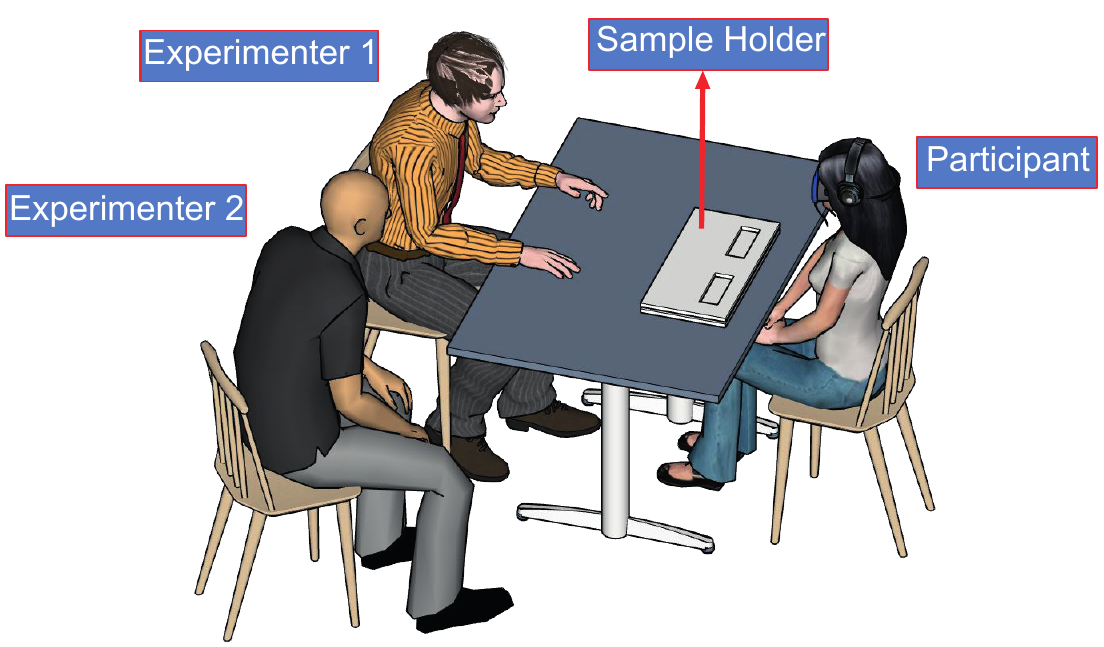}
\caption{a) The set-up used in our psychophysical experiments. A sample-holder made of plexiglass is utilized to secure the sample surfaces for stable tactile exploration. Throughout the whole experiments, participants were asked to wear a headphone displaying white noise and a goggle covering their eyes. Experimenter 1, shown in the rendering, was responsible for starting and stopping each trial by tapping the passive hand of the participant placed on the table and also replacing the samples between the trials. Experimenter 2 was solely responsible for the timing of trials.}
\centering
\label{fig:ch6_PerceptionSetup}
\end{figure*}

We developed a set-up to conduct our tactile perception experiments (Fig. \ref{fig:ch6_PerceptionSetup}). It is comprised of a sample holder designed and cut from plexiglass with two housings for placing a pair of sample surfaces side by side on a table, a headphone for playing white noise to the ears of participants, and a goggle to cover the eyes of participants. In this way, we ensured that there was no visual and auditory interference. Hence, the participants were able to make their decision purely based on tactile cues.

\subsubsection{Procedure}\label{sec:ch6_TactilePerceptionProcedure}
The 2AFC method was utilized in this experiment. The samples were displayed in pairs in random order, and the participants were asked to explore both samples consecutively with their index fingers and choose the one that resisted more to sliding. We intentionally did not want to use the adjectives “rough/smooth” or “sticky/slippery” for the following reasons: First, our study investigates which physical cues play a role in tactile discrimination of smooth surfaces and we did not want to bias the participants by hinting those cues. Second, those adjectives are not easy to explain to the subjects since all the samples in our study are glass with extremely smooth surfaces. Our pilot study revealed that the participants indeed had difficulty in understanding and differentiating those adjectives. There were a total of 10 pairs and each pair was displayed 10 times to each participant. Hence, each participant performed 100 trials and there were a total of 800 trials in the experiments (10 pairs $\times$ 10 repetitions$/$pair $\times$ 8 participants). The location of the samples in each pair was randomized such that each sample was displayed five times on each side (left/right) of the sample holder.

A separate training session was conducted for each participant. Before the training session, the experimenter explained the experimental procedure to the participant. During the training session, all samples were displayed to the participants once and they were asked to explore the surfaces with their index fingers to get familiar with them. The participants washed their hands with soap and rinsed them with water before the actual experiment. One experimenter sat in front of the participant to replace the samples and another one held the time. The communication between the experimenter and the participant was ensured by tapping the passive hand of the participant placed on the table. One tap meant the new set of samples was ready for exploration (i.e. new trial), and two taps meant the 15 seconds time limit was reached for exploring the samples and the participant must make a decision. Participants responded verbally by stating \textit{right} or \textit{left}, and their voices were recorded using a microphone for post-processing of the responses.

\subsection{Measurement of Coefficient of Friction}
The goal of the experiment was to measure the CoF between finger and the samples. This was achieved by dividing the measured tangential force to the normal force. 

\subsubsection{Participants}\label{sec:ch6_FrictionParticipants}
Seven subjects (two females and seven males; mean age = 24.71, SD = 3.14) participated in the friction experiments. 

\subsubsection{Apparatus}\label{sec:ch6_FrictionApparatus}
We utilized the set-up presented in Section \ref{sec:ch3_ExperimentalMethods} for all friction measurements.

\subsubsection{Procedure}\label{sec:ch6_FrictionProcedure}
Each participant completed the experiment in one session. They were instructed to put their index finger inside the hand support to keep it stationary. During the experiments, the sample placed under their finger was moved in tangential direction and the normal and tangential forces acting on the finger were measured. Participants were asked to minimize their body movement during the measurements as much as possible to reduce any possible noise in the data. Using a micropipette, 1 $\mu$L liquid vaseline was injected on the surface of the participants' fingertip to eliminate the stick-to-slip behavior. Before collecting data, the sample surface was moved back and forth ten times beneath the finger under a constant normal force and sliding velocity to establish a homogeneous sliding path. The experiment was repeated three times for each sample.

\subsection{Calculation of Surface Free Energy}\label{sec:ch6_SurfaceFreeEnergy}
\subsubsection{Concept}\label{sec:ch6_SurfaceFreeEnergyConcept}
Surface free energy determines how materials adhere to each other \cite{little2006final}. Adhesive forces are generated at the interface due to the molecular interactions between surfaces. For example, a strong adhesive force tends a liquid to spread over a smooth surface making a small angle of contact. 

According to the acid-base theory, nonpolar and polar components constitute the total surface free energy of any material. The nonpolar component (Lifshitz-van der Waals) is denoted by $\gamma^{LW}$ and the polar component has two sub-components: Lewis acid and Lewis base denoted by $\gamma^{+}$ and $\gamma^{-}$, respectively. Hence, the total surface free energy of any material is calculated using the following equation \cite{al2017evaluation}:
\begin{equation}\label{eq:total surface energy}
\gamma^{Total} = \gamma^{LW} + 2\sqrt{\gamma^{+} \gamma^{-}}
\end{equation}
If the surface free energy components of a liquid (L) and a smooth surface (S) in contact are known in advance, the work required to separate them (i.e. adhesive work) can be calculated as:
\begin{equation}\label{eq:WAB}
    W_{LS} = 2 \sqrt{\gamma_L^{LW} \gamma_S^{LW}} + 2 \sqrt{\gamma_L^{+} \gamma_S^{-}} + 2 \sqrt{\gamma_L^{-} \gamma_S^{+}}
\end{equation}
On the other hand, if the contact angle ($\theta$) between a liquid and a smooth surface is known, the adhesive work between them can also be estimated by:
\begin{equation}\label{eq:WLS}
    W_{LS} = \gamma_L \left(1+\cos{\theta}\right)
\end{equation}
Hence, the following equality is obtained:
\begin{equation}\label{eq:nonlinear surface energy equation}
    \gamma_L \left(1+\cos{\theta}\right) = 2 \sqrt{\gamma_S^{LW} \gamma_L^{LW}} + 2 \sqrt{\gamma_S^{+} \gamma_L^{-}} + 2 \sqrt{\gamma_S^{-} \gamma_L^{+}}
\end{equation}

Little and Bhasin \cite{little2006final} suggested that three unknown components of the surface free energy for any solid surface ($\gamma_S^{LW}$, $\gamma_S^{+}$, and $\gamma_S^{-}$) can be calculated by measuring the contact angles between the surface and at least three liquids with known surface free energy components and then solving for Eq. \ref{eq:nonlinear surface energy equation}.

\subsubsection{Measurement of Contact Angle}\label{sec:ch6_SurfaceFreeEnergyProcedure}
\begin{figure*}[t!]
\centering
\includegraphics[width=0.82\linewidth]{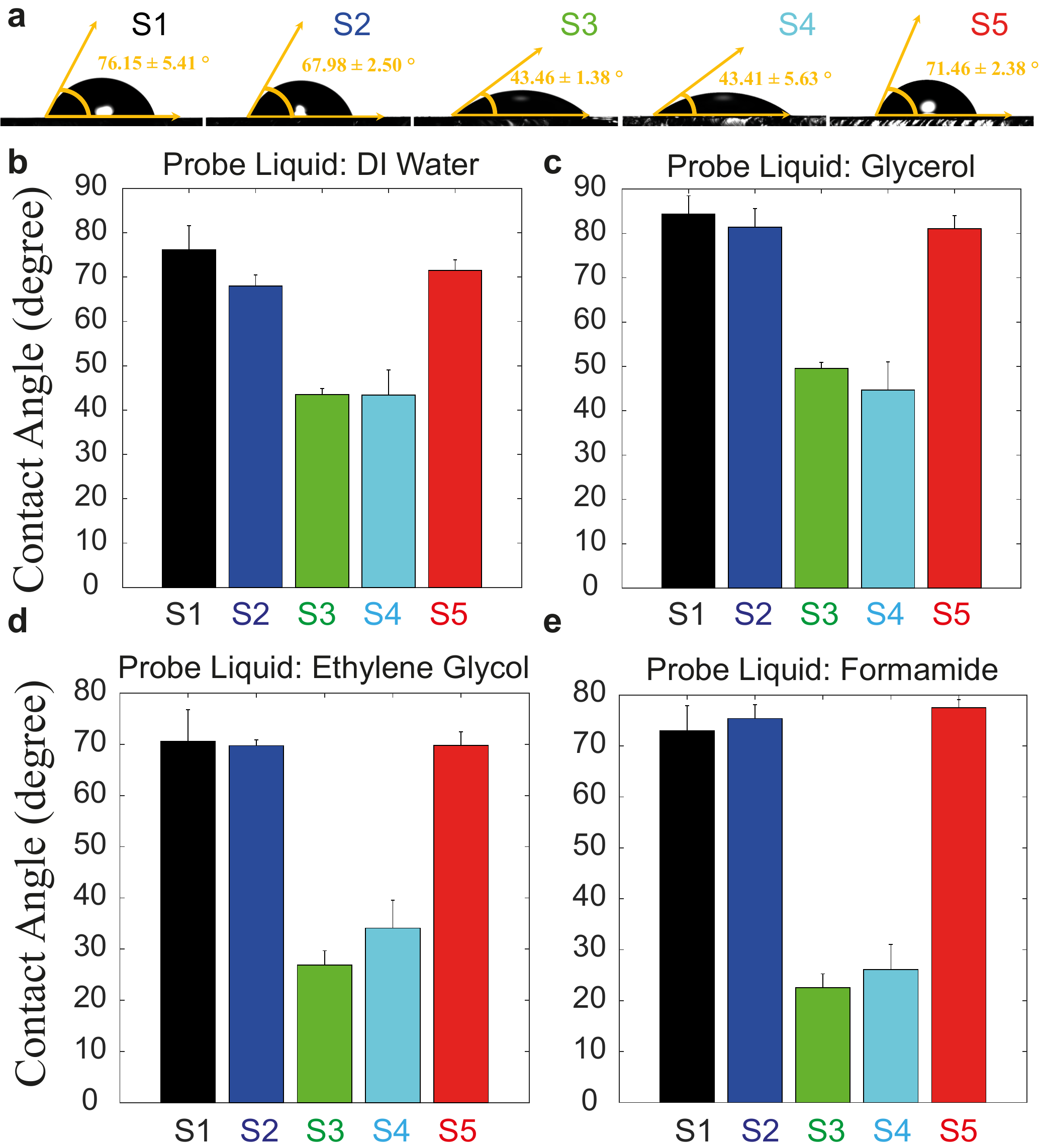}
\caption{a) Images of DI water dropped on the sample surfaces S1-S5 showing their wettability. Measured static contact angles between the sample surfaces and a) DI water, b) Glycerol, c) Ethylene Glycol, and d) Formamide.}
\centering
\label{fig:ch6_StaticContactAngle}
\end{figure*}

The static contact angles between the sample surfaces and 4 different liquids were measured by the Sessile drop test. Using a contact angle meter (Attension ThetaLite TL101-Auto1, Biolin Scientific Inc.), we put a droplet of liquid on each sample surface and then measured the contact angle via image processing techniques. For this purpose, we utilized 4 probe liquids: DI water, Glycerol, Ethylene Glycol, and Formamide. The surface free energy components of the selected probe liquids are tabulated in Table \ref{table:ch6_SurfaceEnergyProbeLiquids}. Fig. \ref{fig:ch6_StaticContactAngle}a shows the images of a DI water droplet on each sample surface and the contact angles it make with the surfaces.

The dynamic contact angles for a droplet of DI water were measured by inflating and deflating the drop via a micropipette \cite{butt2022contact}. A droplet of DI water (5 $\mu$L) was dispensed on each sample surface and the volume of the droplet was increased until the maximum advancing contact angle was reached; followed by a subsequent decrease in volume until the droplet disappeared. Throughout this process, images of the advancing and receding contact angles were captured by the video camera of the contact angle meter.

\begin{table}[!b]
\renewcommand{\arraystretch}{1.3}
\caption{Surface free energy components of the probe liquids utilized in contact angle measurements. The values are taken from \cite{al2017evaluation}.}
\label{table:ch6_SurfaceEnergyProbeLiquids}
\centering
\begin{tabular}{|l|l|l|l|c|}
\hline
\textbf{Liquid}&\textbf{$\gamma^{LW}$}&\textbf{$\gamma^+$}&\textbf{$\gamma^-$}&\textbf{$\gamma^{Total}$}\\ \hline\hline
DI Water&21.80&25.50&25.50&72.80\\ \hline
Glycerol&34.00&3.92&57.40&64.00\\ \hline
Ethylene Glycol&29.00&1.92&47.00&47.99\\ \hline
Formamide&39.00&2.28&39.60&58.00\\
\hline
\end{tabular}
\end{table}

\subsection{Measurement of Surface Roughness}\label{sec:ch6_SurfaceRoughnessMeasurement}\label{surface roughness}
The surface roughness of each sample was measured using an AFM having a resolution of 0.1 nm along the direction of surface normal (Dimension Icon SPM, Bruker Inc.).

\section{Results}\label{sec:ch6_Results}
\subsection{Tactile Perception Experiments}
\begin{figure*}[b!]
\centering
\includegraphics[width=0.6\linewidth]{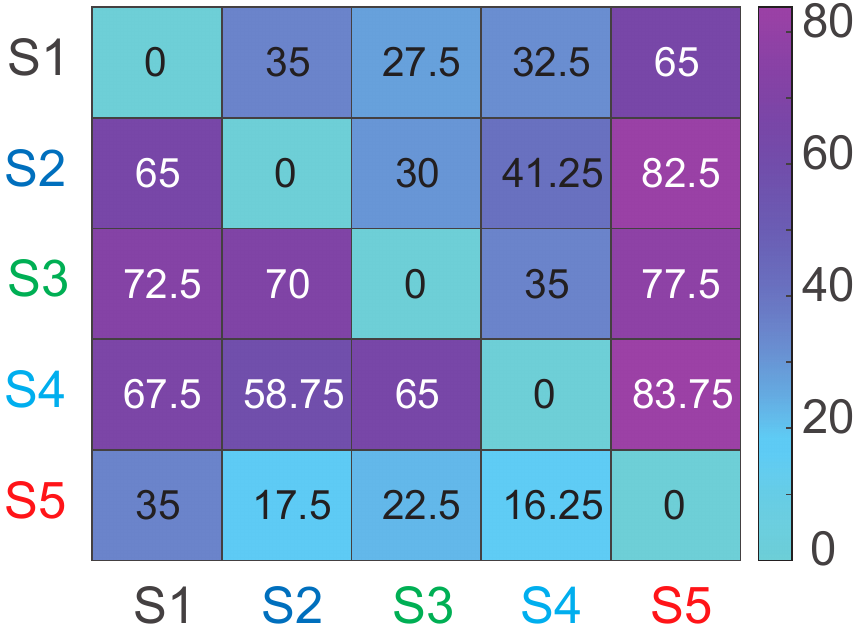}
\caption{The confusion matrix based on the averaged responses of participants; each row represents the percentage of a particular sample as being felt more resistant to sliding compared to the others.}
\centering
\label{fig:ch6_PerceptionResults}
\end{figure*}

We recorded the responses of all participants and generated a confusion matrix based on the average of those responses (see Fig. \ref{fig:ch6_PerceptionResults}). Each row of the confusion matrix shows the percentage of a particular sample as being felt more resistant to sliding compared to the others. According to this matrix, the samples can be ranked based on their resistivity to sliding as S5 (the least), S1, S2, S3, S4 (the most).

\subsection{Measurement of Coefficient of Friction}
\begin{figure*}[t!]
\centering
\includegraphics[width=1\linewidth]{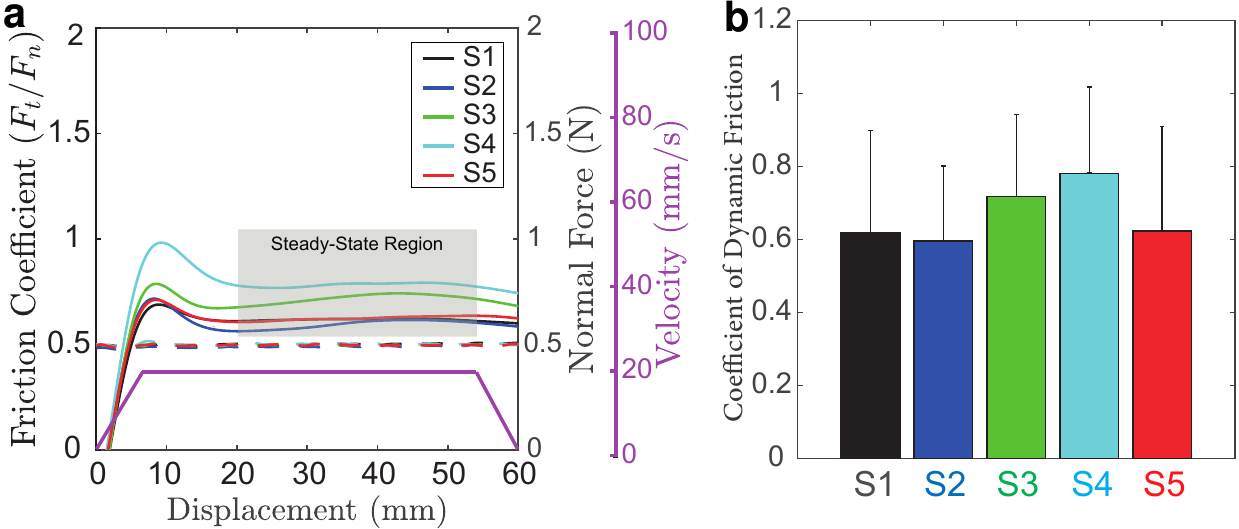}
\caption{The results of our friction experiments; a) Coefficient of friction (CoF), normal force, and the velocity profile of the horizontal stage as a function of relative displacement between finger and the sample surface. b) Mean steady-state values of dynamic CoF with standard error of means.}
\centering
\label{fig:ch6_FrictionResults}
\end{figure*}

The CoF curve reported in Fig. \ref{fig:ch6_FrictionResults}a are the mean values of 21 trials recorded for each sample (3 trials$/$participant $\times$ 7 participants). The same figure also shows the average normal force for each sample and the horizontal stage's velocity profile as a function of displacement. The steady-state region for all CoF curves was taken as the interval from 20 mm to 50 mm and the mean values of dynamic CoF together with their standard error of means are presented in Fig. \ref{fig:ch6_FrictionResults}b. One-way ANOVA showed significant effect of coating material (G1, G2, G3) on dynamic CoF ($F(2, 102) = 4.34$, $p = 0.015$). Post hoc comparisons using the Tukey HSD test showed that G2 is significantly higher than G1 in terms of dynamic CoF ($p < 0.017$).

\subsection{Measurement of Contact Angle and Calculation of Surface Free Energy}
For each sample, we measured the static contact angles (between the sample surfaces and the probe liquids) at three different locations on the surface. The measured contact angle for each location is the arithmetic mean of the angles formed on the right and left-hand sides of the droplet. The measured contact angles between the surfaces of samples and the four probe liquids are presented in Fig. \ref{fig:ch6_StaticContactAngle}b, c, d, and e.

\begin{figure*}[t!]
\centering
\includegraphics[width=0.6\linewidth]{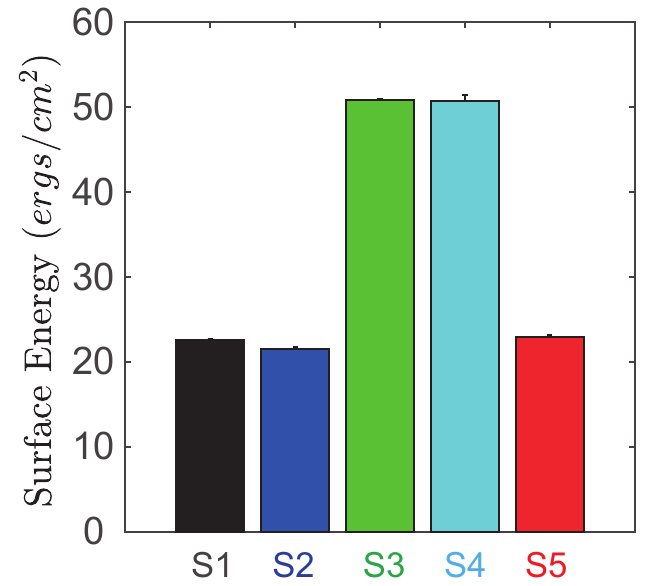}
\caption{The surface free energy of the samples calculated from the measured static contact angles.}
\centering
\label{fig:ch6_SurfaceEnergy}
\end{figure*}

Using the measured contact angles, the surface free energy components of the 4 liquids tabulated in Table \ref{table:ch6_SurfaceEnergyProbeLiquids}, and Eq.\ref{eq:nonlinear surface energy equation}, we estimated the surface free energy of the samples. For this purpose, we first solved a set of 4 nonlinear and over-determined equations in order to calculate the components of the surface free energy for each sample. Then, the total surface energy for each sample was calculated by Eq. \ref{eq:total surface energy} and the results are reported in Fig. \ref{fig:ch6_SurfaceEnergy}. One-way ANOVA showed significant effect of coating material on surface free energy ($F(2, 12) = 5549.57$, $p < 0.001$). Post hoc comparisons using the Tukey HSD test showed that G2 is significantly higher than G1 ($p < 0.001$) and G3 ($p < 0.001$).

\begin{figure*}[t!]
\centering
\includegraphics[width=1\linewidth]{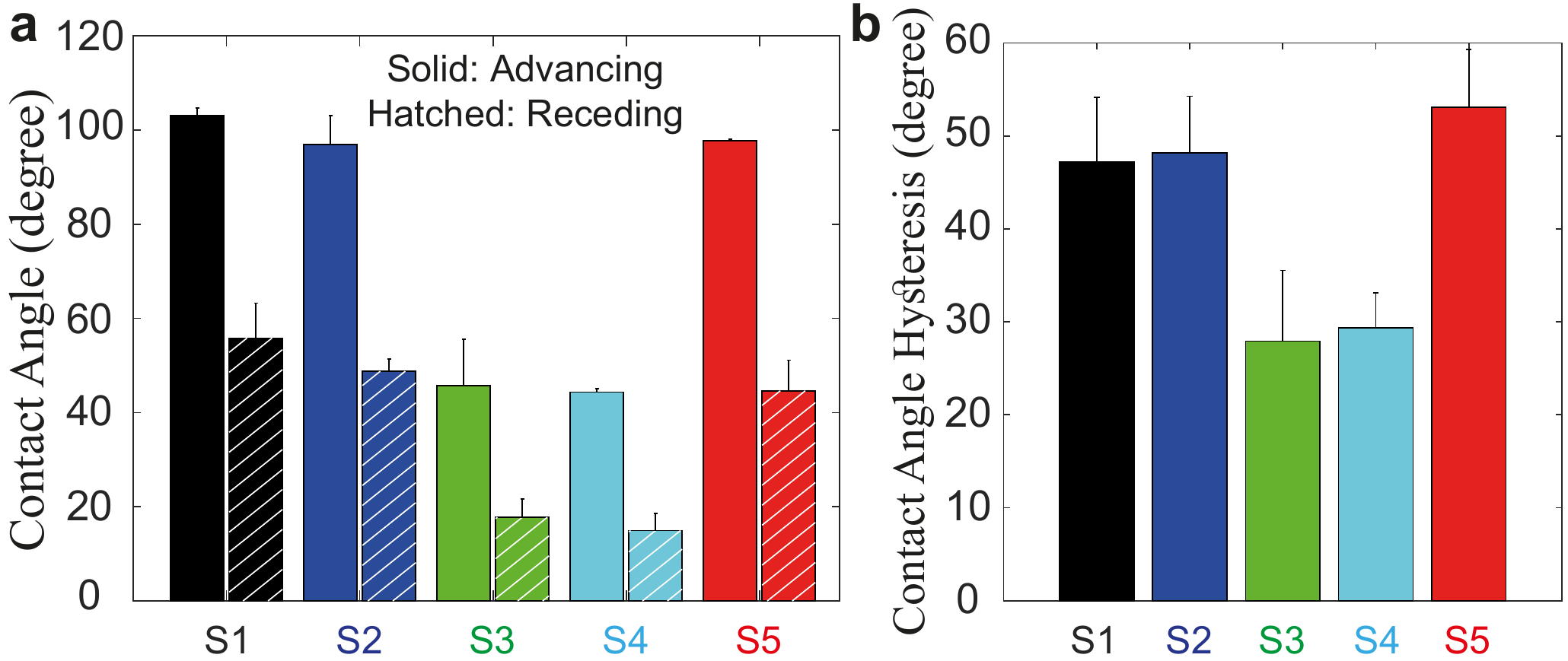}
\caption{a) The advancing and receding contact angles of the samples for DI water, and b) the contact angle hysteresis of the samples, which was calculated as the difference between the advancing and receding contact angles.}
\centering
\label{fig:ch6_DynamicContactAngle}
\end{figure*}

The contact angle hysteresis was calculated as the difference between the maximum advancing and minimum receding contact angles. The measurements were repeated 3 times and the mean values with their corresponding standard deviations for the advancing and receding contact angles and the contact angle hysteresis are presented in Fig. \ref{fig:ch6_DynamicContactAngle}a and \ref{fig:ch6_DynamicContactAngle}b, respectively. One-way ANOVA showed significant effect of coating material on contact angle hysteresis ($F(2, 12) = 24.62$, $p < 0.001$). Post hoc comparisons using the Tukey HSD test showed that G2 is significantly lower than G1 ($p < 0.001$) and G3 ($p < 0.001$)

\subsection{Measurement of Surface Roughness}
\begin{figure*}[!t]
\centering
\includegraphics[width=1\linewidth]{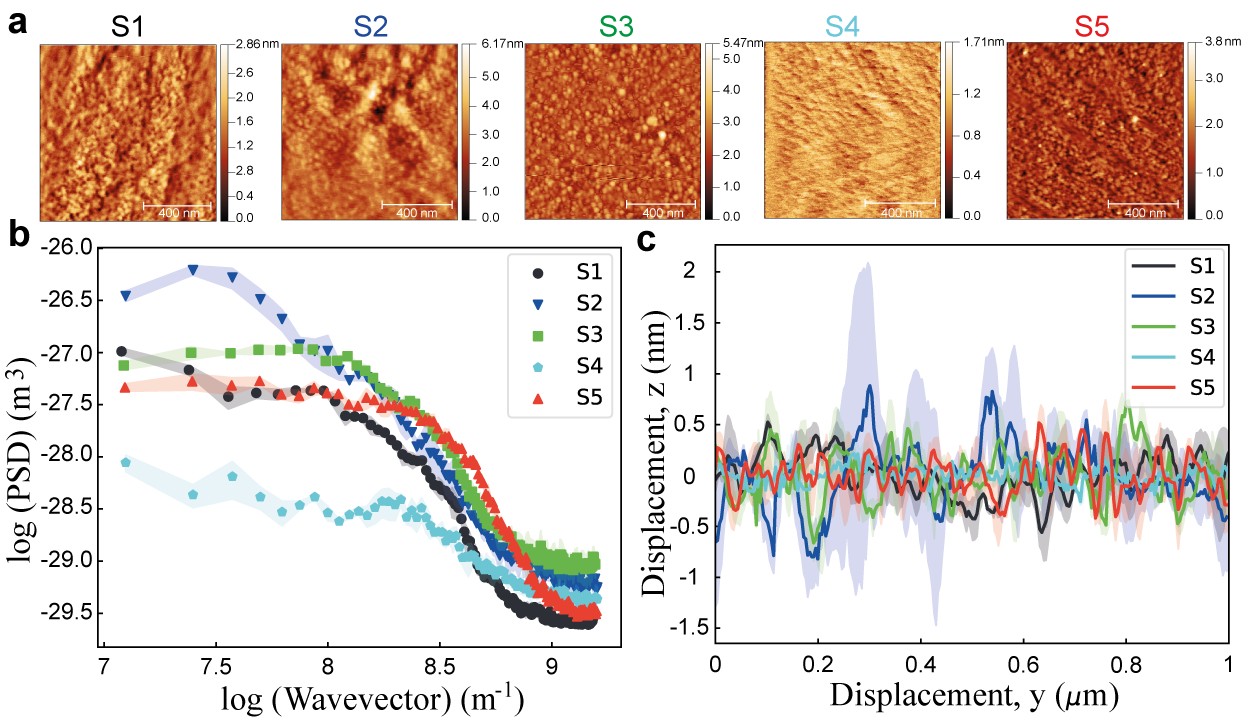}
\caption{Results of AFM measurements performed at two sites on each sample surface. a) Topographic 2D images, b) average 1D vertical power spectrum densities with their corresponding standard deviations (shaded areas), and c) average line scans with their standard deviations (shaded areas).}
\label{fig:ch6_SurfaceRoughness}
\end{figure*}

We measured the surface roughness of sample surfaces by AFM in tapping mode at two different sites on the surface for an area of 1 $\mu$m $\times$ 1 $\mu$m. The sample surfaces were transferred directly from the coating lab to the measurement room inside a dust-free sample box. Special care was given to keep the surfaces clean and untouched. The results of the AFM measurements are tabulated in Table \ref{table:ch6_SurfaceRoughness}. These results are the mean values and standard deviations of 2 measurements for each sample. Fig. \ref{fig:ch6_SurfaceRoughness} presents the topographic 2D images (a), the average 1D vertical power spectrum densities (b), and the average line scans (c) of the sample surfaces.

\section{Discussion}\label{sec:ch6_Discussion}
The results of our tactile perception experiments show that the participants can successfully discriminate the samples based on the tactile cues alone. According to the results reported in Fig. \ref{fig:ch6_PerceptionResults}, we conclude that the sample surface S5 (G3) felt least resistive to the sliding finger of participants, followed by the samples in group G1 (S1 and S2). In general, the samples in group G2 (S3 and S4) are felt most resistive to the sliding finger. Since all the surfaces in our study are extremely smooth with surface roughness magnitudes in the order of a few nanometers as measured by AFM and there is no clear correlation between the AFM metrics (see Table. \ref{table:ch6_SurfaceRoughness}) and the results of our tactile perception study, the tactile discrimination ability of the participants is more likely due to the surface chemistry and not the surface topography. Our results are in line with those of the earlier studies suggesting surface chemistry could play a role in tactile perception of smooth surfaces (\cite{skedung2018feeling, carpenter2018human}).

\begin{table}[!t]
	\centering
	\caption{Results of surface roughness measurements for the samples used in this study (S$_q$: RMS Roughness, S$_a$: Average Roughness, S$_{sk}$: Skewness, S$_{ku}$: Kurtosis, S$_z$: Maximum Height)}
	\label{table:ch6_SurfaceRoughness}
	\begin{tabular}{|l|l|l|l|l|l|l|c|}
		\hline
		\textbf{Parameter}&\textbf{S1}&\textbf{S2}&\textbf{S3}&\textbf{S4}&\textbf{S5}&\textbf{Unit}\\ \hline\hline
		\textbf{S$_q$}&0.33$\pm$0.02&0.60$\pm$0.04&0.46$\pm$0.02&0.15$\pm$0.02&0.37$\pm$0.03&$nm$\\ \hline
		\textbf{S$_a$}&0.26$\pm$0.01&0.44$\pm$0.05&0.36$\pm$0.02&0.11$\pm$0.01&0.29$\pm$0.02&$nm$\\ \hline
        \textbf{S$_{sk}$}&0.02$\pm$0.04&0.48$\pm$0.40&0.57$\pm$0.12&-0.26$\pm$0.03&0.14$\pm$0.07&-\\ \hline
        \textbf{S$_{ku}$}&3.10$\pm$0.002&5.86$\pm$2.34&4.09$\pm$0.46&3.48$\pm$0.06&3.70$\pm$0.30&-\\ \hline
        \textbf{S$_z$}&2.83$\pm$0.03&7.07$\pm$1.27&5.45$\pm$0.03&1.60$\pm$0.15&3.91$\pm$0.15&$nm$\\ \hline
	\end{tabular}
\end{table}

Our physical measurements performed with the same samples support the results of our tactile perception experiments. Our friction measurements reveal that the samples in group G2 (S3 and S4) have higher CoF compared to those in group G1 (S1 and S2) and group G3 (S5), while the differences between G1 and G3 are not significant. However, we should mention that a small amount of vaseline was applied to the surfaces and the interface was in mixed boundary lubrication, and hence, dry friction models might not be sufficient to describe the contact mechanics between the finger and surfaces. Similarly, the results of the Sessile drop tests show that the samples in G2 (S3 and S4) make significantly lower contact angles with the 4 liquids used in the measurements, indicating higher surface energies. Again, the differences between G1 (S1 and S2) and G3 (S5) are not significant. The results also show that the contact angle hysteresis for the samples in G1 and G3 are higher than those in G2. This result suggests that either the surface roughness is higher or the surface chemistry is inhomogeneous for the samples in G1 and G3, compared to G2. Since the surface roughness of our samples is very low (in the order of a few nanometers), it is more likely that inhomogeneous surface chemistry affects the tactile perception of the participants.

All these results suggest that the type of coating material used at the top layer of the surfaces has an influence on our tactile perception (though the coating layer below the top one does not seem to have any influence). In this regard, the sample with Zirconium Oxide coating (S5) appears to be the least resistive to sliding finger, followed by the samples with Titanium Oxide coatings (S1, S2). The samples with Silicon-based coatings (S3, S4) are the most resistive.

%////////////////////////////////////////////////////////////////////////////////////////////////////////////////////////////////////////////////////////////////////////////////////////////////////

\chapter{Conclusions and Future Directions}\label{chapter:ConclusionFutureDirections}
\section{Conclusions}
Due to its unique features, electroadhesıon technology has already been utilized in many applications in different domains so far. However, our understanding of the physics behind it is still highly limited. Although controlling the magnitude of attractive electrostatic force that holds two surfaces together is critical in many applications of electroadhesion, the effects of the input voltage signal (amplitude and frequency) and several other internal (e.g., electrical and mechanical properties of the contacting surfaces) and external (e.g., humidity and temperature) factors on this force are not fully known yet. Furthermore, since the contacting surfaces have finite roughness, there is an air gap between them, which varies from a few nanometers to a few micrometers across the contact area and makes it difficult to measure the electrostatic forces directly.

To understand the physics of electrical and mechanical interactions between human finger and touchscreen under electroadhesion, we investigated the frequency-dependent frictional response of human finger under electroadhesion by focusing on the transfer of induced charges between finger and touchscreen in Chapter \ref{chapter:FrequencyDependentElectroadhesion}. In particular, the electrostatic force between finger and touchscreen was measured and modeled for frequencies ranging from 1 to 10$^6$ Hz using the principles of electric fields and Persson's contact mechanics theory. In the case of DC input, the induced charges accumulate at the interfaces of SiO$_2$-air and air-SC and some portion of them drifts to the surface of touchscreen. However, in the AC case, since the polarity of the input voltage signal alternates frequently at high frequencies of stimulation, the charges are not able to gather at the interfaces and hence, there is no leakage after approximately 250 Hz. On the other hand, more charges are able to leak to the surface of the touchscreen at lower frequencies, and as a result, the electric field and hence the electrostatic force decrease (Fig. \ref{fig:ch3_ElectrostaticForce}). Our experimental data and modeling results reveal that the electrostatic force shows an inverted parabolic behavior where the force increases with increasing frequency until 250 Hz and decreases afterward. We suggest that lower electrostatic attraction force at low and high frequencies is due to the charge leakage and the frequency-dependent electrical properties of the SC, respectively. Interestingly, we observed a relatively high value for the electrostatic force at approximately 100 kHz in measurements and our model captured this behavior surprisingly well. Further investigation revealed that the dielectric loss of the SC at this frequency has the lowest value (see purple curve in Fig. \ref{fig:ch3_ElectrostaticForce}a). In other words, the ability of the SC to store electrical potential energy becomes maximum at approximately 100 kHz and drops afterward.

In Chapter \ref{chapter:ExperimentalElectrostaticForcesFromImpedance}, we proposed a systematic approach based on the measurement of electrical impedances to experimentally infer the air gap thickness first and then the magnitude of electrostatic forces between the finger and the touchscreen. We measured the electrical impedances of finger skin and touchscreen and subtracted them from the total sliding impedance to obtain the remaining impedance. Then, we extracted the air gap impedance from the remaining impedance by removing the undesired effects of electrode polarization impedance. We showed that the electrode polarization impedance is effective at low frequencies, and it is not possible to estimate the electrostatic forces correctly without filtering out its effect from the remaining impedance. Although we demonstrated the application of our proposed approach in contact interactions of the human finger with a tactile surface, we should emphasize that electrode polarization is not exclusive to the electrical measurements performed on biological structures only, such as human skin in our application. It can also occur in all other electroadhesıon applications involving two contacting surfaces with different electrical properties. 

Our electrical impedance measurements revealed new information about the physics of finger interactions with a tactile surface under electroadhesion, which has not been published before. In Chapter \ref{chapter:FrequencyDependentElectroadhesion}, we claimed that the charge leakage from the finger to the surface of the touchscreen is the main cause of the reduction in electrostatic forces for frequencies below 30 Hz. Chapter \ref{chapter:ExperimentalElectrostaticForcesFromImpedance} provides experimental evidence for this claim. As shown in Fig. \ref{fig:ch4_ElectrostaticForce}, the electrostatic forces inferred by the impedance measurements are low (high) when the charge leakage is taken (not) into account. We argue that an EDL builds up at the contact interface of the finger surface (Fig. \ref{fig:ch4_Introduction}c) and causes the electrons to leak to the surface of the touchscreen. Furthermore, our analysis based on magnitude ratio (MR) and phase synchronization (PPS) shows that finger skin (touchscreen) contributes most to the total impedance at 250 Hz (10 kHz) frequencies (Fig. \ref{fig:ch4_FurtherAnalysis}c and d). Surprisingly, the electrostatic forces inferred from the friction measurements also attain peak values at those frequencies (see the black-colored solid lines in Fig. \ref{fig:ch4_ElectrostaticForce}). This suggests that not only the electrical properties of the finger skin but also those of the insulator layer of the touchscreen play a role in the magnitude of electrostatic forces. Although the earlier modeling studies\cite{vezzoli2014electrovibration, vardar2016effect, forsbach2021rigorous} took into account the frequency-dependent permittivity and resistivity of SC, constant values were considered for the insulator layer of touchscreen (SiO$_2$). However, our impedance measurements performed on the touchscreen showed otherwise (see Fig. \ref{fig:ch4_ResultsTouchscreen}).

We investigated the effect of input signal type (DC vs. AC) and moisture on our tactile perception in Chapter \ref{chapter:RoleMoistureElectroadhesion}. The detection threshold for tactile stimuli under the AC condition was found to be significantly lower than that of the DC condition for all participants. The FFT analysis of the experimental data and the proposed circuit model in Chapter \ref{chapter:ExperimentalElectrostaticForcesFromImpedance} provided a deeper understanding of the physical mechanisms behind the difference in tactile perception of DC vs. AC stimulation. The model shows that the interface becomes purely resistive and cannot successfully store electrical energy as the frequency approaches zero (DC), resulting in a weak electrical field and electrostatic forces. Moreover, we observed that moisture has an adverse effect on the tactile perception of electroadhesion. This result is in line with our earlier experimental study \cite{sirin2019fingerpad}, which showed that the relative increase in the CoF is smaller for higher levels of fingertip skin moisture. The earlier studies in the literature have already reported that finger moisture affects the contact dynamics under tangential loading even when there is no electroadhesion \cite{andre2010fingertip,andre2011effect}. Our measurements further showed that the magnitude of electrical impedance drops significantly when sweat exists at the interface between the finger and the touchscreen (see Fig. \ref{fig:ch5_ImpedanceResults}).

Human tactile perception of a touch surface under electroadhesion is governed by the sensation of friction due to the surface itself and the additional friction due to electroadhesion. Although human tactile perception of electroadhesion was investigated in a few studies \cite{bau2010teslatouch,vardar2016effect}, there are not any studies reporting the effect of insulating material of a touchscreen on our tactile perception with and without electroadhesion. Within the time frame of this thesis, we focused on the latter (no electroadhesion). For this purpose, in Chapter \ref{chapter:TactilePerceptionTouchscreen}, we produced five samples with surface roughness below 1 nm using coating techniques and studied human tactile perception of those surfaces. The results of our study showed that human finger is an exceptionally good sensor capable of detecting contact forces at nanometer scale to discriminate surface chemistry. Although a few earlier studies, similar to ours, have already investigated human tactile perception of extremely smooth surfaces having an average roughness at nanometer levels, the contact mechanism underlying our ability to discriminate these surfaces is still not fully understood. Moreover, the literature is missing physical measurements supporting human tactile perception studies to reveal the details behind this discrimination ability. For example, although it is known that adhesion plays a major role in our tactile interactions with smooth surfaces, we still do not know how much each adhesive force component (such as van der Waals, electrostatic, hydrogen bonding) contributes to the sliding friction between the finger and a smooth surface coated with a certain type of material. Furthermore, finger moisture, humidity and temperature of the air, and surface treatment applied to the coated surface to make it hydrophobic and/or oleophobic are the additional factors that contribute to the contact interactions between finger and a touchscreen, which require a systematic and interdisciplinary approach involving tribology, psychophysics, and material science to tackle them. We believe that our work constitutes the initial steps towards this aim.

\section{Future Directions}
The electro-mechanical model proposed in this thesis does not take into account the effect of capillary bridges. It influences the thickness of the air gap, real contact area, and electrostatic force. In fact, it varies during sliding due to the multi-scale roughness of the human fingerpad. A systematic study is required to investigate the effect of moisture and lubrication on the contact mechanics between human finger and touchscreen with and without electroadhesion. This study can also be extended to investigate the effect of the touchscreen's insulating layer and additional processes applied to the top layer to make it more hydrophobic/oleophobic on electroadhesion.  

In 2016, Abie et al. \cite{abie2016universality} conducted an investigation into the potential presence of scaling properties in the AC conductance of human hair. This was accomplished through the utilization of electrical impedance spectroscopy measurements and the subsequent scaling of the measurement outcomes. The conductance-frequency curves were standardized, with each measured value being divided by the DC conductance, and the frequency being normalized with respect to both the DC conductance and the prevailing temperature or relative humidity. The findings suggest the potential existence of the universality nature in the AC conductance of human hair. Furthermore, it is plausible that this same universality nature extends to the human skin, which could be studied in the context of finger-touchscreen applications to scale the results based on varying levels of temperature and humidity.

% //////////////////////////////////////////////////////////////////////////////////////////////////////////////////////////////////////////////////////////
% References don't have to be double spaced either
%\setstretch{1.2}
\bibliographystyle{apalike}
\bibliography{references}

\appendix
\chapter{Charge Continuity Equations for Interfaces}\label{appendix:charge continuity}

\begin{figure}[b!]
\centering
\includegraphics[width=0.8\linewidth]{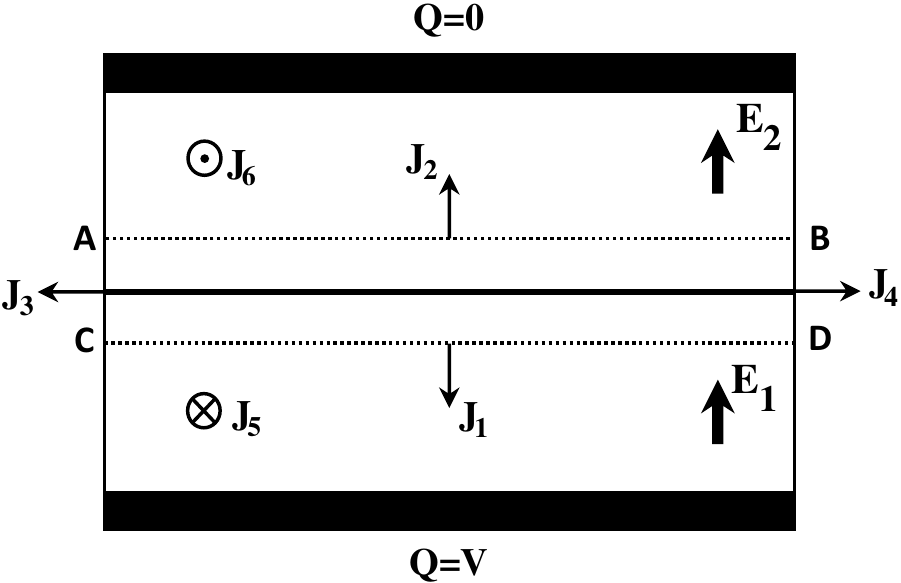}
\caption{Schematic representation of two solids in contact with each other and their interface enclosed by the volume of A-B-C-D.}
\label{fig:Appendix_TwoSolids}
\end{figure}

Let's consider that two solids are in contact as shown in Fig. \ref{fig:Appendix_TwoSolids} and there is a potential difference between them. We also consider a volume enclosed by the surfaces A-B-C-D as shown in the figure. For simplicity, we assume that the areas of the opposite sides of the volume are equal to each other, and define $S_1$, $S_2$, and $S_3$ as the areas of the top and bottom surfaces, the areas of the left and right surfaces, and the areas of the front and back surfaces, respectively. According to the charge conservation law \cite{cheng1989field}, the amount of charge entering our specified volume must be equal to the amount of charge leaving it. In this case, the integral form of the charge continuity equation can be written as:
\begin{equation}
    \oint J.\,dS=J_1S_1+J_2S_1+J_3S_2+J_4S_2+J_5S_3+J_6S_3=-\frac{d\rho_v}{dt}
\end{equation}
where, $\rho_v$ is the charge enclosed by the volume and J$_i$ is the current density leaving from each side of the volume (see the figure). By reducing the space between the top and the bottom surfaces to zero, the integration over the left, right, front, and back sides of the volume becomes zero and the enclosed charge $\rho_v$ simply becomes the surface charge between the solids, $\rho_s$, and the charge continuity equation takes the following form:
\begin{equation}
    J_1+J_2=-\frac{d\rho_s}{dt}
\end{equation}
This equation is utilized to express the charge continuity at the interfaces of SiO$_2$-air and air-SC.

\chapter{Charge Densities in Time Domain}\label{appendix:charge densities}
For an input voltage signal as $V(t) = \, V_0\cos(\omega t)$, the time domain solutions of Eqs. \ref{eqLap1} and \ref{eqLap2} are:
\begin{equation}\begin{split}\label{eq rho1}
    \rho_1(t)&= V_0 \biggl[\alpha_1 \cos(\omega t)+\alpha_2 \sin(\omega t)+\eta {\mathrm{e}}^{-t\biggl(\ddfrac{a}{2}+\ddfrac{e}{2}\biggr)}\mathrm{cosh}(\theta t) \\
    &\qquad +\biggl(\lambda-\eta\biggl(\ddfrac{a}{2}+\ddfrac{e}{2}\biggr)\biggr) {\mathrm{e}}^{-t\biggl(\ddfrac{a}{2}+\ddfrac{e}{2}\biggr)}\ddfrac{\mathrm{sinh}(\theta t)}{\theta} \,
   \Biggr]                   
\end{split}
\end{equation}
\begin{equation}\begin{split}\label{eq rho2}
    \rho_2(t)&= V_0\biggl[\beta_1 \cos(\omega t)+\beta_2\sin(\omega t)+\kappa {\mathrm{e}}^{-t\biggl(\ddfrac{a}{2}+\ddfrac{e}{2}\biggr)}\mathrm{cosh}(\theta t) \\
    & \qquad +\biggl(\delta-\kappa\biggl(\ddfrac{a}{2}+\ddfrac{e}{2}\biggr)\biggr) {\mathrm{e}}^{-t\biggl(\ddfrac{a}{2}+\ddfrac{e}{2}\biggr)}\ddfrac{\mathrm{sinh}(\theta t)}{\theta} \,
    \biggr]
    \end{split}
\end{equation}
where, the coefficients are defined as:
\begin{flalign*}
    & \alpha_1 = -\ddfrac{(a c+b f)\omega^2+a c e^2+b^2 d f-a b e f-b c d e}{\psi} &\\
    & \alpha_2 = -\ddfrac{c\omega^3 +(c e^2-a b f+b c d-b e f)\omega}{\psi} &\\
    & \beta_1 = -\ddfrac{(c d+e f)\omega^2+b c d^2+a^2e f-a b d f-a c d e}{\psi} &\\
    & \beta_2 = -\ddfrac{f \omega^3+(f a^2-a c d+b d f-c d e)\omega}{\psi} &\\
    & \eta = \ddfrac{\left(a c+b f\right)\omega^2+a c e^2+d f b^2-a b e f-b c d e}{\psi} &\\
    & \kappa = \ddfrac{\left(c d+e f\right)\omega^2+b c d^2+e f a^2-a b d f-a c d e}{\psi} &
    \end{flalign*}
    \begin{flalign*}
    & \lambda = \ddfrac{(c a e-c b d)\omega^2+c a e^3-c b d e^2-f a b e^2-f b e a^2}{\psi} &\\
    & +\ddfrac{f a d b^2+c a b d e-c b^2 d^2+f d e b^2}{\psi} &\\
    & \delta = \ddfrac{\left(f a e-f b d\right)\omega^2+f e a^3-f b d a^2-c d e a^2-c a d e^2}{\psi} &\\
    & +\ddfrac{c a b d^2+f a b d e-f b^2 d^2+c b e d^2}{\psi} &\\
    & \theta = \sqrt{\left(\ddfrac{a}{2}-\ddfrac{e}{2}\right)^2+b d} &\\
    & \psi = \omega^4+(a^2+e^2+2 b d)\omega^2+(a e-b d)^2 &
\end{flalign*}
In Fig. \ref{fig:ch3_ACCharges}, we showed that the interface charges become zero at higher frequencies. Here, we perform a limit analysis on time domain charge equations to prove this fact mathematically; as $\omega \xrightarrow[]{} \infty$ all coefficients and hence both $\rho_1$ and $\rho_2$ become zero. It means that the electrical charges do not have enough time to accumulate at the interfaces at higher frequencies and therefore, the net charges become zero.

\chapter{Electrical Contact Conductivity}\label{appendix:electrical contact conductivity}
 When the nominal contact pressure is not too high, the electrical contact conductivity $\alpha$ can be calculated as \cite{persson2018dependency,barber2003bounds}:
\begin{equation}\label{eq alpha}
    \alpha=\ddfrac{2\sigma^*p}{Y^*L_0}
\end{equation}
where, $\sigma^*$ and Y$^*$ are the effective conductivity and the effective elastic modulus, respectively. They are defined as:
\begin{equation}
    \frac{1}{\sigma^*}=\frac{1}{\sigma_1}+\frac{1}{\sigma_2}
\end{equation}
\begin{equation}
    \ddfrac{1}{Y^*} =\frac{1-\nu_1^2}{Y_1}+\frac{1-\nu_2^2}{Y_2}
\end{equation}
where, Y and $\nu$ are the elastic moduli and Poisson's ratios of contacting solids 1 and 2, respectively. 
According to Persson's contact mechanics theory, $L_0$ in Eq. \ref{eq alpha} is a characteristic length parameter and since human skin is self-affine fractal, it can be defined as \cite{persson2010heat}:
\begin{equation}
    L_0\approx \left(\ddfrac{2(1-H)}{\pi H} \right)^{1/2} h_{rms} \left[r(H)-\left(\ddfrac{q_0}{q_1}\right)^H\right]
\end{equation}
where,
\begin{equation}
    r(H)=\ddfrac{H}{2(1-H)}\int_{1}^{\infty} dx \left(x-1\right)^{-1/2}x^{-1/2(1-H)}
\end{equation}
Here, h$_{rms}$ is the rms amplitude of surface roughness for fingerpad, H is the Hurst exponent, x is the integration factor, and q$_0$ and q$_1$ are the long-distance roll-off wavevector and the short-distance cut-off wavevector, respectively.

\chapter{Persson's Contact Mechanics Theory}\label{appendix:persson theory}
In any contact problem, the surfaces make contact only at a few points and at higher magnifications of $\zeta$, those points appear to have partial contact. Hence, multiple levels of magnification are taken into account in Persson's contact mechanics theory and the magnification is defined as $\zeta=\,q/q_0$, where $q$ is the wavevector that varies from q$_L$ (the shortest wavevector) to q$_1$ \cite{persson2001theory,persson2006contact}. In addition, this theory utilizes the surface roughness power spectrum C(q) as a useful mathematical tool in characterizing a surface with different length scales of roughness, which is defined for a self-affine fractal surface as follows \cite{persson2007relation,persson2014fractal}:
\begin{equation}\label{eq Cq}
    C(q) = \ddfrac{H}{\pi}\ddfrac{h_{rms}^2}{q_0^2}\left(\ddfrac{q}{q_0}\right)^{-2\left(1+H\right)}
\end{equation}
The ratio of the real area to apparent contact area at the magnification $\zeta$ is defined as \cite{almqvist2011interfacial,yang2008contact}:
\begin{equation}\label{eq Area}
    \ddfrac{A_{real}(\zeta)}{A_0}=\ddfrac{1}{\left(\pi G\right)^{1/2}} \int_{0}^{p} d\sigma e^{-\sigma^2/4G}=\erf{\left(\ddfrac{p}{2G^{1/2}}\right)}
\end{equation}
where, $\sigma$ denotes stress and $G$ is:
\begin{equation}\label{eq G}
    G(\zeta)=\ddfrac{\pi}{4}{Y}^{\ast^2}\int_{q_0}^{\zeta q_0} dq\,q^3 C(q)
\end{equation}

Now, we consider u$_1(\zeta)$ as the average height separating the surfaces which appear to move out of contact when the magnification increases from $\zeta$ to $\zeta+\Delta \zeta$, where $\Delta \zeta$ is a small infinitesimal change in the magnification, and defined:
\begin{equation}\label{eq u1}
    u_1(\zeta)=\Bar{u}(\zeta)+\Bar{u}^\prime(\zeta)A_{real}(\zeta)/A_{real}^\prime(\zeta)
\end{equation}
Here, $\Bar{u}(\zeta)$ is the average interfacial separation in which all the roughness at the magnification $\zeta$ is considered. It is calculated by Eq. \ref{eq ubar}, where the term $[\gamma+3(1-\gamma ) P^2(q,p^\prime,\zeta)]$ is derived from a correction factor which is introduced in \cite{persson2008elastic}. Almqvist et al. \cite{almqvist2011interfacial} suggest that using $\gamma=0.45$ provides good agreement between experimental and numerical solutions in contact mechanics problems involving elastic contacts.
\begin{equation}\label{eq ubar}
\begin{aligned}
\begin{split}
    \Bar{u}(\zeta)&= \sqrt{\pi}\int_{\zeta q_0}^{q_1} dq\,q^2C(q)w(q)\\
    &\qquad \int_{p(\zeta)}^{\infty} dp^\prime \ddfrac{1}{p^\prime}\left[\gamma+3\left(1-\gamma\right)P^2(q,p^{\prime},\zeta)\right] e^{-\left[w(q,\zeta)p^\prime/Y^*\right]^2}
    \end{split}
    \end{aligned}
\end{equation}
where, $p(\zeta)=\,p A_0/A_{real}(\zeta)$. Defining $s(q)=\,w(q,\zeta)/Y^*$ we have:
\begin{equation}\label{eq P prime}
    P(q,p^{\prime},\zeta) = \ddfrac{2}{\sqrt{\pi}}\int_{0}^{s(q)p} dx e^{-x^2}
\end{equation}
\begin{equation}\label{eq w}
    w(q,\zeta) = \left(\pi \int_{\zeta q_0}^{q} dq^\prime\, {q^\prime}^3 C(q^\prime)\right)^{-1/2}
\end{equation}
Eventually, the probability distribution of interfacial separations is defined as:
\begin{equation}\label{eq separations distribution}
\begin{aligned}
\begin{split}
    P(u)&\approx \ddfrac{1}{A_0}\int_{1}^{\infty} d\zeta \left[-A^\prime(\zeta)\right]\ddfrac{1}{\left(2\pi h_{rms}^2(\zeta)\right)^{1/2}}\\
    &\qquad \left[\exp\left(-\ddfrac{\left(u-u_1(\zeta)\right)^2}{2h_{rms}^2(\zeta)}\right)+\exp\left(-\ddfrac{\left(u+u_1(\zeta)\right)^2}{2h_{rms}^2(\zeta)}\right) \right]
    \end{split}
    \end{aligned}
\end{equation}
where, the root mean square roughness amplitude for any magnification is:
\begin{equation}\label{eq hrms}
    h_{rms}(\zeta) = \left(2\pi \int_{\zeta q_0}^{q_1} dq\, qC(q)\right)^{1/2}
\end{equation}
Note that the prime sign in the above equations is used to denote the derivative.

\chapter{Electro-osmotic and Electrodermal Activity of Human Skin}\label{appendix:electro-osmosis and electrodermal}
Grimnes \cite{grimnes1983skin} applied voltage signals to the ventral forearm, hand dorsal, and upper arm using three different types of electrodes and measured the current passing through the skin. The results showed that the current signal lags the voltage signal and the current passing through the skin for the voltage signal with negative polarity is more than that of the positive polarity. This finding suggested that the conductivity of the skin was high for the voltage signal with negative polarity compared to the positive polarity. He explained this disparity by a phenomenon called “electro-osmosis”. Electro-osmosis is the bulk movement of a liquid inside a capillary in response to an applied electric field across a charged surface \cite{reppert2002frequency}. In the case of Grimnes's experiments with human skin, it corresponds to the movement of sweat inside the sweat ducts. When a voltage signal with a negative polarity is applied to the skin surface, the negative charges attract the positive ions of the EDL, resulting in the bulk movement of sweat towards the pores. Repeating the experiments in different frequencies, Grimnes also showed that the electro-osmosis in human skin was frequency-dependent and its effect decreased with increasing frequency. In our experiments, we also observed the effect of electro-osmosis at low frequencies but not at high frequencies (Figure 3a). The voltage signal with negative polarity applied to the touchscreen resulted in a lower impedance magnitude in the skin measurements than that of the positive polarity due to the increase in skin conductance caused by electro-osmosis.

Electrodermal activity (EDA), also known as skin conductance or galvanic skin response, refers to the changes in the electrical conductance of the skin that occur due to the activation of the sympathetic nervous system. EDA has been widely used in psychophysiological research as a non-invasive measure of sympathetic arousal, emotional states, and stress response \cite{boucsein2012electrodermal}. The electrical conductance of the skin is influenced by the activity of sweat glands, which are innervated by sympathetic nerve fibers. When a person experiences emotional or physiological arousal, such as fear or excitement, the sympathetic nervous system is activated, leading to an increase in sweat gland activity and a subsequent increase in skin conductance. Conversely, when a person is relaxed or calm, the sympathetic nervous system activity decreases, leading to a reduction in skin conductance. The measurement of EDA involves the use of electrodes placed on the surface of the skin, typically on the fingers or palms of the hands. The electrodes are designed to detect the changes in the electrical conductance of the skin, which are amplified and recorded. The recorded EDA signal can then be analyzed to determine the level of sympathetic nervous system activity. The relation between EDA and skin electrical conductance is based on the fact that the electrical conductance of the skin is directly related to the number of active sweat glands. When the sweat glands are activated, they release electrolytes onto the surface of the skin, which increases the skin's electrical conductance. Therefore, an increase in EDA indicates an increase in sweat gland activity, a reliable indicator of sympathetic nervous system activation.

\chapter{Polarization and Dispersion in Dielectrics}\label{sec:Polarization and Dispersion}\label{appendix:polarization and dispersion}
Polarization occurs when an electric field is applied to a dielectric. As the dielectric polarizes, its capacitance increases. At low frequencies of electrical stimulation, more molecules align with the applied electric field, which results in an increase in the energy storage capacity of the dielectric, and hence more capacitive behavior is observed in the measurements as in our study. As the frequency increases, due to the frequent change in the polarity of the input signal, some molecules cannot align well with the applied external electric field, and the capacitance of the dielectric decreases. In a dielectric material, different types of polarization can be observed. Orientational, ionic (also known as atomic polarization), and electronic polarizations are typical examples \cite{kuang1998low}.

An electric field causes a dielectric to polarize and the direction of polarization changes as the field direction alternates. However, it takes some time for dipoles to rotate and charges to travel inside the dielectric, and hence the polarization cannot happen instantly. For example, when the electric field is switched on, the orientational polarization occurs over a certain period of time, known as the relaxation time. As a result, if the electric field alternates at a frequency greater than the time constant of this relaxation, the dipole orientation is unable to keep up with the field (i.e. the polarization direction is unable to remain aligned with the field) and hence this polarization mechanism fails to contribute to the polarization of the dielectric. Similarly, each polarization mechanism contributes to the overall polarization of the dielectric for different ranges of frequencies. The change in polarization of dielectric material as a function of frequency can be observed by the permittivity measurements. Dispersion is manifested in the permittivity plot as the changes in the slope of the curve.

\begin{figure}[t]
   \centering
   \includegraphics[width=1\linewidth]{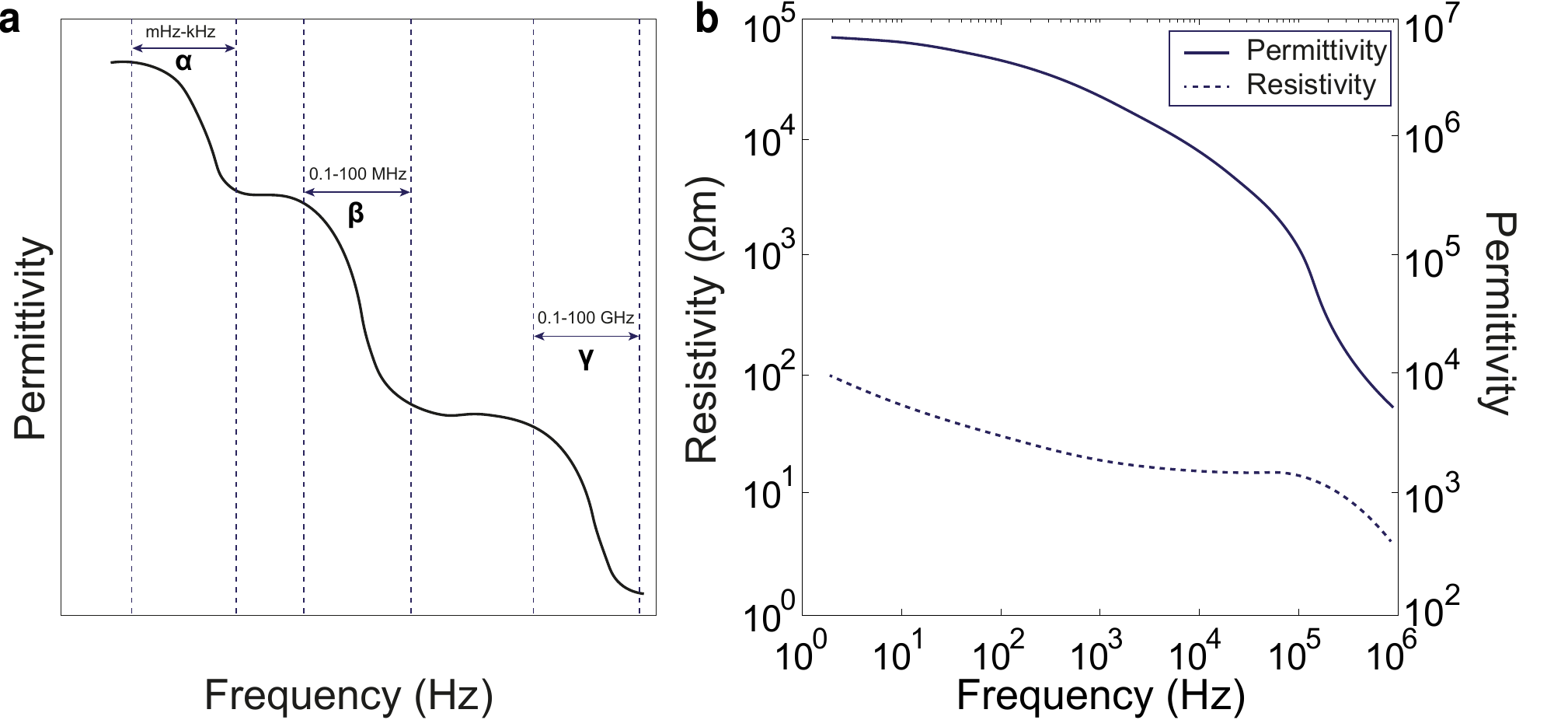}
    \caption{A) Three different types of dispersion are typically observed in permittivity measurements of biological materials (adopted from \cite{schwan1994electrical}). B) Measured electrical properties of the SC as a function of frequency (adopted from \cite{yamamoto1976dielectric}).}
    \label{fig:Appendix_Dielectric Biological}
\end{figure}

Fig. \ref{fig:Appendix_Dielectric Biological}a shows the change in permittivity (and hence the capacitance) of a typical biological material as a function of frequency, in which the $\alpha$, $\beta$, and $\gamma$ dispersions are observed. A relatively weak $\alpha$-dispersion for human skin \cite{pethig1984dielectric} was based on the measured electrical properties of the SC \cite{yamamoto1976dielectric} (Fig. \ref{fig:Appendix_Dielectric Biological}b). The capacitance data for the skin in our measurements also show at least one type of dispersion. Since the $\alpha$ and $\beta$-dispersions occur in biological materials for the frequencies below a few kHz and between a few kHz to several MHz, respectively \cite{kuang1998low}, we observe the end of $\alpha$-dispersion and beginning of $\beta$-dispersion in our measurements.

\vita{Easa AliAbbasi is a Ph.D. candidate at the computational sciences and engineering department of Koc University. He is a member of the Robotics and Mechatronics Laboratory (RML) and currently, his research focuses on surface haptics. He attended University of Oslo for a period of two months as a visiting researcher and learned the basics of electrical impedance and bioimpedance measurements. He successfully completed his master's studies at University of Tabriz with a mechatronics engineering degree. His research interests include haptics, mechatronics, and physics-based modeling.

}
\end{document}